\documentclass[11pt]{article}
\usepackage{graphicx,color}
\usepackage{relsize}
\usepackage{appendix}
\usepackage{latexsym,amsmath,amssymb,graphicx,booktabs}
\usepackage{epsfig,latexsym,cite}
\usepackage{float}
\usepackage{hyperref}
\usepackage{cleveref}
\numberwithin{equation}{section}

\input epsf
\usepackage{amssymb}
\usepackage{amsmath}
\usepackage{amsfonts}
\usepackage{upgreek}
\usepackage{latexsym}

\newcommand{\unit}{\hat{1\!\!1}}

\graphicspath{ {./plots/} }
\begin{document}

\begin{titlepage}

\vspace*{2cm}

\centerline{\Large \bf Spatial correlations of dark energy}

\centerline{\Large \bf from quantum fluctuations in inflation}

\vspace{5mm}

\vskip 0.4cm
\vskip 0.7cm
\centerline{\large Enis Belgacem$^a~\footnote{e-mail: e.belgacem@uu.nl}$, Tomislav Prokopec$^a$~\footnote{e-mail: t.prokopec@uu.nl}}
\vskip 0.3cm
\centerline{\em $^a$Institute for Theoretical Physics and EMME$\Phi$, Faculty of Science, Utrecht University,}
\centerline{\em Princetonplein 5, 3584 CC Utrecht, The Netherlands}

\vskip 1.9cm

\begin{abstract}
\noindent
This paper contains a detailed study of the properties of a simple model attempting to explain dark energy as originated from quantum fluctuations of a light spectator scalar field in inflation. In~\cite{Belgacem:2021ieb} we recently outlined how Starobinsky's stochastic formalism can be used to study the spatial correlations imprinted on dark energy by its quantum origin in this model and we studied their possible role in relieving the Hubble tension. Here we provide a more comprehensive derivation of the results in~\cite{Belgacem:2021ieb} and we refine some of our estimates, comparing to the approximate results obtained previously. 
Among the main results, we analyze the non-coincident correlators predicted by a full field theoretical treatment and their relation with those computed within the stochastic formalism. We find that in the region where stochastic theory predicts significant sub-Hubble correlators it is in disagreement with field theoretical predictions. However, agreement can be restored by introducing a reduced speed of sound for the scalar field.
We also discuss an alternative approach to the problem of studying correlators within the stochastic formalism based directly on the evolution of probability distributions. We find that the two approaches give the same answer for 2-point functions of the field, but not for 4-point functions relevant to density correlators and we discuss the behaviour of the two methods with respect to Wick's theorem.

\end{abstract}

\end{titlepage}

\newpage

\tableofcontents

\newpage

\section{Introduction}
\label{sect:intro}

The discovery of the accelerated expansion of the Universe in 1998 by the Supernova Cosmology Project~\cite{SupernovaCosmologyProject:1998vns} and the High-Z Supernova Search Team~\cite{SupernovaSearchTeam:1998fmf} has triggered the quest for the origin of cosmic acceleration. The simplest explanation, namely a cosmological constant, equivalent to a perfect fluid with equation of state parameter $-1$, has proven to be able to explain a diversified number of observations. Therefore it has been fully enrolled in the standard cosmological model, which also requires cold dark matter, to give the current $\Lambda$CDM model. However, both the theoretical shortcomings (the cosmological constant problem~\cite{Weinberg:1988cp, Padilla:2015aaa, Koksma:2011cq, Lucat:2018slu}) and the tensions between the values of cosmological parameters predicted by probing different ages and scales of the Universe, most notably in $H_0$ (Hubble tension, e.g. between CMB data~\cite{Planck:2018vyg} and supernovae luminosity distance measurements~\cite{Riess:2019cxk}) and in the parameters ($\Omega_M$, $\sigma_8$) affecting galaxy number counts (see~\cite{Douspis:2018xlj}), have stimulated the search for alternatives to $\Lambda$CDM. In such models, the accelerated expansion at late times is not merely due to a cosmological constant, but to a dynamical dark energy. A typical investigation line is given by quintessence models (see~\cite{Tsujikawa:2013fta} for a review), where a field (typically scalar) evolves in such a way that its energy density becomes relevant, and eventually lead the expansion, at late times. This desired behavior can be achieved by the existence of a tracker solution~\cite{Peebles:1987ek, Ratra:1987rm, Zlatev:1998tr}, in which the energy density of the scalar field decreases slightly less rapidly than radiation during the radiation-dominated epoch and, after matter-radiation equality, starts decreasing less rapidly than matter density and eventually behaves like dark energy.
In quintessence models, the potential energy of the scalar field is the essential ingredient for the late-time acceleration, but it is also possible to obtain dark energy from a scalar field with non-quadratic kinetic terms, in the so-called ${\it k}$-essence models \cite{Armendariz-Picon:1999hyi, Armendariz-Picon:2000nqq}.
Quintessence and ${\it k}$-essence models are examples of scalar-tensor theories of gravity with self-accelerated solutions. A larger set of scalar-tensor theories used in dark energy studies is the Horndeski class \cite{Horndeski:1974wa, Charmousis:2011bf, Kobayashi:2011nu}, which is made by the most general covariant scalar-tensor theories with second-order equations of motion.
Examples of theories in the Horndeski class are Brans-Dicke theory~\cite{Brans:1961sx}, galileons~\cite{Nicolis:2008in} and $f(R)$ gravity~\cite{DeFelice:2010aj}.
It has also been shown that a wider class of theories (beyond Horndeski~\cite{Zumalacarregui:2013pma, Gleyzes:2014dya}, in turn included in the class of degenerate higher-order scalar-tensor (DHOST) theories~\cite{Langlois:2015cwa}, see also~\cite{Kobayashi:2019hrl} for a review) can evade Ostrogradsky instabilities~\cite{Ostrogradsky:1850fid}, despite leading to equations of motion of higher order.
Quintessence models (and the more general class of theories mentioned before) attempting to explain dark energy without a cosmological constant do not properly address the {\it initial conditions} issue about the value the quintessence field needs to start from. Furthemore, they usually assume perfect correlations of dark energy on large scales, typically without any justification.

An alternative line of research is based on the possibility that dark energy could emerge at late times from the amplification of inflationary quantum fluctuations inherited in the subsequent phases of the Universe evolution~\cite{Ringeval:2010hf, Glavan:2013mra, Glavan:2014uga}. This was shown to be possible with a non-minimally coupled light scalar field. In such a model, inflation itself, by amplifying infrared quantum fluctuations, provides the natural initial conditions for the evolution of the field which will later backreact on the cosmological expansion, manifesting as dark energy.
\noindent In~\cite{Glavan:2017jye}, based on the result of~\cite{Glavan:2015cut, Glavan:2013mra, Glavan:2014uga} that the quantum backreaction can be largely ascribed to infrared (IR) modes, a suitable application of Starobinsky's stochastic formalism was used to study the time evolution of the energy density and pressure of a light non-minimally coupled spectator scalar field. The conclusions  agree with the field theoretic treatment in~\cite{Glavan:2015cut}, where the renormalized stress-energy tensor was evaluated.
The approach in~\cite{Glavan:2017jye} required evaluation of coincident 2-point IR correlators. Later, it was also shown in~\cite{Belgacem:2021ieb} that the same quantum fluctuations also imply unperfect spatial correlations of dark energy over a length scale determined by the comoving Hubble horizon at the beginning of inflation and that they are significant today for a duration of inflation of the order of $60$ e-foldings. In~\cite{Belgacem:2021ieb}, energy density correlations were expressed in terms of non-coincident 4-point functions, whose evolution was again studied with stochastic formalism. An interesting application of dark energy spatial correlations is the study of the Hubble tension and the same work~\cite{Belgacem:2021ieb} showed a remarkable reduction of the tension from $4.4~\sigma$ in $\Lambda$CDM down to possibly $1~\sigma$ in the simple dark energy model considered. Working out the observational predictions of the model is clearly of fundamental importance to test in and compare it to $\Lambda$CDM. It was found in~\cite{Demianski:2019vmq} that the model is slightly favored with respect to $\Lambda$CDM, although not at a statistically significant level.
More recently, the work in~\cite{inprep:2022lumdist} has studied a class of phenomenological models of dark energy which exhibit spatial correlations and their implications on the observed luminosity distances of supernovae, presenting a discussion on the detectability of such effects. The result is that the ongoing Dark Energy Survey (DES)~\cite{Bernstein:2011zf} is not able to detect the expected signal, but the upcoming LSST survey~\cite{LSSTDarkEnergyScience:2018jkl} can succeed, thanks to its large sky coverage which reduces the effect of cosmic variance.

In Section~\ref{sect:model} we summarize the techniques and slightly extend the results of~\cite{Belgacem:2021ieb} and~\cite{Glavan:2017jye} to compute non-coincident 2-point and 4-point functions within the stochastic approach.
\noindent Starobinsky's stochastic formalism~\cite{Starobinsky:1982ee, Starobinsky:1986fx} manages to describe the dynamics of the infrared (long) modes in terms of classical stochastic equations, where the coupling between long and short modes appears as a noise source. The stochastic formalism has been used in the long-wavelength approach of~\cite{Salopek:1990jq}. Its assumptions have been better understood and the theory was used in the separate universe picture~\cite{Salopek:1990re, Wands:2000dp} and developed into the stochastic $\Delta N$ formalism, see e.g.~\cite{Vennin:2015hra, Prokopec:2019srf, Ando:2020fjm}.
One of its prominent applications is in understanding formation of primordial back-holes \cite{Pattison:2017mbe, Ezquiaga:2019ftu, Figueroa:2021zah}.

It is remarkable that stochastic theory maps the quantum problem into a classical stochastic description. However, this is only possible in certain regimes. In particular, as we will see in Section~\ref{sec: QFT}, when considering non-coincident field correlators, only the behavior for super-Hubble separations is captured by the stochastic formalism, while shorter scales require a full field theoretic treatment.
Stochastic theory has been tested against quantum field theory for a scalar field in an exact de Sitter background, without taking into account the backreaction of the scalar field on the metric, which breaks the exact de Sitter assumption~\cite{Finelli:2008zg, Garbrecht:2014dca}.

One of the main goals of the present paper is to test the stochastic formalism predictions for non-coincident correlators of a spectator field, not only in inflation, but also at later times (radiation and matter epochs).
In Section~\ref{sec: QFT}, we prove that, even though there is agreement at the end of inflation, stochastic formalism and quantum field theory disagree at near-Hubble and sub-Hubble scales in matter-dominated epoch, with stochastic theory predicting a steeper decrease of correlators with distance compared to the full field theoretic answer. This disagreement can be attributed to the neglect of spatial gradients by the stochastic formalism, which play an important role in the post-inflationary evolution of near-Hubble modes.
We propose and show that agreement can be restored by introducing a reduced speed of sound $c_s\leq1$. Such a feature is quite common in Horndeski scalar-tensor theories and it generally appears in the effective field theory of dark energy (cf. eq. (39) of~\cite{Gubitosi:2012hu} or, similarly in the context of inflation, eq. (38) of~\cite{Cheung:2007st}).
Likewise, we can assume the speed of sound $c_s$ as an extra free parameter in the quantum dark energy model considered here.

Section~\ref{sec: QFT} also shows that 4-point functions predicted by the full field theory treatment are related to 2-point functions by Wick's theorem\footnote{Wick's theorem applies to perturbative quantum field theory and is a consequence of the Gaussian nature of the free field appearing in the unperturbed Hamiltonian. This is the case for eqs.~(\ref{eq:action})--(\ref{eq:action FLRW}) as long as the quantum backreaction effects, which make the metric a quantum field, are negligible and $\hat{\Phi}$ is a free scalar field with a time-dependent mass determined by the classical background metric. At late times, due to the backreaction of the field on cosmological expansion, non-Gaussianities are unavoidably generated because the problem cannot be seen anymore as the evolution of a free field in a given FLRW background. Indeed, the quantum nature of the field necessarily produces local fluctuations of the Hubble rate, thus the field and Hubble rate dynamics have to be solved consistently. Dark energy is contained in the expectation value of these quantum effects. This idea will be made more precise in Section~\ref{sect: qbackr}, where a first approximation is applied to refine the matching between the model and the current cosmological parameters.}. On the contrary, the 4-point functions evaluated according to the formulation of stochastic formalism in Section~\ref{sect:model} violate Wick's theorem (more precisely the formulation there only agrees with Wick's theorem in the coincident regime and for large spatial separations, but it fails to reproduce the correct behavior at intermediate scales). Starting from this disagreement, in Section~\ref{sec: FP stochastic} we develop an alternative way to apply the stochastic formalism to extract non-coincident correlations functions, applying the techniques in~\cite{Starobinsky:1994bd} to the non-equilibrium regime. This approach is based on the time evolution of classical probability distributions via Fokker-Planck equations, rather than the coupled equations for IR correlators in Section~\ref{sect:model}. We find that the formulation of stochastic theory in Section~\ref{sec: FP stochastic} agrees with Wick's theorem and, since this was also the outcome of the quantum field theory treatment, we expect that the stochastic theory answer for the 4-point functions in Section~\ref{sec: FP stochastic} is more trustworthy than the corresponding results in Section~\ref{sect:model}.

In Section~\ref{sect: qbackr}, we extend the matching of the model parameters to the usual cosmological parameters $H_0$, $\Omega_M$ and $\Omega_\Lambda$ presented in~\cite{Belgacem:2021ieb} and~\cite{Glavan:2017jye}, by taking into account the quantum backreaction effect of the scalar field on the Hubble rate, which is relevant in the most recent stages of the Universe evolution.

Then, in Section~\ref{sect: Hubble tension}, we discuss how to apply the results from previous sections to the Hubble tension problem. This part of the paper is meant to extend the discussion presented in Section III of~\cite{Belgacem:2021ieb} and to go beyond the approximations used there. It also provides a possible path for a more consistent numerical calculation of the Hubble tension probability and the key ideas can be applied to any model predicting spatial fluctuations of dark energy with known probability distribution.

In Section~\ref{sect: concl} we conclude by discussing the main results of the manuscript.

\section{Quantum dark energy model} 
\label{sect:model}

The simple dark energy model that we want to consider is based on the following action for a light non-minimally coupled scalar field $\Phi$ in $D=4$ spacetime dimensions.
\begin{equation}
\boxed{
S[\Phi] = \int\! d^4\!x \, \sqrt{-g} \, 
	\biggl\{ -\frac{1}{2} g^{\mu\nu} \partial_\mu \Phi \partial_\nu \Phi
		- \frac{1}{2}m^2\Phi^2 - \frac{1}{2}\xi R \Phi^2 \biggr\}
}\,.
\label{eq:action}
\end{equation}
For a given background metric $g_{\mu\nu}$, the non-minimal coupling term between the field $\Phi$ and the metric with Ricci scalar $R$ contributes to the effective mass of the field $M^2\equiv m^2+\xi R$.

Following~\cite{Belgacem:2021ieb} and~\cite{Glavan:2017jye}, we specialize the action~(\ref{eq:action}) to a FLRW background with metric element $ds^2=-dt^2+a^2(t)d\vec{x}^2$, where $t$ is cosmological time, $\vec{x}$ are the comoving coordinates and $a(t)$ is the scale factor. Then the field depends on cosmological time and comoving coordinates, which we denote altogether as $x$, so that $\Phi(x)=\Phi(t,\vec{x})$. If $\dot{\Phi}$ is the time derivative of $\Phi$ and $\vec{\nabla}\Phi$ its spatial gradient, then the action~(\ref{eq:action}) specializes to

\begin{equation}
\label{eq:action FLRW}
S[\Phi]\equiv\int\! dt~d^3x \, \mathcal{L}_\Phi=\int\! dt~d^3x \, a^3\left\{\frac12 \dot{\Phi}^2-\frac{\bigl(\vec{\nabla}\Phi\bigr)^2}{2a^2}-\frac12 \left[m^2+6\xi (2-\epsilon) H^2\right]\Phi^2\right\} \,,
\end{equation}
where $H(t)\equiv\dot{a}(t)/a(t)$ is the (global) Hubble expansion rate and $\epsilon(t)\equiv-\dot{H}(t)/H^2(t)$ is the principal slow-roll parameter. The (squared) effective mass is
\begin{equation}
\label{eq: effective mass squared}
M^2(t)=m^2 +6\xi \left(2-\epsilon(t)\right) H^2(t)\,.
\end{equation}
The assumption of light field means that $m/H(t)<1$ throughout the history of the Universe. In inflation and radiation epochs we can actually assume $m/H(t)\ll1$, but in matter-dominated epoch the mass $m$ plays an important role because, as we will see, it is necessary for generating a contribution to the energy-momentum tensor scaling as dark energy and eventually leading the expansion, which is the ultimate effect we are looking for.

Canonical momentum $\Pi(x)$ is defined as
\begin{equation}
\Pi(x) \equiv \frac{\partial \mathcal{L}_\Phi}{\partial \dot{\Phi}(x)} =
	a^3 \dot{\Phi}(x) \,.
\end{equation}

We can then quantize the scalar field in the classical FLRW background metric by canonical quantization, promoting $\Phi(x)$ and $\Pi(x)$ to quantum operators $\hat{\Phi}(x)$ and $\hat{\Pi}(x)$, satisfying equal-time canonical commutation rules
\begin{equation}
\label{comm-fields}
\bigl[ \hat{\Phi}(t,\vec{x}), \hat{\Pi}(t,\vec{x}^{\,\prime}) \bigr]
	= i \delta^{3}(\vec{x}\!-\!\vec{x}^{\,\prime}) \, ,
\qquad
 \bigl[ \hat{\Phi}(t,\vec{x}), \hat{\Phi}(t,\vec{x}^{\,\prime}) \bigr] =0
	= \bigl[ \hat{\Pi}(t,\vec{x}), \hat{\Pi}(t,\vec{x}^{\,\prime}) \bigr] \,.
\end{equation}
The classical Hamilton equations are mapped into the Heisenberg equations
\begin{eqnarray}
\frac{d}{dt} \hat{\Phi}(t,\vec{x}) - a^{-3}(t) \, \hat{\Pi}(t,\vec{x}) = 0 \, ,
\label{FullEOM1}
\\
a^{-3}(t)\, \frac{d}{dt} \hat{\Pi}(t,\vec{x})
	- \frac{\nabla^2}{a^2(t)} \hat{\Phi}(t,\vec{x})
	+ M^2(t)\, \hat{\Phi}(t,\vec{x}) = 0 \, ,
\label{FullEOM2}
\end{eqnarray}
which can be combined to get the second order equation of motion\footnote{Here we assume that the backreaction of $\hat{\Phi}$ on the expansion of the Universe is negligible, and therefore $H(t)$ is the global (classical) expansion rate, defined as $H^2(t)\equiv\langle\Omega|\hat{H}^2\left(t,\vec{x}\right)|\Omega\rangle$, where the vacuum state $|\Omega\rangle$ obeys eq.~(\ref{eq: vacuum state}). The global Hubble rate $H(t)$ does not depend on $\vec{x}$ because of the spatial homogeneity of the background metric and vacuum state. The local (squared) Hubble rate operator $\hat{H}^2\left(t,\vec{x}\right)$ is discussed in Section~\ref{sect: qbackr}.} for the quantum field $\hat{\Phi}(t,\vec{x})$,
\begin{equation}
\label{eq: EOM Phi 2nd order}
\frac{d^2}{dt^2} \hat{\Phi}(t,\vec{x})+3H(t)\frac{d}{dt} \hat{\Phi}(t,\vec{x})
-\frac{\nabla^2}{a^2(t)} \hat{\Phi}(t,\vec{x})+ M^2(t)\, \hat{\Phi}(t,\vec{x}) = 0\,.
\end{equation}

In Section~\ref{sec: QFT} we will be interested in a slight generalization of the equation of motion, with the introduction of a speed of sound $c_s\leq1$, assumed constant for simplicity, namely

\begin{equation}
\label{eq: EOM Phi 2nd order cs<1}
\frac{d^2}{dt^2} \hat{\Phi}(t,\vec{x})+3H(t)\frac{d}{dt} \hat{\Phi}(t,\vec{x})
-c_s^2\frac{\nabla^2}{a^2} \hat{\Phi}(t,\vec{x})+ M^2(t)\, \hat{\Phi}(t,\vec{x}) = 0\,.
\end{equation}
The original model~(\ref{eq:action}) has $c_s=1$ but, as already discussed in the Introduction~\ref{sect:intro}, several scalar-tensor theories of gravity predict a reduced speed of sound, notably galileons and more generally theories in the Horndeski class.

Unless explicitly specified, we always assume $c_s=1$ in this Section.
The introduction of creation/annihilation operators and the construction of the Fock space is well known and the relevant parts have also been reviewed in~\cite{Belgacem:2021ieb} and~\cite{Glavan:2017jye}. Therefore we just limit ourselves to summarize some of the steps, starting by moving to momentum space and expanding the field and canonical momentum operators in creation/annihilation operators $\hat{b}(\vec{k})$ and $\hat{b}^{\dag}(\vec{k})$ for the Fourier mode $\vec{k}$:
\begin{subequations}
\begin{eqnarray}
\hat{\Phi}(t,\vec{x}) & = &
	\mathlarger{\int}\!\! \frac{d^{3}k}{(2\pi)^{\frac{3}{2}}}
	\biggl\{ e^{i\vec{k}\cdot\vec{x}} \varphi(t,k) \, 
			\hat{b}(\vec{k}) 
		+ e^{-i\vec{k}\cdot\vec{x}} \varphi^*(t,k) \, 
			\hat{b}^{\dag}(\vec{k})  \biggr\} \, ,\label{eq: canon Phi}\\
\hat{\Pi}(t,\vec{x}) & = & a^3(t)
	\mathlarger{\int}\!\! \frac{d^{3}k}{(2\pi)^{\frac{3}{2}}}
	\biggl\{ e^{i\vec{k}\cdot\vec{x}} \dot{\varphi}(t,k) \, 
			\hat{b}(\vec{k}) 
		+ e^{-i\vec{k}\cdot\vec{x}} \dot{\varphi}^*(t,k) \, 
			\hat{b}^{\dag}(\vec{k})  \biggr\} \, .\label{eq: canon Pi}
\end{eqnarray}
\end{subequations}
The mode function $\varphi(t,k)$ depends only on $t$ and on the norm of the comoving momentum $k\!=\!\|\vec{k}\|$, due to spatial homogeneity and isotropy of the FLRW background. The canonical commutation rules~(\ref{comm-fields}) are imposed by requiring that
\begin{equation}
\label{eq: b bdag commutation rules}
\bigl[ \hat{b}(\vec{k}) , \hat{b}^{\dag}(\vec{k}^{\,\prime}) \bigr]
	= \delta^{3}(\vec{k}\!-\!\vec{k}^{\,\prime}) \, , 
\qquad
\bigl[ \hat{b}(\vec{k}) , \hat{b}(\vec{k}^{\prime}) \bigr] = 0
	= \bigl[ \hat{b}^{\dag} (\vec{k}) , 
		\hat{b}^{\dag}(\vec{k}^{\prime}) \bigr] \,,
\end{equation}
with the Wronskian normalization condition for the mode function
\begin{equation}
\label{Wronskian}
\varphi(t,k) \dot{\varphi}^*(t,k)
	- \dot{\varphi}(t,k) \varphi^*(t,k) = i~a^{-3}(t)\, .
\end{equation}
The equation of moton~(\ref{eq: EOM Phi 2nd order}) implies for the mode function $\varphi(t,k)$,
\begin{equation}
\label{modeEOMreduced}
\ddot{\varphi}(t,k) + 3H(t)\dot{\varphi}(t,k)
	+ \left[\frac{k^2}{a^2(t)}+ M^2(t)\right]\, \varphi(t,k) = 0\,,
\end{equation}
or, for a generic speed of sound $c_s$ in eq.~(\ref{eq: EOM Phi 2nd order cs<1}),
\begin{equation}
\label{modeEOMreduced cs}
\ddot{\varphi}(t,k) + 3H(t)\dot{\varphi}(t,k)
	+ \left[c_s^2\frac{k^2}{a^2(t)}+ M^2(t)\right]\, \varphi(t,k) = 0\,.
\end{equation}

The energy-momentum tensor is evaluated from the variation of the action~(\ref{eq:action}) with respect to the metric. The corresponding quantum operator $\hat{T}_{\mu\nu}$ is
\begin{equation}
\label{energy_momentum}
\hat{T}_{\mu\nu} = \partial_\mu \hat{\Phi} \, \partial_\nu\hat{\Phi}
		- \frac{1}{2} g_{\mu\nu} g^{\alpha\beta} 
			\partial_\alpha \hat{\Phi} \, \partial_\beta \hat{\Phi}
		- \frac{m^2}{2} g_{\mu\nu} \hat{\Phi}^2
		+ \xi \bigl[ G_{\mu\nu}  - \nabla_\mu \nabla_\nu 
			+ g_{\mu\nu} \square \bigr]
			\hat{\Phi}^2 \,,
\end{equation}
where $G_{\mu\nu}$ is the Einstein tensor, $\nabla_\mu$ denotes covariant derivative and $\square\equiv g^{\alpha\beta}\nabla_\alpha \nabla_\beta$ is the covariant d'Alembert operator.
On the FLRW background metric, one can determine the quantum energy density $\rho_Q  \equiv \langle\hat{\rho}_Q\rangle\equiv  - \langle \hat{T}^0{}_0\rangle$ and pressure $p_Q \delta^i_j  \equiv  \langle \hat{T}^i{}_j \rangle$ by taking expectation values on the homogeneous and isotropic vacuum state $|\Omega\rangle$ annihilated by all operators $\hat{b}(\vec{k})$, which means
\begin{equation}
\label{eq: vacuum state}
\hat{b}(\vec{k})|\Omega\rangle=0\,.
\end{equation}
When taking into account the aforementioned homogeneity and isotropy, the final results are
\begin{equation}
\label{eq: rho_Q}
\rho_Q=\frac{H^2}{2}\left\{\left[ \left(\frac{m}{H}\right)^2 +6\xi \right] \Bigl\langle\hat{\Phi}^2\Bigr\rangle+6\xi\frac{\Bigl\langle\left\{\hat{\Phi},\hat{\Pi}\right\}\Bigr\rangle}{a^3 H}+\frac{\Bigl\langle\hat{\Pi}^2 \Bigr\rangle}{a^6 H^2}+\frac{\Bigl\langle\left(\vec{\nabla}\hat{\Phi}\right)^2\Bigr\rangle}{a^2 H^2}\right\}\,,
\end{equation}
\begin{eqnarray}
p_Q=\frac{H^2}{2}\left\{\left[-2\xi (3-2\epsilon)-\left(\frac{m}{H}\right)^2 (1-4\xi)+24\xi^2 (2-\epsilon)\right] \Bigl\langle\hat{\Phi}^2\Bigr\rangle+2\xi\frac{\Bigl\langle\left\{\hat{\Phi},\hat{\Pi}\right\}\Bigr\rangle}{a^3 H}\right.\,\nonumber\\
\left.+(1-4\xi)\frac{\Bigl\langle\hat{\Pi}^2\Bigr\rangle}{a^6 H^2}-\frac{1-12\xi}{3}\frac{\Bigl\langle\left(\vec{\nabla}\hat{\Phi}\right)^2\Bigr\rangle}{a^2 H^2}\right\}\,.\label{eq: p_Q}
\end{eqnarray}
The correlators in eqs.~(\ref{eq: rho_Q}--\ref{eq: p_Q}) are the coincident $\Bigl\langle \hat{\Phi}^2 (t,\vec{x}) \Bigr\rangle$, $\Bigl\langle \hat{\Pi}^2 (t,\vec{x}) \Bigr\rangle$, $\Bigl\langle \bigl\{ \hat{\Phi}(t,\vec{x}) ,\hat{\Pi}(t,\vec{x}) \bigr\} \Bigr\rangle$ (with the usual definition of anticommutator $\{\hat{A},\hat{B}\}\equiv\hat{A}\hat{B}+\hat{B}\hat{A}$) and $\Bigl\langle\left(\vec{\nabla}\hat{\Phi}(t,\vec{x})\right)^2\Bigr\rangle$.
They only depend on time $t$ and not on the comoving position $\vec{x}$, again because of the assumed spatial homogeneity and isotropy of the background metric $g_{\mu\nu}(t)$ and vacuum state $|\Omega\rangle$.
\noindent When one replaces the fields/canonical momenta in eqs.~(\ref{eq: rho_Q}--\ref{eq: p_Q}) by their free field expansions in eqs.~(\ref{eq: canon Phi}--\ref{eq: canon Pi}) (or eqs.~(\ref{PhiLong}--\ref{PiLong}) in the stochastic approximation that we will discuss), one obtains one-loop results for the energy density and pressure.
\noindent In preparation for subsection~\ref{subsect:corr 4pt}, we also write explicitly the energy density operator (which can be obtained from eq.~(\ref{eq: rho_Q}) by removing the expectation values):
\begin{eqnarray}
\hat{\rho}_Q\left(t,\vec{x}\right)=\frac{H^2(t)}{2}\left\{\left[ \left(\frac{m}{H(t)}\right)^2 +6\xi \right] \hat{\Phi}^2\left(t,\vec{x}\right)+6\xi\frac{\left\{\hat{\Phi}\left(t,\vec{x}\right),\hat{\Pi}\left(t,\vec{x}\right)\right\}}{a^3(t) H(t)}\right.\,\nonumber\\
\left.+\frac{\hat{\Pi}^2\left(t,\vec{x}\right) }{a^6(t) H^2(t)}+\frac{\left(\vec{\nabla}\hat{\Phi}\left(t,\vec{x}\right)\right)^2}{a^2(t) H^2(t)}\right\}\,.\label{eq: rho_Q operator}
\end{eqnarray}

In the next subsection~\ref{subsec: stoch IR}, when studying the dynamics of the infrared (IR) modes which are super-Hubble, following~\cite{Belgacem:2021ieb} and~\cite{Glavan:2017jye} we will neglect the contribution of spatial gradients $\left(\vec{\nabla}\hat{\Phi}(t,\vec{x})\right)^2$. This is justified because, for a given mode with comoving wavenumber $k$, the expectation values of spatial gradient terms entering in eqs.~(\ref{eq: rho_Q}--\ref{eq: p_Q}) are suppressed by a factor $k^2/(aH)^2\ll1$.
We will comment again on the role of spatial gradients in subsection~\ref{subsubsec: stoch vs field plot} when discussing the comparison between stochastic formalism (briefly revised in subsection~\ref{subsec: stoch IR}) and the full quantum field theory (QFT) results. We will see that they are actually relevant after matter-radiation equality, but their contribution can be made small by a reduced speed of sound $c_s<1$.

\subsection{Stochastic formalism}
\label{subsec: stoch IR}

\noindent 
As anticipated in the Introduction~\ref{sect:intro}, Starobinsky's stochastic formalism has been applied in~\cite{Glavan:2017jye} to determine the time evolution of the coincident correlators (which are 2-pt functions of field/canonical momentum) appearing in eqs.~(\ref{eq: rho_Q}--\ref{eq: p_Q}). The reason lies upon the findings in~\cite{Glavan:2015cut, Glavan:2013mra, Glavan:2014uga} that the energy-momentum tensor of a very light non-minimally coupled scalar field, backreacting on the cosmological expansion, is dominated by the infrared (IR) modes. The same idea has been applied in~\cite{Belgacem:2021ieb} to study non-coincident 4-pt functions entering the density-density correlator. Here we review those results and include in our treatment the non-coincident 2-pt functions. This will also allow us to comment on the validity or violation of Wick's theorem expected for Gaussian fields and it will partly serve as a basis for the investigations in the next Sections.

At cosmological time $t$, we set the separation of modes at a scale $\mu a(t) H(t)$ where $0<\mu<1$ is a dimensionless constant. Then modes with comoving wavenumber $k<\mu a(t) H(t)$ are said to be long-wavelength modes (they are super-Hubble) while those with $k>\mu a(t) H(t)$ are short (sub-Hubble).

\noindent The long-wavelength parts of the field and canonical momentum operators, denoted by $\hat{\phi}(t,\vec{x})$ and $\hat{\pi}(t,\vec{x})$ in contrast to the full field  $\hat{\Phi}(t,\vec{x})$ and canonical momentum $\hat{\Pi}(t,\vec{x})$ operators in eqs.~(\ref{eq: canon Phi}--\ref{eq: canon Pi}), are defined as
\begin{subequations}
\begin{eqnarray}
\hat{\phi}(t,\vec{x}) & = & 
	\mathlarger{\int}\!\! \frac{d^{3}k}{(2\pi)^{\frac{3}{2}}} \,
	\theta\left( \mu a(t) H(t) \!-\! \| \vec{k} \| \right)
	\biggl\{ e^{i\vec{k}\cdot\vec{x}} \varphi(t,k) \, 
			\hat{b}(\vec{k}) 
		+ e^{-i\vec{k}\cdot\vec{x}} \varphi^*(t,k) \, 
			\hat{b}^{\dag}(\vec{k})  \biggr\} \, ,
\label{PhiLong}
\\
\hat{\pi}(t,\vec{x}) & = &
	a^3(t)\mathlarger{\int}\!\! \frac{d^{3}k}{(2\pi)^{\frac{3}{2}}} \,
	\theta\left( \mu a(t) H(t) \!-\! \| \vec{k} \| \right)
	\biggl\{ e^{i\vec{k}\cdot\vec{x}}\dot{\varphi}(t,k) \, 
			\hat{b}(\vec{k}) 
		+ e^{-i\vec{k}\cdot\vec{x}}\dot{\varphi}^*(t,k) \, 
			\hat{b}^{\dag}(\vec{k})  \biggr\} \,. \qquad
\label{PiLong}
\end{eqnarray}
\end{subequations}
In eqs.~(\ref{PhiLong}--\ref{PiLong}), the separation between short and long modes has been set by a top-hat function (Heaviside step function) with transition at $k=\mu a(t) H(t)$ where $0<\mu\ll1$ is a dimensionless factor that selects the super-Hubble UV cutoff of the stochastic theory. The introduction of $\mu\neq1$ allows for control of the dependence of physical quantities on the UV cutoff $\mu H(t)$ of the stochastic theory, which limits physical wavenumbers to $\frac{k}{a(t)}\leq\mu H(t)$.

\noindent The dynamics of the long modes is described by a modified version of the Heisenberg equations of motion,
\begin{eqnarray}
\frac{d}{dt} \hat{\phi}(t,\vec{x}) 
	- a^{-3}(t)\, \hat{\pi}(t,\vec{x})
	= \hat{f}_\phi(t,\vec{x}) \, ,
\label{EOMdPhiLong}
\\
a^{-3}(t)\, \frac{d}{dt} \hat{\pi}(t,\vec{x})
	- \frac{\nabla^2}{a^2(t)} \hat{\phi}(t,\vec{x})
	+ M^2(t)\, \hat{\phi}(t,\vec{x})
	= a^{-3}(t) \hat{f}_\pi (t,\vec{x}) \, ,
\label{EOMdPiLong}
\end{eqnarray}
where the ``stochastic forces" on the right-hand sides are due to modes crossing the separation scale $\mu a H$ and they are given by (see~\cite{Glavan:2017jye}):

\begin{eqnarray}
\hat{f}_\phi(t,\vec{x}) & = & 
	\mu a H^2 \left(1-\epsilon\right)\mathlarger{\int}\!\! \frac{d^{3}k}{(2\pi)^{\frac{3}{2}}} \,
	\delta\left(\|\vec{k}\|-\mu a H \right)
	\biggl\{ e^{i\vec{k}\cdot\vec{x}} \varphi(t,k) \, 
			\hat{b}(\vec{k}) 
		+ e^{-i\vec{k}\cdot\vec{x}} \varphi^*(t,k) \, 
			\hat{b}^{\dag}(\vec{k})  \biggr\} \, ,
\label{forcePhi}
\\
\hat{f}_\pi(t,\vec{x}) & = &
	\mu a^4 H^2 \left(1-\epsilon\right)\mathlarger{\int}\!\! \frac{d^{3}k}{(2\pi)^{\frac{3}{2}}} \,
	\delta\left(\|\vec{k}\|-\mu a H \right)
	\biggl\{ e^{i\vec{k}\cdot\vec{x}}\dot{\varphi}(t,k) \, 
			\hat{b}(\vec{k}) 
		+ e^{-i\vec{k}\cdot\vec{x}}\dot{\varphi}^*(t,k) \, 
			\hat{b}^{\dag}(\vec{k})  \biggr\} \,. \qquad
\label{forcePi}
\end{eqnarray}

\subsection{Non-coincident 2-point IR correlators}\label{subsect:corr}
When trading the expectation values in eqs.~(\ref{eq: rho_Q}--\ref{eq: p_Q}) for their respective long-wavelength parts and neglecting spatial gradients, the problem of determining the time evolution of energy density and pressure reduces to the study of IR coincident correlators (rescaled to have the same dimensions) $\Delta_{\phi,\phi}(t)$, $\Delta_{\phi,\pi}(t)$ and $\Delta_{\pi,\pi}(t)$ defined in eqs. (30)-(32) of~\cite{Glavan:2017jye}.

Similarly to~\cite{Belgacem:2021ieb}, we can generalize them to the following non-coincident equal-time 2-point correlators, which only depends on time $t$ and relative comoving separation $r=\|\vec{x}_2-\vec{x}_1\|$ between the points considered\footnote{The correlator $\Delta_{\phi,\pi}(t,r)$ in eq.~(\ref{DeltaPhiPi}) can be equivalently defined replacing $\Bigl\langle \bigl\{ \hat{\phi}(t,\vec{x}_1) ,\hat{\pi}(t,\vec{x}_2) \bigr\} \Bigr\rangle$ by $\Bigl\langle \hat{\phi}(t,\vec{x}_1)\hat{\pi}(t,\vec{x}_2)+\hat{\pi}(t,\vec{x}_1)\hat{\phi}(t,\vec{x}_2) \Bigr\rangle$, because it only depends on $\vec{x}_1$ and $\vec{x}_2$ through their relative distance $r$. These properties follow from the homogeneity and isotropy of the metric $g_{\mu\nu}(t)={\rm diag}\left(-1, a^2(t),a^2(t),a^2(t)\right)$ and the state $|\Omega\rangle$.}:
\begin{eqnarray}
\Delta_{\phi,\phi}(t,r) & \equiv & 
	\Bigl\langle \hat{\phi}(t,\vec{x}_1) \hat{\phi}(t,\vec{x}_2) \Bigr\rangle \, ,
\label{DeltaPhiPhi}
\\
\Delta_{\phi,\pi}(t,r) & \equiv & 
	\frac{1}{a^3(t) H(t)}\Bigl\langle \bigl\{ \hat{\phi}(t,\vec{x}_1) ,
						\hat{\pi}(t,\vec{x}_2) \bigr\} \Bigr\rangle \, ,
\label{DeltaPhiPi}
\\
\Delta_{\pi,\pi}(t,r) & \equiv & 
	\frac{1}{a^6(t) H^2(t)} \Bigl\langle \hat{\pi}(t,\vec{x}_1) \hat{\pi}(t,\vec{x}_2) \Bigr\rangle \, .
\label{DeltaPiPi}
\end{eqnarray}

The coincident correlators in eqs. (30)-(32) of~\cite{Glavan:2017jye} are obtained by setting $r=0$ in eqs.~(\ref{DeltaPhiPhi}--\ref{DeltaPiPi}) and then energy density and pressure are approximated starting from~(\ref{eq: rho_Q}--\ref{eq: p_Q}) as

\begin{equation}
\label{eq: rho_Q long}
\rho_Q(t)\approx\frac{H^2}{2}\left\{\left[ \left(\frac{m}{H}\right)^2 +6\xi \right]\Delta_{\phi,\phi}(t,0) +6\xi\Delta_{\phi,\pi}(t,0)+\Delta_{\pi,\pi}(t,0)\right\}\,,
\end{equation}
\begin{eqnarray}
p_Q(t)\approx\frac{H^2}{2}\left\{\left[-2\xi (3-2\epsilon)-\left(\frac{m}{H}\right)^2 (1-4\xi)+24\xi^2 \left(2-\epsilon\right)\right] \Delta_{\phi,\phi}(t,0)\right.\,\nonumber\\
\left.+2\xi\Delta_{\phi,\pi}(t,0)+(1-4\xi)\Delta_{\pi,\pi}(t,0)\right\}\,.
\label{eq: p_Q long}
\end{eqnarray}

The time evolution of the general non-coincident 2-pt correlators in eqs.~(\ref{DeltaPhiPhi}--\ref{DeltaPiPi}) is conveniently described by switching from cosmological time $t$ to the number of e-foldings
\begin{equation}
\label{eq: efolds definition}
N(t)\equiv\ln{\left(\frac{a(t)}{a_{\rm in}}\right)}\,
\end{equation}
measured from the beginning of inflation, where $a_{\rm in}$ is the scale factor at the beginning of inflation (corresponding to $N=0$).

Formally the time evolution equations are the same as in eqs. (33)-(35) given in~\cite{Glavan:2017jye} for coincident correlators,

\begin{eqnarray}
\frac{\partial}{\partial N}\Delta_{\phi,\phi} - \Delta_{\phi,\pi}=n_{\phi,\phi} \,,
\label{eomPhiPhi}
\\
\frac{\partial}{\partial N}\Delta_{\phi,\pi}+(3-\epsilon)\Delta_{\phi,\pi}+2\Bigl( \frac{M}{H} \Bigr)^{\!2} \Delta_{\phi,\phi}-2\Delta_{\pi,\pi}=n_{\phi,\pi} \,,
\label{eomPhiPi}
\\
\frac{\partial}{\partial N}\Delta_{\pi,\pi}+2(3-\epsilon)\Delta_{\pi,\pi}+\Bigl( \frac{M}{H} \Bigr)^{\!2} \Delta_{\phi,\pi}=n_{\pi,\pi} \,,
\label{eomPiPi}
\end{eqnarray}
but now the noise sources $n$'s on the right-hand sides are functions of $t$ and $r$, so eqs. (36)-(38) of~\cite{Glavan:2017jye}, connecting the noise sources to the stochastic forces~(\ref{forcePhi}--\ref{forcePi}), are generalized by

\begin{eqnarray}
n_{\phi,\phi}(t,r)  &=&  \frac{1}{H(t)} \Bigl\langle 
	\left\{ \hat{f}_{\phi}(t,\vec{x}_1) , \hat{\phi}(t,\vec{x}_2)\right\} \Bigr\rangle \, ,\nonumber \\
n_{\phi,\pi}(t,r)  &=&  \frac{1}{a^3(t)H^2(t)} \Bigl\langle 
	\left\{ \hat{f}_{\phi}(t,\vec{x}_1) ,\hat{\pi}(t,\vec{x}_2) \right\} +
           \left\{ \hat{f}_{\pi}(t,\vec{x}_1) ,\hat{\phi}(t,\vec{x}_2) \right\} \Bigr\rangle \, ,\nonumber \\
n_{\pi,\pi}(t,r)  &=&  \frac{1}{a^6(t)H^3(t)} \Bigl\langle 
	\left\{ \hat{f}_{\pi}(t,\vec{x}_1) , \hat{\pi}(t,\vec{x}_2)\right\} \Bigr\rangle \,.
\label{eq: sources 2pt general}
\end{eqnarray}
Their expressions in terms of the mode function $\varphi(t,k)$, which generalize eqs. (39)-(42) of~\cite{Glavan:2017jye}, are:
\begin{eqnarray}
n_{\phi,\phi} &=& \frac{1}{2\pi^2} (\mu aH)^3 (1-\epsilon)\left[|\varphi(t,k)|^2\right]_{k=\mu aH} j_0(\mu aHr) \,, \nonumber \\
n_{\phi,\pi} &=& \frac{1}{2\pi^2} \mu^3 a^3 H^2 (1-\epsilon)\left[\frac{\partial}{\partial t}|\varphi(t,k)|^2\right]_{k=\mu aH} j_0(\mu aHr) \,, \nonumber \\
n_{\pi,\pi} &=& \frac{1}{2\pi^2} \mu^3 a^3 H (1-\epsilon)\left[|\dot{\varphi}(t,k)|^2\right]_{k=\mu aH} j_0(\mu aHr) \,,
\label{eq: sources 2pt mode function}
\end{eqnarray}
where $j_0(z)\equiv\frac{\sin(z)}{z}$ is the $0$-th order spherical Bessel function.
Comparing with eqs. (39)-(42) of~\cite{Glavan:2017jye}, one immediately realizes that the only modification with respect to the coincident case is the factor $j_0(\mu aHr)$ in the stochastic sources~(\ref{eq: sources 2pt mode function}), which correctly reduces to $1$ when $r\to0$.

Similarly to~\cite{Glavan:2017jye}, we solve the system of equations (\ref{eomPhiPhi}--\ref{eomPiPi}) in the three epochs of de Sitter inflation ($\epsilon=0$), radiation-domination ($\epsilon=2$) and matter-domination ($\epsilon=3/2$).
The most important modification is that here we are considering non-coincident 2-pt correlators instead of the coincident ones of~\cite{Glavan:2017jye}, but this is just a simple version of the technically more complicated problem of non-coincident 4-pt functions already solved in~\cite{Belgacem:2021ieb}. Therefore the strategy is already set up and we can rely on it to construct the solution to eqs.~(\ref{eomPhiPhi}--\ref{eomPiPi}).
In Appendix~\ref{app: mat+CC}, we also discuss a refinement for the last stages of evolution by studying correlators in a Universe containing both matter and a cosmological constant. This  is still an approximation, but it is supposed to (partially) take into account in analytical form the backreaction of the scalar field on the FLRW background in the most recent e-foldings of evolution before the current time.
As discussed in Section IV of~\cite{Glavan:2017jye}, evolving the correlators in a given classical FLRW background metric is legitimate as long as the quantum backreaction on cosmological expansion is small with respect to the classical sources of energy density guiding the expansion. This is the case throughout all epochs except for recent times when quantum backreaction manifests as dark energy and later takes the lead of cosmological expansion. We begin with the evolution in a de Sitter inflationary epoch.
\subsubsection{de Sitter inflation}
\label{subsubsec: 2pt de Sitter}

\noindent Following~\cite{Glavan:2017jye} and~\cite{Belgacem:2021ieb}, in de Sitter inflation we use the Chernikov-Tagirov-Bunch-Davies (CTBD) mode function. We also include the possibility of a reduced speed of sound $c_s$: looking at the difference between eqs.~(\ref{modeEOMreduced}) and~(\ref{modeEOMreduced cs}), it just amounts to replacing $k\to c_s k$ in eq. (45) of~\cite{Glavan:2017jye}. Therefore the de Sitter mode function is
\begin{equation}
\varphi(t,k) = \sqrt{\frac{\pi}{4a^3(t)H_I}}~H_{\nu_I}^{(1)} \left(\frac{c_s k}{a(t)H_I}\right)\, ,
\qquad
\nu_I \equiv \sqrt{\tfrac{9}{4}-\bigl( \tfrac{M}{H_I} \bigr)^{\!2} } \, ,
\label{mode function}
\end{equation}
where we denoted by $H_I$ is the constant Hubble rate in inflation and by $M^2=m^2 +12\xi H_I^2$ the constant effective squared mass, while $H_{\nu_I}^{(1)}$ the Hankel function of the first kind.
Eq.~(\ref{mode function}) satisfies the Wronskian condition~(\ref{Wronskian}).

As in~\cite{Glavan:2017jye}, since we are dealing with long modes $k<\mu a H_I\ll a H_I$, we can simplify the mode function to
\begin{equation}
\varphi(t,k) \approx -\frac{i}{\sqrt{\pi}} 2^{\nu_I-1} \Gamma(\nu_I)a(t)^{\nu_I-3/2} H_I^{\nu_I-1/2} c_s^{-\nu_I} k^{-\nu_I}\,.
\label{eq: phi infl approx}
\end{equation}
For $(m/H_I)^2\ll1$ and $|\xi|\ll1$, working at leading order in
\begin{equation}
\label{eq: def X inflation}
X\equiv M^2/H_I^2=(m/H_I)^2+12\xi\,,
\end{equation}
one gets similarly to~\cite{Glavan:2017jye} (but with the extra factor $j_0(\mu aH_I r)$) the expression for the noise sources in de Sitter inflation, which we write in (row) vector form, to leading order in $X$, as\footnote{We used $\nu_I=\sqrt{\frac94 -X}\simeq\frac32-\frac{X}{3}$ and just kept $\nu_I\simeq\frac 32$ in the power of $c_s$.}
\begin{equation}
\label{eq: source 2pt de Sitter}
\left(n_{\phi,\phi},n_{\phi,\pi},n_{\pi,\pi}\right)\approx c_s^{-3}\frac{H_I^2}{4\pi^2}j_0(\mu aH_I r)~\left(1, -\frac23 X, \frac19 X^2\right)\,.
\end{equation}

We discuss in detail the solution of eqs.~(\ref{eomPhiPhi}--\ref{eomPiPi}) for the non-coincident 2-pt correlators.
First, eqs.~(\ref{eomPhiPhi}--\ref{eomPiPi}) can be written in matrix form by introducing:

\noindent1) the column vector $\Delta_{(\rm 2)}$ of correlators whose corresponding row form (transpose vector $\Delta_{(\rm 2)}^{\rm T}$) is
\begin{equation}
\label{eq: def Delta2}
\Delta_{(\rm 2)}^{\rm T}=\left(\Delta_{\phi,\phi},\Delta_{\phi,\pi},\Delta_{\pi,\pi}\right)\,,
\end{equation}
\noindent2) the column vector $n_{(\rm 2)}$ of noise sources whose corresponding row form (transpose vector $n_{(\rm 2)}^{\rm T}$) is
\begin{equation}
\label{eq: def n2}
n_{(\rm 2)}^{\rm T}=\left(n_{\phi,\phi},n_{\phi,\pi},n_{\pi,\pi}\right)\,,
\end{equation}
\noindent 3) the inflationary (de Sitter) constant matrix evolution for 2-pt functions (using $\epsilon=0$)
\begin{equation}
A_{(\rm 2),I}=
\begin{pmatrix}
   0 & -1 & 0 \\
 2X & 3 & -2 \\
 0 & X & 6\\
\end{pmatrix}\,,
\end{equation}
where the subscript ``$(\rm 2),I$" is meant  to remind that this matrix refers to the evolution of 2-pt correlators in inflation.
These definitions lead to the following equivalent form of eqs.~(\ref{eomPhiPhi}--\ref{eomPiPi}) in de Sitter,
\begin{equation}
\frac{\partial}{\partial N}\Delta_{(\rm 2)}(N,r)+A_{(\rm 2),I}~\Delta_{(\rm 2)}(N,r)=n_{(\rm 2)}(N,r)\,.
\label{system_deSitter 2pt}
\end{equation}

The general solution is simple because of the constancy of the matrix $A_{(\rm 2)}$ in de Sitter. It is
\begin{equation}
\Delta_{({\rm 2})}(N,r)=\exp{\left[-A_{(\rm 2),I}N\right]}\Delta_{({\rm 2})}(0,r)+\int_0^N dN'~\exp{\left[A_{(\rm 2),I}(N'-N)\right]}~n_{({\rm 2})}(N',r)\,.
\label{sol_deSitter 2pt 1}
\end{equation}
As in~\cite{Glavan:2017jye} and~\cite{Belgacem:2021ieb}, we assume zero initial conditions for the correlators because we are interested in their growth purely generated by the inflationary expansion, thus $\Delta_{({\rm 2})}(0,r)=0$. We briefly discuss the effect of non-zero initial conditions inherited from a pre-inflationary epoch in Appendix~\ref{app: non-zero ICs}.

The computation of~(\ref{sol_deSitter 2pt 1}) is conveniently done by diagonalizing the matrix $A_{(\rm 2),I}$. Let us call $B$ the change-of-basis matrix (i.e. the matrix whose columns are the eigenvectors of $A_{(\rm 2),I}$) and by $B^{-1}$ its inverse. 
If $\lambda_j,~j=1,2,3$ are the eigenvalues of $A_{(\rm 2)}$, then the $i$-th 2-pt correlator $\Delta_{(\rm 2),i}(N,r)$ is given by
\begin{equation}
\Delta_{({\rm 2}),i}(N,r)=\sum_{j=1}^3 B_{ij}e^{-\lambda_j N}\int_0^N dN'~e^{\lambda_j N'}\sum_{k=1}^3 (B^{-1})_{jk}~n_{({\rm 2}),k}(N',r)\,.
\label{sol_deSitter 2pt 2}
\end{equation}
Starting from eq.~(\ref{eq: source 2pt de Sitter}), we write
\begin{equation}
n_{({\rm 2}),k}(N',r)\approx c_s^{-3}\frac{H_I^2}{4\pi^2}~j_0\left(\mu a_{\rm in}H_I {\rm e}^{N'} r\right)~\alpha_{k}
\end{equation}
where $\alpha$ is the constant vector with components $\left(1,-\frac23 X,\frac19 X^2\right)$ and then eq.~(\ref{sol_deSitter 2pt 2}) gives
\begin{equation}
\Delta_{({\rm 2}),i}(N,r)\approx c_s^{-3}\frac{H_I^2}{4\pi^2}~\sum_{j=1}^3 \sum_{k=1}^3 B_{ij}~(B^{-1})_{jk}~\alpha_{k}~F_j(N,r)\,,
\label{sol_deSitter 2pt 3}
\end{equation}
where
\begin{equation}
\label{eq: definition Fj}
F_j(N,r)=e^{-\lambda_{j} N}\int_0^N dN'~e^{\lambda_{j} N'}~j_0\left(\mu a_{\rm in}H_I r~{\rm e}^{N'}\right)
\end{equation}
Using the approximation\footnote{We have checked numerically that it is a good approximation.} $j_0(z)\approx\theta\left(1-z\right)$ (see also~\cite{Belgacem:2021ieb}) and introducing the definition
\begin{equation}
\label{eq: def r0 infl}
r_0\equiv\left(\mu a_{\rm in}H_I\right)^{-1}\,,
\end{equation}
 it is straightforward to show that
\begin{equation}
\label{eq: F_j}
F_j(N,r) \approx
\begin{cases}
         \frac{1-e^{-\lambda_{j} N}}{\lambda_{j}} & \text{if $r<r_0~e^{-N}$}\\
        \frac{e^{-\lambda_{j} N}}{\lambda_{j}}~\left[\left(\frac{r}{r_0}\right)^{-\lambda_{j}}-1\right]& \text{if $r_0 e^{-N}<r<r_0$}\\
      0 & \text{if $r>r_0$}\,.
\end{cases}
\end{equation}
The eigenvalues $\lambda_{j}$ of the matrix $A_{(\rm 2)}$ are $\left\{\lambda_{1},\lambda_{2},\lambda_{3}\right\}=\left\{3-\sqrt{9-4X},3,3+\sqrt{9-4X}\right\}$ while the change-of-basis matrix can be chosen (up to any constant factor, which cancels in eq.~(\ref{sol_deSitter 2pt 3}) as

\begin{equation}
B=
\begin{pmatrix}
 1   & 1 & 1 \\
 -\lambda_{1} & -3 & -\lambda_{3} \\
 \frac14 \lambda_{1}^2 & X & \frac14 \lambda_{3}^2\\
\end{pmatrix}\,.
\end{equation}

At leading order in $X$, the approximated set of egenvalues is $\left\{\frac23 X,3,6-\frac23 X\right\}$.
Due to the result for $F_j(N,r)$ in~(\ref{eq: F_j}), one can realize that, after a few e-foldings, the sum in eq.~(\ref{sol_deSitter 2pt 3}) is dominated by the smallest eigenvalue and, with some algebra, the result for the correlators in eq.~(\ref{sol_deSitter 2pt 3}), at leading order in $X$, is
\begin{equation}
\label{sol_deSitter 2pt 4}
\begin{pmatrix}
\Delta_{\phi,\phi}(N,r)\\
\Delta_{\phi,\pi}(N,r)\\
\Delta_{\pi,\pi}(N,r)\\
\end{pmatrix}
\approx c_s^{-3}\frac{H_I^2}{4\pi^2}
\begin{pmatrix}
1\\
-\frac23 X\\
\frac19 X^2\\
\end{pmatrix}
\times
\begin{cases}
         \frac{3}{2X}\left(1-e^{-\frac23 X N}\right) & \text{if $r<r_0~e^{-N}$}\\
        \frac{3}{2X}~e^{-\frac23 X N}~\left[\left(\frac{r}{r_0}\right)^{-\frac23 X}-1\right]& \text{if $r_0 e^{-N}<r<r_0$}\\
      0 & \text{if $r>r_0$}\,.
\end{cases}
\end{equation}
It is clear from eq.~(\ref{sol_deSitter 2pt 4}) that the amplification of inflationary quantum fluctuations works at its best for $X<0$ so that the factor $e^{-\frac23 X N}$ grows exponentially with $N$. This was already pointed out in~\cite{Glavan:2017jye} and~\cite{Belgacem:2021ieb}; the best conditions for a negative $X$ are a negative non-minimal coupling $\xi<0$ and a very light field so that $(m/H_I)^2\ll|\xi|\ll 1$, where the last condition ensures consistency of the approximation $|X|\ll1$ used so far (because $X\simeq -12|\xi|$). The requirement $X<0$ for the best enhancement of fluctuations (and therefore a smaller number of inflationary e-foldings required to match the dark energy content of the Universe today) can be understood from the effective potential of the scalar field. We comment again on this in Section~\ref{subsec: 1-pt PDF} after eq.~(\ref{eq: 1pt prob sol}).

With $X\simeq -12|\xi|$ the amplitude of correlators in~(\ref{sol_deSitter 2pt 4}) grows exponentially. At the end of inflation, lasting $N_I$ e-foldings, and assuming $e^{8|\xi|N_I}\gg1$, the 2-pt correlators are

\begin{equation}
\label{sol_deSitter 2pt 5}
\boxed{
\begin{pmatrix}
\Delta_{\phi,\phi}(N_I,r)\\
\Delta_{\phi,\pi}(N_I,r)\\
\Delta_{\pi,\pi}(N_I,r)\\
\end{pmatrix}
\simeq c_s^{-3}\frac{H_I^2}{32\pi^2|\xi|}~e^{8|\xi|N_I}
\begin{pmatrix}
1\\
8|\xi|\\
16 \xi^2\\
\end{pmatrix}
s_{(\rm 2)}(r)\,}\,,
\end{equation}
where
\begin{equation}
\label{eq: profile 2-pt s_2(r)}
\boxed{
s_{(\rm 2)}(r)\simeq
\begin{cases}
        1 & \text{if $r<r_0~e^{-N_I}$}\\
        1-\left(\frac{r}{r_0}\right)^{8|\xi|}& \text{if $r_0~e^{-N_I}<r<r_0$}\\
        0 & \text{if $r>r_0$}\,
\end{cases}
}\,.
\end{equation}
The spatial profile at the end of inflation is described by the function $s_{(\rm 2)}(r)$ (the subscript ``$(\rm 2)$" stands for the 2-pt IR correlators). It depends on the scale $r_0=\left(\mu a_{\rm in}H_I\right)^{-1}$ defined in eq.~(\ref{eq: def r0 infl}), which is (up to $\mu^{-1}$) the comoving Hubble length at the beginning of inflation.

\noindent The results~(\ref{sol_deSitter 2pt 5}--\ref{eq: profile 2-pt s_2(r)}) generalize to the non-coincident case those in eq.~(58) of~\cite{Glavan:2017jye}.
The one-loop energy density $\rho_Q(N_I)$ and pressure $p_Q(N_I)$ at the end of inflation can be found from the coincident correlators $\Delta_{\phi,\phi}(N_I,0)$, $\Delta_{\phi,\pi}(N_I,0)$ and $\Delta_{\pi,\pi}(N_I,0)$ using eqs.~(\ref{eq: rho_Q long})--(\ref{eq: p_Q long}) with $\epsilon=0$ and $\xi=-|\xi|<0$. The correlators $\Delta_{\phi,\pi}(N_M,0)$ and $\Delta_{\pi,\pi}(N_M,0)$ can been neglected in eqs.~(\ref{eq: rho_Q mat})--(\ref{eq: p_Q mat}) because their contribution to energy density and pressure is suppressed by another factor of $|\xi|$ with respect to $\Delta_{\phi,\phi}(N_M,0)$. At leading order in $|\xi|$ and $(m/H)^2$,
\begin{equation}
\label{eq: rho_Q mat}
\rho_Q(N_I)\approx\left(\frac{m^2}{2}-3|\xi|H_I^2\right)\Delta_{\phi,\phi}(N_I,0)\,,
\end{equation}
\begin{eqnarray}
p_Q(N_M)\approx\left(-\frac{m^2}{2}+3|\xi|H_I^2\right)\Delta_{\phi,\phi}(N_I,0)\,.
\label{eq: p_Q mat}
\end{eqnarray}
Eqs.~(\ref{eq: rho_Q mat})--(\ref{eq: p_Q mat}) show a cosmological constant type contribution $p_Q\approx-\rho_Q$ and they agree with eq. (59) of~\cite{Glavan:2017jye}.
Of course, the non-coincident correlators in eq.~(\ref{sol_deSitter 2pt 5}) contain more information than simply $\rho_Q$ and $p_Q$, because they describe spatial dependence.

\subsubsection{Radiation epoch}
\label{subsubsec: 2pt rad}

\noindent The 2-pt IR correlators at the end of inflation are then inherited in radiation epoch. In the subsequent evolution, like in~\cite{Glavan:2017jye} and~\cite{Belgacem:2021ieb}, the contribution of noise sources appearing in the right-hand side of eqs.~(\ref{eomPhiPhi})--(\ref{eomPiPi}) is negligible with respect to the initial conditions which have been amplified by inflation with a factor $\sim e^{8|\xi|N_I}/|\xi|$. In post-inflationary epochs the stochastic problem reduces to the classical evolution of the stochastic initial conditions for the field inherited from inflation. When neglecting noise sources, the set of eqs.~(\ref{eomPhiPhi})--(\ref{eomPiPi}) becomes a linear homogeneous system of equations. This implies that the problem can be reduced to the evolution of coincident correlators, because the spatial dependence $s_{(\rm 2)}(r)$ in eq. (\ref{eq: profile 2-pt s_2(r)}) inherited from inflation cannot be further modified by the noise sources. The evolution of coincident correlators in radiation and matter epochs was solved in~\cite{Glavan:2017jye}. An alternative way to find them can be obtained by adapting the matrix notation used in eq. (\ref{system_deSitter 2pt}) to radiation and matter epochs.

In radiation epoch, let us start measuring the number of e-foldings from its beginning (i.e. from the end of inflation, assuming an instantaneous reheating).
In the limit\footnote{Corrections due to the mass $m$ are evaluated in eqs. (76)-(78) of~\cite{Glavan:2017jye}. They are not necessary to obtain the results in eqs. (\ref{eq: rho_Q end of rad})--(\ref{eq: p_Q end of rad}), which agree with~\cite{Glavan:2017jye}.} of very light field $(m/H)^2\ll|\xi|\ll 1$, we can study the evolution of correlators neglecting the mass $m$. Then, in radiation epoch, the 2-pt correlators $\Delta_{(\rm 2)}(N,r)$ evolve from their initial value $\Delta_{(\rm 2)}(N,r)$ as
\begin{equation}
\Delta_{({\rm 2})}(N,r)=\exp{\left[-A_{(\rm 2),R}N\right]}\Delta_{({\rm 2})}(0,r)\,,
\label{sol_rad 2pt 1}
\end{equation}
where $A_{(\rm 2),R}$ is the constant (in the massless limit) evolution matrix in radiation epoch ($\epsilon=2$)
\begin{equation}
A_{(\rm 2),R}=
\begin{pmatrix}
   0 & -1 & 0 \\
 0 & 1 & -2 \\
 0 & 0 & 2\\
\end{pmatrix}\,.
\end{equation}
From the upper triangular form of $A_{(\rm 2),R}$ one can immediately read the eigenvalues $\{0,1,2\}$, then diagonalize the matrix and compute
\begin{equation}
\exp{\left[-A_{(\rm 2),R}N\right]}=
\begin{pmatrix}
   1 & 1-e^{-N} & \left(1-e^{-N}\right)^2 \\
 0 & e^{-N} & 2e^{-N}\left(1-e^{-N}\right) \\
 0 & 0 & e^{-2N}\\
\end{pmatrix}\,.
\end{equation}
At the end of radiation epoch (matter-radiation equality), lasting $N_R\approx 50$ e-foldings, this gives
\begin{equation}
\exp{\left[-A_{(\rm 2),R}N_R\right]}\approx
\begin{pmatrix}
   1 & 1 & 1 \\
 0 & e^{-N_R} & 2e^{-N_R} \\
 0 & 0 & e^{-2N_R}\\
\end{pmatrix}\,.
\end{equation}

Taking into account the initial conditions inherited from inflation in eq.~(\ref{sol_deSitter 2pt 5}) and their hierarchy $\Delta_{\phi,\phi}(N_I,r)\gg\Delta_{\phi,\pi}(N_I,r)\gg\Delta_{\pi,\pi}(N_I,r)$ for $|\xi|\ll1$, the result of eq.~(\ref{sol_rad 2pt 1}) at the end of radiation epoch (matter-radiation equality) is
\begin{equation}
\label{sol_rad 2pt 2}
\boxed{
\begin{pmatrix}
\Delta_{\phi,\phi}(N_R,r)\\
\Delta_{\phi,\pi}(N_R,r)\\
\Delta_{\pi,\pi}(N_R,r)\\
\end{pmatrix}
\simeq 
\begin{pmatrix}
\Delta_{\phi,\phi}(N_I,r)\\
\Delta_{\phi,\pi}(N_I,r)~e^{-N_R}\\
\Delta_{\pi,\pi}(N_I,r)~e^{-2N_R}\\
\end{pmatrix}
\simeq 
\begin{pmatrix}
\Delta_{\phi,\phi}(N_I,r)\\
0\\
0\\
\end{pmatrix}
}\,.
\end{equation}
In the last equality we highlighted that $\Delta_{\phi,\pi}(N_R,r)$ and $\Delta_{\pi,\pi}(N_R,r)$ are completely irrelevant at matter-radiation equality (and even much earlier).
We call $H_{\rm eq}\equiv H\left(N_R\right)$ the Hubble rate at matter-radiation equality.

The corresponding energy density and pressure are obtained from the coincident correlators by eqs.~(\ref{eq: rho_Q long})--(\ref{eq: p_Q long}) with $\epsilon=2$ and they can be approximated as
\begin{equation}
\label{eq: rho_Q end of rad}
\rho_Q(N_R)\approx\left(\frac{m^2}{2}-3|\xi|H_{\rm eq}^2\right)\Delta_{\phi,\phi}(N_R,0)\,,
\end{equation}
\begin{eqnarray}
p_Q(N_R)\approx\left(-\frac{m^2}{2}-|\xi|H_{\rm eq}^2\right)\Delta_{\phi,\phi}(N_R,0)\,.
\label{eq: p_Q end of rad}
\end{eqnarray}
We identify a term behaving like a cosmological constant (CC) and another like a negative radiation contribution in agreement with eqs. (79), (80) of~\cite{Glavan:2017jye}. Note that the mass $m$ is fundamental to obtain the CC-like contribution with opposite energy density and pressure. This is not in contradiction with the fact that we neglected it to compute the result~(\ref{sol_rad 2pt 1}), because this assumption was only used to further simplify the time evolution of correlators, which then enter the energy-momentum tensor via eqs.~(\ref{eq: rho_Q long})--(\ref{eq: p_Q long}).

\subsubsection{Matter epoch}
\label{subsubsec: 2pt mat}

\noindent The evolution in matter epoch can be determined with the same strategy as for radiation epoch, namely by neglecting the stochastic sources in the right-hand sides of eqs.~(\ref{eomPhiPhi})--(\ref{eomPiPi}) and using the left-hand sides of the same equations to evolve the 2-pt IR correlators from their initial conditions at matter-radiation equality given in eq.~(\ref{sol_rad 2pt 2}).
Again, we neglect the mass $m$ contribution (see Section IV.C of~\cite{Glavan:2017jye} for the corrections due to it) when evolving the correlators. However the mass $m$ will still play a fundamental role for the energy density and pressure, just like we already discussed after eqs.~(\ref{eq: rho_Q end of rad})--(\ref{eq: p_Q end of rad}).

Let us start measuring the number of e-foldings in matter era from its beginning, namely from matter-radiation equality.
When neglecting the mass $m$ in the time evolution of correlators, the structure of the solution is analogous to eq.~(\ref{sol_rad 2pt 1}) obtained in radiation epoch. It takes the form
\begin{equation}
\Delta_{({\rm 2})}(N,r)=\exp{\left[-A_{(\rm 2),M}N\right]}\Delta_{({\rm 2})}(0,r)\,,
\label{sol_mat 2pt 1}
\end{equation}
where $A_{(\rm 2),M}$ is now the constant matrix (using $\epsilon=3/2$ in matter era)
\begin{equation}
A_{(\rm 2),M}=
\begin{pmatrix}
   0 & -1 & 0 \\
 -6|\xi| & 3/2 & -2 \\
 0 & -3|\xi| & 3\\
\end{pmatrix}\,.
\end{equation}
The eigenvalues of the matrix $A_{(\rm 2),M}$ are $\left\{\frac{3-\sqrt{9+48|\xi|}}{2},\frac32,\frac{3-\sqrt{9+48|\xi|}}{2}\right\}$ or, at leading order in $|\xi|\ll1$, the set $\left\{-4|\xi|,\frac32,3+4|\xi|\right\}$.
Introducing the diagonal matrix $\Lambda_M\equiv{\rm diag}\left(\lambda_1,\lambda_2,\lambda_3\right)\equiv{\rm diag}\left(-4|\xi|,\frac32,3+4|\xi|\right)$, we write $A_{(\rm 2),M}=B_M\Lambda_M B_M^{-1}$, where $B_M$ is the change-of-basis matrix given, at leading order in $|\xi|$, by
\begin{equation}
B_M=
\begin{pmatrix}
   0 & -1 & 0 \\
 -6|\xi| & 3/2 & -2 \\
 0 & -3|\xi| & 3\\
\end{pmatrix}\,.
\end{equation}
Then the entries $(ij)$ of the matrix exponential appearing in eq.~(\ref{sol_mat 2pt 1}) are
\begin{equation}
\left(\exp{\left[-A_{(\rm 2),M}N\right]}\right)_{ij}=\sum_{k=1}^3 \left(B_M\right)_{ik}~e^{-\lambda_k N}~\left(B_M^{-1}\right)_{kj}\,.
\end{equation}
The sum above is soon dominated by the smallest eigenvalue $\lambda_1\approx-4|\xi|<0$, which gives an exponential growth, while the other eigenvalues give an exponential decay. Using this observation and the initial condition (\ref{sol_rad 2pt 2}) with negligible $\Delta_{\phi,\pi}$ and $\Delta_{\pi,\pi}$ components, one easily finds that after $N_M$ e-foldings in matter era the 2-pt correlators given by eq.~(\ref{sol_mat 2pt 1}) are

\begin{equation}
\label{sol_mat 2pt 2}
\boxed{
\begin{pmatrix}
\Delta_{\phi,\phi}(N_M,r)\\
\Delta_{\phi,\pi}(N_M,r)\\
\Delta_{\pi,\pi}(N_M,r)\\
\end{pmatrix}
\simeq 
\begin{pmatrix}
1\\
4|\xi|\\
4\xi^2\\
\end{pmatrix}
e^{4|\xi|N_M}~\Delta_{\phi,\phi}(N_R,r)}\,.
\end{equation}

The exponential growth as $e^{4|\xi|N_M}$ of correlators in matter era is confirmed by the quantum field theory calculation in eq.~(\ref{eq: coinc matter}) of this paper, as well as by~\cite{Glavan:2015cut}, but it was missed in eqs. (87)-(89) of~\cite{Glavan:2017jye}. The result~(\ref{sol_mat 2pt 2}) serves to reaffirm this growth in matter era also in the context of stochastic formalism. As we will see in Section~\ref{sec: QFT} within the field theoretic treatment, the fact that both inflation and matter era give rise to an exponential growth of correlators for $\xi<0$ and $\left(m/H\right)^2\ll|\xi|\ll1$ is due to the similarity of their mode functions in the massless limit: in inflation and in matter era the CTBD mode functions are Hankel functions of the same order.

The coincident correlators in matter era $\Delta_{\phi,\phi}(N_M,0)$, $\Delta_{\phi,\pi}(N_M,0)$ and $\Delta_{\pi,\pi}(N_M,0)$ determine the energy density and pressure via eqs.~(\ref{eq: rho_Q long})--(\ref{eq: p_Q long}), using $\epsilon=3/2$. At leading order in $|\xi|$ and $\left(m/H(N_M)\right)^2$, they are approximated\footnote{Similarly to inflation, the correlators $\Delta_{\phi,\pi}(N_M,0)$ and $\Delta_{\pi,\pi}(N_M,0)$ have been neglected in eqs.~(\ref{eq: rho_Q mat})--(\ref{eq: p_Q mat}) because their contribution to energy density and pressure is suppressed by another factor of $|\xi|$ with respect to $\Delta_{\phi,\phi}(N_M,0)$.} by
\begin{equation}
\label{eq: rho_Q mat}
\rho_Q(N_M)\approx\left(\frac{m^2}{2}-3|\xi|H^2(N_M)\right)\Delta_{\phi,\phi}(N_M,0)\,,
\end{equation}
\begin{eqnarray}
p_Q(N_M)\approx-\frac{m^2}{2}\Delta_{\phi,\phi}(N_M,0)\,.
\label{eq: p_Q mat}
\end{eqnarray}
By inspection of eqs.~(\ref{eq: rho_Q mat})--(\ref{eq: p_Q mat}), we can see a CC-like (cosmological constant) term and a negative matter-like contribution, in agreement with eq. (90) of~\cite{Glavan:2017jye}.
This result can be used to match the model to the current cosmological parameters and reproduce the right amount of dark energy. The consequences of this step are revised in the subsection~\ref{subsec: match}. Before doing that, we discuss the other important feature of the dark energy model of quantum origin treated in this paper, namely spatial correlations. For this purpose, a study of non-coincident 4-pt functions is needed.

\subsection{Non-coincident 4-point IR correlators}\label{subsect:corr 4pt}
Since in the model described in this paper dark energy is a consequence of quantum fluctuations amplified by inflation, one can expect that important physical information can be extracted not only from the expectation value of the energy-momentum tensor at a single point (as in~eqs.~(\ref{eq: rho_Q long})--(\ref{eq: p_Q long})), but also from correlations between the energy-momentum tensor at two different space points. Thus, a source of information on the features of dark energy predicted by the simple action~(\ref{eq:action}) is in the correlator $\langle \hat{\rho}_Q(t,\vec{x}_1) \hat{\rho}_Q(t,\vec{x}_2)\rangle$ between energy densities\footnote{The correlator $\langle\hat{\rho}_Q(x)\hat{\rho}_Q(x')\rangle$ is an essential ingredient in building the $\langle\hat{T}_{\mu\nu}(x)\hat{T}_{\rho\sigma}(x')\rangle$ correlator which, together with the local contribution from the graviton 4-pt vertex, builds the one-loop graviton self-energy. Therefore, studying the $\langle\hat{\rho}_Q(x)\hat{\rho}_Q(x')\rangle$ correlator teaches us something on the off-coincident one-loop graviton self-energy.} at the two comoving positions $\vec{x}_1$ and $\vec{x}_2$, evaluated at the same time $t$. This was one of the main subjects of study in~\cite{Belgacem:2021ieb}. Here we want to complete the presentation given in that paper and show all the definitions/equations needed, which were omitted in~\cite{Belgacem:2021ieb} for brevity.
In stochastic formalism one approximates the occurrences of field/canonical momentum in correlators with their long-wavelength parts. Starting from eq.~(\ref{eq: rho_Q operator}) and neglecting spatial gradients, the density-density correlator (at equal times) is expressed in terms of six 4-point IR correlators as follows:
\begin{align}
\langle \hat{\rho}_Q(t,\vec{x}_1) \hat{\rho}_Q(t,\vec{x}_2)\rangle  \approx \frac{H^4}{4} \biggl\{ \Bigl[ \Bigl( \frac{m}{H} \Bigr)^{\!2} +6\xi \Bigr]^2 \Delta_{\phi^2,\phi^2}(t,r)+6\xi \Bigl[ \Bigl( \frac{m}{H} \Bigr)^{\!2} +6\xi \Bigr]\Delta_{\phi^2,\phi\pi}(t,r)\,\nonumber\\
+36\xi^2\Delta_{\phi\pi,\phi\pi}(t,r)+\Bigl[ \Bigl( \frac{m}{H} \Bigr)^{\!2} +6\xi \Bigr]\Delta_{\phi^2,\pi^2}(t,r)+6\xi\Delta_{\phi\pi,\pi^2}(t,r)+\Delta_{\pi^2,\pi^2}(t,r) \biggr\} \,.
\label{rho-rho_correlator}
\end{align}

The complete definitions of the 4-point functions needed\footnote{The definition of the first correlator $\Delta_{\phi^2,\phi^2}(t,r)$ was already given in eq. (9) of~\cite{Belgacem:2021ieb}.} are:
\begin{eqnarray}
\Delta_{\phi^2,\phi^2}(t,r) & \equiv & 
	\Bigl\langle \hat{\phi}^2 (t,\vec{x}_1) \hat{\phi}^2 (t,\vec{x}_2) \Bigr\rangle \, ,
\label{DeltaPhi2,Phi2}
\\
\Delta_{\phi^2,\phi\pi}(t,r) & \equiv & 
	\frac{1}{a^{3}(t) H(t)}\Bigl\langle \hat{\phi}^2 (t,\vec{x}_1)\left\{\hat{\phi} (t,\vec{x}_2),\hat{\pi} (t,\vec{x}_2)\right\}+\left\{\hat{\phi} (t,\vec{x}_1),\hat{\pi} (t,\vec{x}_1)\right\}\hat{\phi}^2 (t,\vec{x}_2) \Bigr\rangle \, ,
\label{DeltaPhi2,PhiPi}
\\
\Delta_{\phi\pi,\phi\pi}(t,r) & \equiv & 
	\frac{1}{a^{6}(t) H^2(t)}\Bigl\langle \left\{\hat{\phi} (t,\vec{x}_1),\hat{\pi} (t,\vec{x}_1)\right\} \left\{\hat{\phi} (t,\vec{x}_2),\hat{\pi} (t,\vec{x}_2)\right\} \Bigr\rangle \, ,
\label{DeltaPhiPi,PhiPi}
\\
\Delta_{\phi^2,\pi^2}(t,r) & \equiv & 
	\frac{1}{a^{6}(t) H^2(t)}\Bigl\langle \hat{\phi}^2 (t,\vec{x}_1) \hat{\pi}^2 (t,\vec{x}_2)+\hat{\pi}^2 (t,\vec{x}_1) \hat{\phi}^2 (t,\vec{x}_2) \Bigr\rangle \, ,
\label{DeltaPhi2,Pi2}
\\
\Delta_{\phi\pi,\pi^2}(t,r) & \equiv & 
	\frac{1}{a^{9}(t) H^3(t)}\Bigl\langle \hat{\pi}^2 (t,\vec{x}_1)\left\{\hat{\phi} (t,\vec{x}_2),\hat{\pi} (t,\vec{x}_2)\right\}+\left\{\hat{\phi} (t,\vec{x}_1),\hat{\pi} (t,\vec{x}_1)\right\}\hat{\pi}^2 (t,\vec{x}_2) \Bigr\rangle \, ,
\label{DeltaPhiPi,Pi2}
\\
\Delta_{\pi^2,\pi^2}(t,r) & \equiv & 
	\frac{1}{a^{12}(t) H^4(t)} \Bigl\langle \hat{\pi}^2 (t,\vec{x}_1) \hat{\pi}^2 (t,\vec{x}_2) 
	\Bigr\rangle \, .
\label{DeltaPi2,Pi2}
\end{eqnarray}
where, as always, $r\equiv\|\vec{x}_2-\vec{x}_1\|$ is the relative comoving distance between the two points.
Similarly to the 2-pt IR correlator, also these non-coincident 4-pt IR correlators evolve in time under the effect of noise sources. In terms of the number of e-foldings~(\ref{eq: efolds definition}), the full system of equations describing the process (including eqs. (10), (11) in~\cite{Belgacem:2021ieb}) is

\begin{eqnarray}
\frac{\partial}{\partial N} \Delta_{\phi^2,\phi^2} - \Delta_{\phi^2,\phi\pi} = n_{\phi^2,\phi^2} \, ,
\label{eomPhi2,Phi2}
\\
\frac{\partial}{\partial N} \Delta_{\phi^2,\phi\pi} + (3-\epsilon) \Delta_{\phi^2,\phi\pi}
	- 2 \Delta_{\phi\pi,\phi\pi}-2\Delta_{\phi^2,\pi^2}+ 4 \Bigl( \frac{M}{H} \Bigr)^{\!2} 
		\Delta_{\phi^2,\phi^2} = n_{\phi^2,\phi\pi} \, ,
\label{eomPhi2,PhiPi}
\\
\frac{\partial}{\partial N} \Delta_{\phi\pi,\phi\pi} +2(3-\epsilon) \Delta_{\phi\pi,\phi\pi}- 2\Delta_{\phi\pi,\pi^2}+2 \Bigl( \frac{M}{H} \Bigr)^{\!2} 
		\Delta_{\phi^2,\phi\pi} = n_{\phi\pi,\phi\pi} \, ,
\label{eomPhiPi,PhiPi}
\\
\frac{\partial}{\partial N} \Delta_{\phi^2,\pi^2} + 2(3-\epsilon) \Delta_{\phi^2,\pi^2}
	-  \Delta_{\phi\pi,\pi^2}+ \Bigl( \frac{M}{H} \Bigr)^{\!2} 
		\Delta_{\phi^2,\phi\pi} = n_{\phi^2,\pi^2} \, ,
\label{eomPhi2,Pi2}
\\
\frac{\partial}{\partial N} \Delta_{\phi\pi,\pi^2} +3(3-\epsilon) \Delta_{\phi\pi,\pi^2}- 4\Delta_{\pi^2,\pi^2}+2 \Bigl( \frac{M}{H} \Bigr)^{\!2} 
		\Delta_{\phi^2,\pi^2}+2 \Bigl( \frac{M}{H} \Bigr)^{\!2} 
		\Delta_{\phi\pi,\phi\pi} = n_{\phi\pi,\pi^2} \, ,
\label{eomPhiPi,Pi2}
\\
\frac{\partial}{\partial N} \Delta_{\pi^2,\pi^2} + 4(3-\epsilon) \Delta_{\pi^2,\pi^2}
	+ \Bigl( \frac{M}{H} \Bigr)^{\!2} 
		\Delta_{\phi\pi,\pi^2} = n_{\pi^2,\pi^2} \, .
\label{eomPi2,Pi2}
\end{eqnarray}
The sources appearing at the right-hand sides of the equations above can be expressed in terms of the field operator $\hat{\phi}$, canonical momentum operator $\hat{\pi}$ and stochastic forces $\hat{f}_{\phi}$, $\hat{f}_{\pi}$. They are given by eq.~(\ref{eq: noise def}) in Appendix~\ref{app: 4-pt stoch sources}. Starting from those definitions, they can be written in momentum space in terms of integrals involving the mode function $\varphi(t,k)$. The resulting expressions are listed in eq.~(\ref{eq: noise}).

One can then study how the 4-pt correlators evolve during different cosmological epochs (de Sitter, radiation domination and matter domination) via their coupled equations~(\ref{eomPhi2,Phi2})--(\ref{eomPi2,Pi2}), following essentially the same steps that we already explained for the evolution of 2-pt correlators in eqs.~(\ref{eomPhiPhi})--(\ref{eomPiPi}). Since this was already done in~\cite{Belgacem:2021ieb}) (see in particular its Appendix A), we recall the results from there and make additional comments.

\subsubsection{de Sitter inflation}

Similarly to the definition~(\ref{eq: def Delta2}), let us collect the 4-pt functions~(\ref{DeltaPhi2,Phi2})--(\ref{DeltaPi2,Pi2}) in a column vector $\Delta_{(\rm 4)}$, whose corresponding row form (transpose vector $\Delta_{(\rm 4)}^{\rm T}$) is
\begin{equation}
\Delta_{(\rm 4)}^{\rm T}\equiv \left(\Delta_{\phi^2,\phi^2},\Delta_{\phi^2,\phi\pi},\Delta_{\phi\pi,\phi\pi},\Delta_{\phi^2,\pi^2},\Delta_{\phi\pi,\pi^2},\Delta_{\pi^2,\pi^2}\right)\,,
\end{equation}
and we can do the same thing for the noise sources defining a column vector $n_{(\rm 4)}$, such that
\begin{equation}
n_{(\rm 4)}^{\rm T}\equiv \left(n_{\phi^2,\phi^2},n_{\phi^2,\phi\pi},n_{\phi\pi,\phi\pi},n_{\phi^2,\pi^2},n_{\phi\pi,\pi^2},n_{\pi^2,\pi^2}\right)\,.
\end{equation}
The system~(\ref{eomPhi2,Phi2})--(\ref{eomPi2,Pi2}) in de Sitter inflation takes the compact form
\begin{equation}
\frac{\partial}{\partial N}\Delta_{(\rm 4)}(N,r)+A_{(\rm 4),I}~\Delta_{(\rm 4)}(N,r)=n_{(\rm 4)}(N,r)\,,
\label{system_deSitter 4pt}
\end{equation}
where the constant matrix $A_{(\rm 4),I}$ and the noise functions $n_{(\rm 4)}(N,r)$ were evaluated in eqs. (A4) and (A5) of~\cite{Belgacem:2021ieb}).
The matrix $A_{(\rm 4),I}$, in terms of $X\equiv(M/H_I)^2=(m/H_I)^2+12\xi$, is
\begin{equation}
A_{(\rm 4),I}=
\begin{pmatrix}
   0 & -1 & 0 & 0 & 0 & 0 \\
 4X & 3 & -2 & -2 & 0 & 0 \\
 0 & 2X & 6 & 0 & -2 & 0 \\
 0 & X & 0 & 6 & -1 & 0 \\
 0 & 0 & 2X & 2X & 9 & -4 \\
 0 & 0 & 0 & 0 & X & 12 \\
\end{pmatrix}\,.
\end{equation}

We also rewrite here the expression for the noise sources (as a row vector) including the effect of a general speed of sound, which was not studied in~\cite{Belgacem:2021ieb}:
\begin{equation}
n_{(\rm 4)}^{\rm T}(N,r)=c_s^{-6}\left(\frac{H_I}{2\pi}\right)^4 \left[1+2j_0(\mu a_{\rm in}{\mathrm e}^N H_I r)\right]\left(\frac{1}{4|\xi|},4,16|\xi|,8|\xi|,64\xi^2,64|\xi|^3\right)\,.
\label{eq: noise_deSitter 4pt}
\end{equation}
Using a method perfectly analogous to that in subsection~\ref{subsubsec: 2pt de Sitter}, one can prove (see eq. (A7) of~\cite{Belgacem:2021ieb}) that, assuming zero initial conditions at the beginning of inflation and working with $\xi<0$ and $(m/H_I)^2\ll|\xi|\ll 1$, then at the end of inflation the non-coincident 4-pt correlators evolve into

\begin{equation}
\label{4pt_inflation}
\boxed{
\Delta_{(\rm 4)}^{\rm T}(N_I,r) \simeq c_s^{-6}\frac{H_I^4}{1024\pi^4\xi^2}e^{16|\xi|N_I}s_{(\rm 4)}(r)~(1,16|\xi|,64\xi^2,32\xi^2,256|\xi|^3,256\xi^4)}\,,
\end{equation}
where the function $s_{(\rm 4)}(r)$ containing the spatial dependence is:

\begin{equation}
\label{spatial_inflation}
\boxed{
s_{(\rm 4)}(r)\simeq 
\begin{cases}
      3 & \text{if $r< r_0~e^{-N_I}$} \\
      3-2\left(\frac{r}{r_0}\right)^{16|\xi|} & \text{if $r_0~e^{-N_I}<  r< r_0$} \\
      1 & \text{if $r>r_0$}\,.
\end{cases}
}
\end{equation}
The scale $r_0\equiv\left(\mu a_{\rm in}H_I\right)^{-1}$ was already introduced in eq.~(\ref{eq: def r0 infl}).
It is interesting to compare the spatial dependence of the 4-pt functions with the one of 2-pt functions. For definiteness, let us focus on $\Delta_{\phi^2,\phi^2}(N_I,r)$ and $\Delta_{\phi,\phi}(N_I,r)$ given by the first component of eqs.~(\ref{4pt_inflation}) and~(\ref{sol_deSitter 2pt 5}), respectively.

One would {\it expect} from Wick's theorem (which holds at one-loop level) that
\begin{equation}
\label{eq: Wick ?}
\Delta_{\phi^2,\phi^2}(N_I,r)\stackrel{?}{=}\left[\Delta_{\phi,\phi}(N_I,0)\right]^2+2\left[\Delta_{\phi,\phi}(N_I,r)\right]^2\,.
\end{equation}
Comparing the results
\begin{equation}
\Delta_{\phi^2,\phi^2}(N_I,r) \simeq c_s^{-6}\frac{H_I^4}{1024\pi^4\xi^2}e^{16|\xi|N_I}s_{(\rm 4)}(r)\,
\end{equation}
and
\begin{equation}
\Delta_{\phi,\phi}(N_I,r) \simeq c_s^{-3}\frac{H_I^2}{32\pi^2|\xi|}e^{8|\xi|N_I}s_{(\rm 2)}(r)\,,
\end{equation}
eq.~(\ref{eq: Wick ?}) is equivalent to check whether
\begin{equation}
\label{eq: Wick ? reduced}
s_{(\rm 4)}(r)\stackrel{?}{=}\left[s_{(\rm 2)}(0)\right]^2+2\left[s_{(\rm 2)}(r)\right]^2\,.
\end{equation}
But using eq.~(\ref{eq: profile 2-pt s_2(r)}), the right-hand side of eq.~(\ref{eq: Wick ? reduced}) is
\begin{equation}
\label{eq: RHS Wick fails}
\left[s_{(\rm 2)}(0)\right]^2+2\left[s_{(\rm 2)}(r)\right]^2\simeq 
\begin{cases}
      3 & \text{if $r< r_0~e^{-N_I}$} \\
      3-4\left(\frac{r}{r_0}\right)^{8|\xi|}+2\left(\frac{r}{r_0}\right)^{16|\xi|} & \text{if $r_0~e^{-N_I}<  r< r_0$} \\
      1 & \text{if $r>r_0$}\,.
\end{cases}
\end{equation}
This is clearly different from the function $s_{(\rm 4)}(r)$ in eq.~(\ref{spatial_inflation}). More precisely they are equal only for $r< r_0~e^{-N_I}$ (basically the coincident regime) and $r>r_0$ (large distance regime), but they disagree at intermediate scales $r_0~e^{-N_I}<  r< r_0$ when interpolating between the extreme values of $3$ and $1$.
We infer that Wick's theorem is not satisfied by the result for 4-pt correlators~(\ref{4pt_inflation})--(\ref{spatial_inflation}).
This is a problem because the spectator scalar field $\hat{\Phi}$, described by the action~(\ref{eq:action}), is free\footnote{This is true when quantum gravitational effects are neglected and when the quantum backreaction of the scalar is negligibly small.} when considering a given classical background metric, which is legitimate throughout all the evolution of the Universe, except for the very few most recent e-foldings of evolution in matter era (see also the discussion after eq.~(\ref{eq: Wick contractions})).
We do not fully understand the origin of this problem, but we think it lies in the form of the 4-pt stochastic sources $n_{(\rm 4)}(N,r)$. It is possible that they do not fully catch the interaction between short and long modes.
Section~\ref{subsec: QFT Wick} shows that, as expected for free fields, quantum field theory (QFT) predicts that Wick's theorem is obeyed. QFT says that the spatial dependence $s_{(\rm 2)}(r)$ of 2-pt functions in eq.~(\ref{eq: profile 2-pt s_2(r)}) is correct, while $s_{(\rm 4)}(r)$ for 4-pt functions in eq.~(\ref{spatial_inflation}) is not.
An alternative way to apply (and rescue) stochastic formalism is proposed in Section~\ref{sec: FP stochastic} and it works with classical probability distributions instead of systems of equations for IR correlators. One of the results therein is that the joint probability distribution of fields at two points is Gaussian and therefore correlators evaluated from it obey Wick's theorem, see Fig.~\ref{fig: 4pt comparison}.

\subsubsection{Radiation epoch}
Once inflation ends, the Unverse enters an epoch where expansion is dominated by radiation. In this case $\epsilon=2$ and the quantity $(M/H)^2$ appearing in eqs.~(\ref{eomPhi2,Phi2})--(\ref{eomPi2,Pi2}) is small since it reduces to the mass contribution $(m/H)^2$. For our purposes it can be safely neglected, like we did in subsection~\ref{subsubsec: 2pt rad}. Furthermore, similarly to the case of 2-pt functions, the stochastic sources $n_{(\rm 4)}$ are much less relevant than the initial conditions inherited from inflation, because only initial conditions contain an enhancement factor that is exponential in $N_I$. Then, with the same technique used for 2-pt functions in subsection~\ref{subsubsec: 2pt rad}, the initial conditions~(\ref{4pt_inflation}) are easily evolved through radiation domination and, at the end of it (lasting $N_R\approx50$ e-foldings until matter-radiation equality) the 4-point functions are (see also Appendix A.2 of~\cite{Belgacem:2021ieb})

\begin{equation}
\label{4pt_radiation}
\boxed{
\begin{pmatrix}
\Delta_{\phi^2,\phi^2}(N_R,r)\\
\Delta_{\phi^2,\phi\pi}(N_R,r)\\
\Delta_{\phi\pi,\phi\pi}(N_R,r)\\
\Delta_{\phi^2,\pi^2}(N_R,r)\\
\Delta_{\phi\pi,\pi^2}(N_R,r)\\
\Delta_{\pi^2,\pi^2}(N_R,r)\\
\end{pmatrix}
\simeq 
\begin{pmatrix}
\Delta_{\phi^2,\phi^2}(N_I,r)\\
\Delta_{\phi^2,\phi\pi}(N_I,r)~e^{-N_R}\\
\Delta_{\phi\pi,\phi\pi}(N_I,r)~e^{-2N_R}\\
\Delta_{\phi^2,\pi^2}(N_I,r)~e^{-2N_R}\\
\Delta_{\phi\pi,\pi^2}(N_I,r)~e^{-3N_R}\\
\Delta_{\pi^2,\pi^2}(N_I,r)~e^{-4N_R}\\
\end{pmatrix}
\simeq 
\begin{pmatrix}
\Delta_{\phi^2,\phi^2}(N_I,r)\\
0\\
0\\
0\\
0\\
0\\
\end{pmatrix}
}\,.
\end{equation}

Note that, within the approximations mentioned, $\Delta_{\phi^2,\phi^2}$ is constant during the radiation period, while the other correlators are suppressed by a factor of $e^{-N_R}$ for each occurrence of a canonical momentum in their definition. This is perfectly analogous to the result for 2-point functions in eq.~(\ref{sol_rad 2pt 2}). At the end of radiation domination ($N_R\approx50$), all the correlators involving canonical momentum are negligible, which is the content of the last equality in eq.~(\ref{4pt_radiation}).
Only $\Delta_{\phi^2,\phi^2}$ survives (roughly unchanged) until matter-radiation equality. In matter-dominated epoch it will ``leak" again into the other correlators.
\subsubsection{Matter epoch}
\label{subsubsec: 4pt mat}
At matter-radiation equality, non-relativistic matter takes the lead of cosmological expansion. 
Eqs.~(\ref{eq: rho_Q mat}), (\ref{eq: p_Q mat}) show that the scalar field contributes to the energy-momentum tensor with a cosmological constant (CC) type portion and a negative matter-like portion. The CC-like part is what manifests as dark energy at recent cosmological times. In Section II.C and Appendix A.3 of~\cite{Belgacem:2021ieb}, the evolution of the scalar field was studied in a Universe containing both matter and a cosmological constant, as a way to take into account the backreaction of the scalar field on the expansion. Appendix~\ref{app: mat+CC} can be used to review (and even refine) those results.
Here, for a simple and more fair comparison with the discussion on 2-pt functions in subsection~\ref{subsubsec: 2pt mat}, we give the results for 4-pt functions in pure matter epoch.
Using the same approximations of subsection~\ref{subsubsec: 2pt mat}, one arrives with similar calculations to the following 4-pt IR correlators  in matter era, after $N_M$ e-foldings from matter-radiation equality:

\begin{equation}
\label{4pt_matter}
\boxed{
\Delta_{(\rm 4)}^{\rm T}(N_M,r) \simeq \Delta_{\phi^2,\phi^2}(N_R,r)~e^{8|\xi|N_M}~\left(1,8|\xi|,16\xi^2,8\xi^2,32|\xi|^3,16\xi^4\right)
}\,.
\end{equation}

At leading order in $|\xi|$ and $\left(m/H(N_M)\right)^2$, the corresponding density-density correlator computed from eq.~(\ref{rho-rho_correlator}) is dominated by $\Delta_{\phi^2,\phi^2}(N_M,r)$ and gives
\begin{equation}
\boxed{
\langle \hat{\rho}_Q(N_M,\vec{x}_1) \hat{\rho}_Q(N_M,\vec{x}_2)\rangle\approx \left[\rho_Q (N_M)\right]^2~s_{(\rm 4)}(r)\qquad\text{with~~$r\equiv\|\vec{x}_2-\vec{x}_1\|$}
}\,,
\label{eq: rho-rho matter era}
\end{equation}
where we also made use of eqs.~(\ref{sol_mat 2pt 2}), (\ref{eq: rho_Q mat}) and the function $s_{(\rm 4)}(r)$ was defined in~(\ref{spatial_inflation}).
Once again, the scale of spatial correlations is determined by $r_0$ appearing in $s_{(\rm 4)}(r)$ and therefore by the comoving Hubble length at the beginning of inflation (from eq.~(\ref{eq: def r0 infl})).

\subsection{Matching at current time}
\label{subsec: match}
The model parameters have to be matched to the current cosmological parameters in order to study its actual predictions. In the last few e-foldings the backreaction of the scalar field on the background metric due to its energy-momentum tensor becomes very important, to the point that eventually it becomes a fundamental contribution to the total energy density of the Universe and in the future it will lead the expansion. A full solution of the problem would unavoidably be numerical. However some simplified understanding is possible in analytical form. For the moment we approximate the FLRW background simply as a matter-dominated Universe, but in Section~\ref{sect: qbackr} we will refine the matching by a more consistent account for the quantum backreaction at late times.
As usual in the literature, let us call $\Omega_M$ the energy density fraction today due to non-relativistic matter (baryons and dark matter), $\Omega_R$ the little fraction due to energy density in radiation and, assuming zero spatial curvature\footnote{CMB data from {\it Planck} combined with baryon acoustic oscillations (BAO) give the constraint $\Omega_K=0.0007\pm 0.0019$ on the spatial curvature parameter, see~\cite{Planck:2018vyg}.}, $\Omega_\Lambda\simeq1-\Omega_M$ is the dark energy contribution (in the form of a cosmological constant). The Hubble parameter today is $H_0$.

In the dark energy model studied in this paper, the cosmological constant type contribution to the energy density (due to the scalar field) can be read from the first term of eq.~(\ref{eq: rho_Q mat}) and it is
\begin{equation}
\label{eq: matching DE}
\rho_\Lambda(N_M)\simeq\frac{m^2}{2}\Delta_{\phi,\phi}(N_M,0)\simeq\frac{m^2}{2}c_s^{-3}\frac{H_I^2}{32\pi^2|\xi|}e^{8|\xi|N_I+4|\xi|N_M}
\end{equation}
Today, after $N_M=\ln\left(\frac{\Omega_M}{\Omega_R}\right)$ e-foldings from matter-radiation equality, this quantity has to match the amount of dark energy density $\rho_\Lambda(N_M)=3 M_P^2 H_0^2 \Omega_\Lambda=\frac{3}{8\pi}m_P^2 H_0^2 \Omega_\Lambda$, where $m_P\equiv G^{-1/2}\simeq1.2\times 10^{19}~{\rm GeV}$ is the Planck mass in natural units and $M_P\equiv \left(8\pi G\right)^{-1/2}\simeq2.4\times 10^{18}~{\rm GeV}$ is the reduced Planck mass.

This gives the required length of inflation in terms of the other parameters:
\begin{equation}
N_I = \frac{1}{8|\xi|}  \ln \biggl[ 
	24\pi|\xi| \Bigl( \frac{m_P}{H_I} \Bigr)^2
	\Bigl( \frac{H_{DE}}{m} \Bigr)^2 c_s^3
	\biggr]-\frac{N_M}{2}
\,,
\label{inflation_efolds_relation}
\end{equation}
where $H_{\rm DE}\equiv H_0\sqrt{\Omega_\Lambda}$ is the Hubble parameter when dark energy will completely dominate the cosmological expansion. This situation will be realized provided that the CC-like term dominates over the matter-like term in $\rho_Q$ at late times. From eq.~(\ref{eq: rho_Q mat}), this requirement imposes
\begin{equation}
|\xi|<\frac{1}{6}\left(\frac{m}{H_{\rm DE}}\right)^2
\label{limits_xi}
\end{equation}
with the assumption $\xi<0$.
The hypothesis that the scalar field is light, which was used in all derivations and allows the best enhancement of quantum fluctuations, is true at all stages of cosmological evolution (until late times) if
\begin{equation}
\label{eq: light field}
\frac{m}{H_{\rm DE}}<1\,.
\end{equation}

Eq.~(\ref{inflation_efolds_relation}) refines the estimates in~\cite{Glavan:2017jye} and~\cite{Belgacem:2021ieb}, which did not take into account the growth of correlators in matter era and includes the effect of the speed of sound $c_s$.

\noindent For $\Omega_M=0.3$ and $\Omega_R=9.1\times10^{-5}$, then the number of e-foldings of matter era (from matter-radiation equality until today) is $N_M\simeq8.1$. Therefore, looking at eq.~(\ref{inflation_efolds_relation}) the growth in matter era only decreases the required number of e-foldings by about $4$.

It is convenient to introduce the dimensionless parameter\footnote{The condition~(\ref{eq: light field}) also imposes $\alpha<\frac{1}{6|\xi|}$. Together with the inequality in eq.~(\ref{eq: def alpha 1}), it means that, for a given value of $\xi<0$, the parameter $\alpha$ can take values $1<\alpha<\frac{1}{6|\xi|}$. This requires $|\xi|<\frac16$.} $\alpha$ defined as
\begin{equation}
\label{eq: def alpha 1}
\boxed{
\alpha\equiv\frac{1}{6|\xi|}\left(\frac{m}{H_{\rm DE}}\right)^2>1
}\,
\end{equation}
where the inequality follows from eq.~(\ref{limits_xi}) and we rewrite the result (\ref{inflation_efolds_relation}) as
\begin{equation}
\boxed{
N_I = \frac{1}{8|\xi|}  \ln \biggl[ 
	\frac{4\pi}{\alpha} \Bigl( \frac{m_P}{H_I} \Bigr)^2
	 c_s^3
	\biggr]-\frac{1}{2}\ln \left(\frac{\Omega_M}{\Omega_R}\right)
}\,.
\label{eq: NI inflation simplified}
\end{equation}

\noindent Then, using~(\ref{eq: matching DE}), the energy density today (cosmological time $t_0$) due to quantum fluctuations of the scalar field is found from~(\ref{eq: rho_Q mat}) to be
\begin{equation}
\label{eq:rhoquantum}
\rho_{\rm Q}(t_0)\simeq3M_P^2 H_0^2\left(\Omega_\Lambda-\frac{1}{\alpha}\right)\,.
\end{equation}
The rest of the energy density in the current Universe must be due to the the classical energy density of matter $\rho_{C}(t_0)$ that was used as a background for the evolution of the scalar field, such that $3M_P^2 H_0^2=\rho_{C}(t_0)+\rho_{\rm Q}(t_0)$, giving
\begin{equation}
\label{eq:rhoclassical}
\rho_{C}(t_0)\simeq3M_P^2 H_0^2\left(\Omega_M+\frac{1}{\alpha}\right)\,.
\end{equation}

\noindent In Section~\ref{sect: qbackr}, we study how a simple, but physically more consistent, modelization of quantum backreaction, modifies eqs.~(\ref{eq:rhoquantum})--(\ref{eq:rhoclassical}), see in particular eq.~(\ref{eq: rhoQ br ratio}).

We conclude this section by discussing how the scale of spatial correlations $r_0=\left(\mu a_{\rm in}H_I\right)^{-1}$, governing e.g. density correlations in eq.~(\ref{eq: rho-rho matter era}), is related to the other cosmological/model parameters. Assuming for simplicity an instantaneous reheating after inflation and matching the inflationary Hubble parameter $H_I$ with that at the beginning of the radiation era, we find the length scale of spatial variations in units of the current Hubble horizon $H_0^{-1}$ (see also eq. (20) of~\cite{Belgacem:2021ieb}):
\begin{equation}
\label{spatial_ratio}
\boxed{
r_0 H_0\simeq\mu^{-1}e^{N_I} \left(\frac{H_I}{H_0}\right)^{-\frac12}\Omega_R^{-\frac14}=\mu^{-1}\left(\frac{H_0\sqrt{\Omega_R}}{H_I\Omega_M}\right)^{\frac12}\biggl[ 
	\frac{4\pi}{\alpha} \Bigl( \frac{m_P}{H_I} \Bigr)^2
	 c_s^3
	\biggr]^{\frac{1}{8|\xi|}}
}\,.
\end{equation}
As an example, let us consider $\Omega_M=0.3$, $\Omega_R=9.1\times10^{-5}$, $\Omega_\Lambda\simeq 0.7$, $H_0\simeq 10^{-33}~{\rm eV}$, $H_I\simeq 10^{13}~{\rm GeV}$, $\xi=-0.06$ (which is in the ballpark of the values used for relieving the Hubble tension in~\cite{Belgacem:2021ieb}), $\alpha\simeq1$ (its minimum allowed value) and a speed of sound equal to speed of light $c_s=1$. Then (using $m_P\simeq1.22\cdot10^{19}~{\rm GeV}$) the number of inflationary e-foldings required from eq.~(\ref{eq: NI inflation simplified}) is $N_I\simeq60$, while (taking for definiteness $\mu=1$) the scale $r_0$ from eq.~(\ref{spatial_ratio}) is $r_0\simeq 0.25 H_0^{-1}$.

It is also interesting to comment on the effect of a reduced speed of sound $c_s<1$, with respect to the base model with $c_s=1$. According to eq.~(\ref{eq: NI inflation simplified}), this produces a reduction in the required number of e-foldings to match the desired amount of dark energy today:
\begin{equation}
\label{eq: cs NI}
N_I(c_s)\approx N_I(1)-\frac{3}{8|\xi|}\ln{\left(\frac{1}{c_s}\right)}\,.
\end{equation}
The origin of such a reduction is in the increased amplitude of quantum fluctuations in inflation when $c_s<1$, ultimately determined by the mode function~(\ref{eq: phi infl approx}), proportional to $c_s^{-\nu_I}\approx c_s^{-3/2}$, which enters the noise sources $n_{(\rm 2)}$ for 2-pt correlators. A larger amplitude when $c_s<1$ implies that a smaller number of inflationary e-foldings is sufficient to reach the desired amount of dark energy, as eq.~(\ref{inflation_efolds_relation}) correctly predicts.

As a consequence, from eq.~(\ref{spatial_ratio}), the scale $r_0$ ruling the shape of spatial correlations is affected by $c_s$ as
\begin{equation}
\label{eq: cs r0}
r_0(c_s)\simeq c_s^{\frac{3}{8|\xi|}} r_0(1)\,.
\end{equation}
This effect could be used to increase the detectability of the model by studying fluctuations of luminosity distances from supernovae with surveys like LSST. In~\cite{inprep:2022lumdist} it has been shown that the amplitude of the angular power spectrum for luminosity distances fluctuations is proportional to $\left(r_0 H_0\right)^{-16|\xi|}$. Using eq.~(\ref{eq: cs r0}), the effect of the speed of sound on this amplitude is therefore
\begin{equation}
\left(r_0(c_s) H_0\right)^{-16|\xi|}\simeq c_s^{-6} \left(r_0(1) H_0\right)^{-16|\xi|}\,,
\end{equation}
amounting to an amplification by $c_s^{-6}$, which would augment the chances of detection.

\section{Comparing stochastic formalism and quantum field theory}
\label{sec: QFT}

Stochastic formalism has been mainly studied in inflation, and its predictions have been tested against a full quantum field theory (QFT) treatment only in the coincident limit. Here we want to check the stochastic results against QFT more generally, namely for non-coincident correlators and up to recent epochs. Stochastic theory describes the evolution of super-Hubble modes and, as expected, it agrees with QFT for super-Hubble separations. However, for intermediate non-coincident separations there is no reason a priori to trust the stochastic predictions and we find (in general) disagreement with QFT.
As we are dealing with a free spectator field (in inflation and in subsequent epochs, except for very recent cosmological times, quantum gravitational effects can be neglected and a classical treatment of gravity applies), the stochastic formulation can be tested by comparing with one-loop QFT results.

As the starting point of the field theoretic analysis, let us consider the non-coincident 2-point functions at equal times $\Delta_{\Phi\Phi}(t,r)$, $\Delta_{\Phi\Pi}(t,r)$ and $\Delta_{\Pi\Pi}(t,r)$ defined from the {\it full} free fields/canonical momenta (not their long-wavelength part) as
\begin{eqnarray}
\Delta_{\Phi\Phi}(t,r) & \equiv & 
	\Bigl\langle \hat{\Phi} (t,\vec{x}_1) \hat{\Phi} (t,\vec{x}_2) \Bigr\rangle \,,\nonumber
\\
\Delta_{\Phi\Pi}(t,r) & \equiv & 
	\frac{1}{a^3(t) H(t)}\Bigl\langle \bigl\{ \hat{\Phi}(t,\vec{x}_1) ,
						\hat{\Pi}(t,\vec{x}_2) \bigr\} \Bigr\rangle \,,\nonumber
\\
\Delta_{\Pi\Pi}(t,r) & \equiv & 
	\frac{1}{a^6(t) H^2(t)} \Bigl\langle \hat{\Pi} (t,\vec{x}_1) \hat{\Pi} (t,\vec{x}_2) \Bigr\rangle \,,
\label{Delta full}
\end{eqnarray}
where $r\equiv\|\vec{x}_2-\vec{x}_1\|$ as usual.
The correlators in~(\ref{Delta full}) can be obtained from Wightman functions defined as
\begin{equation}
i\Delta^{(+)}\left(x;x'\right)\equiv\langle\hat{\Phi}(x)\hat{\Phi}(x')\rangle\equiv\left[i\Delta^{(-)}\left(x;x'\right)\right]^*\,,\label{eq:Wightman def}
\end{equation}
by the following relations\footnote{Both the positive- and negative-frequency Wightman functions can be used for the equal-time correlators in~(\ref{eq: Wightman vs Delta relations}) because they are equal, at equal times, due to spatial homogeneity and isotropy of the FLRW metric and the vacuum state $|\Omega\rangle$. From eq.~(\ref{eq:Wightman def}) it follows that $\left[i\Delta^{(+)}\left(x;x'\right)\right]_{t'=t}=\left[i\Delta^{(-)}\left(x;x'\right)\right]_{t'=t}$ is real, which is consistent with eqs.~(\ref{eq: set corr QFT}).}:
\begin{eqnarray}
\Delta_{\Phi\Phi}(t,r)&=&\left[i\Delta^{(\pm)}\left(x;x'\right)\right]_{t'=t}\,,\nonumber\\
\Delta_{\Phi\Pi}(t,r)&=&\frac{1}{H(t)}\frac{\partial}{\partial t}\left[i\Delta^{(\pm)}\left(x;x'\right)\right]_{t'=t}\,,\nonumber\\
\Delta_{\Pi\Pi}(t,r)&=&\left[\frac{1}{H(t)}\frac{1}{H(t')}\frac{\partial}{\partial t}\frac{\partial}{\partial t'}i\Delta^{(\pm)}\left(x;x'\right)\right]_{t'=t}\,,
\label{eq: Wightman vs Delta relations}
\end{eqnarray}
where $x=\left(t,\vec{x}_1\right)$ and $x'=\left(t',\vec{x}_2\right)$ are spacetime points.

The goal of this Section is to compare off-coincident correlators ($r\neq0$) computed with the stochastic formalism to those evaluated in QFT. Coincident correlators ($r=0$) in stochastic formalism have already been compared with the QFT prediction in~\cite{Glavan:2017jye, Glavan:2015cut}. For the off-coincident correlators, it is sufficient to work in $D=4$ spacetime dimensions, while the coincident ones require to work in general spacetime dimensions $D$ to allow for dimensional regularization of divergences. Therefore, as long as we focus on off-coincident correlators, we work in $D=4$ spacetime dimensions as in previous Sections of this paper. Only in subsection~\ref{subsubsec: coinc limit}, when discussing the results in the coincident limit based on~\cite{Glavan:2017jye, Glavan:2015cut}, we will use general $D$ spacetime dimensions.

\noindent It is straightforward to express the correlators in~(\ref{Delta full}) in terms of the mode function $\varphi(t,k)$ using the decomposition in creation/annihilation operators given in eqs.~(\ref{eq: canon Phi}--\ref{eq: canon Pi}), to get
\begin{eqnarray}
\Delta_{\Phi\Phi}\left(t,r\right)&=&\frac{1}{2\pi^2}\int_{k_0}^{\infty}{\rm d}k~k^2 j_0(kr)|\varphi(t,k)|^2\,,\nonumber\\
\Delta_{\Phi\Pi}\left(t,r\right)&=&\frac{1}{2\pi^2}\int_{k_0}^{\infty}{\rm d}k~k^2 j_0(kr)\frac{1}{H(t)}\frac{\partial}{\partial t}|\varphi(t,k)|^2\,,\nonumber\\
\Delta_{\Pi\Pi}\left(t,r\right)&=&\frac{1}{2\pi^2}\int_{k_0}^{\infty}{\rm d}k~k^2 j_0(kr)\Big|\frac{1}{H(t)}\dot{\varphi}(t,k)\Big|^2\,,
\label{eq: set corr QFT}
\end{eqnarray}
where $j_0$ is the spherical Bessel functions of order $0$, $j_0(z)=\sin(z)/z$.
The infrared (IR) sector of the integral has been regularized with a time-independent comoving cutoff $k_0$. We can anticipate $k_0$ to be of the order of the inverse comoving Hubble horizon at the beginning of inflation $k_0={\mathcal O}(1)a_{\rm in} H_I$. Smaller values of the comoving wavenumber $k<k_0$ (i.e. modes with longer comoving wavelength $\lambda=2\pi/k$) correspond to scales which are already super-Hubble when inflation starts (during inflation the comoving Hubble sphere shrinks, so longer modes are the first to exit the Hubble sphere).

For a comparison between between stochastic theory and quantum field theory predictions, we will focus on the correlator $\Delta_{\Phi,\Phi}\left(t,r\right)$. Following ref. \cite{Glavan:2015cut}, where the authors studied the renormalized energy-momentum tensor for the same model (\ref{eq:action}), we work in conformal time $\eta$, defined from cosmological time $t$ by the relation $d\eta=dt/a(t)$ and rescale the mode function defining
\begin{equation}
U(\eta,k)\equiv a(\eta)~\varphi\left(t(\eta),k\right)\,.
\end{equation}
The non-coincident correlator $\Delta_{\Phi,\Phi}$ written in the first of eqs. (\ref{eq: set corr QFT}) is then
\begin{equation}
\label{eq: Delta U relation}
\Delta_{\Phi\Phi}\left(\eta,r\right)=\frac{1}{2\pi^2 a^2(\eta)}\int_{k_0}^{\infty}{\rm d}k~k^2 j_0(kr)|U(\eta,k)|^2\,.
\end{equation}

Let us focus for the moment on a trivial speed of sound $c_s=1$. We will discuss the effect of a reduced $c_s$ in subsection~\ref{subsubsec: QFT cs<1}.

Starting from eq. (\ref{modeEOMreduced}) and denoting by ``prime" ($^\prime$) a derivative with respect to conformal time, then $U(\eta,k)$ obeys the equation,
\begin{equation}
\label{eq: U eom}
U^{\prime\prime}+\left[k^2+m^2a^2-(1-6\xi)(2-\epsilon){\mathcal H}^2\right]U=0\,,
\end{equation}
and the Wronskian normalization (\ref{Wronskian}) implies
\begin{equation}
\label{eq: U Wronskian}
U~U^{\prime *}-U^*~U^{\prime}=i\,.
\end{equation}
In eqs. (\ref{eq: U eom}--\ref{eq: U Wronskian}), the scale factor $a$, the reduced Hubble parameter ${\mathcal H}\equiv a^\prime/a=aH$ and the parameter $\epsilon\equiv-\dot{H}/H^2=1-{\mathcal H}^\prime/{\mathcal H}^2$ are functions of the conformal time $\eta$, while $U$ is function of $\eta$ and $k$, as already said.

\subsection{Mode functions}
 In the three epochs of de Sitter inflation, radiation domination and matter domination, $\epsilon$ takes the constant values $0$, $2$ and $3/2$, respectively\footnote{When $\epsilon$ is constant, a useful relation for the background evolution is ${\mathcal H}a^{\epsilon-1}={\rm constant}$, which follows from integrating $\epsilon=1-{\mathcal H}^\prime/{\mathcal H}^2$. In particular, in de Sitter inflation ${\mathcal H}/a=H_I={\rm constant}$, in radiation epoch ${\mathcal H}a={\rm constant}$ and in matter epoch ${\mathcal H}^2 a={\rm constant}$.}.
In a period of constant $\epsilon$ the solutions for the mode functions $U(\eta,k)$ are known exactly in the massless limit. 
The massless approximation is definitely good for the very light non-minimally coupled scalar we are dealing with, except for the very few most recent e-foldings in matter era when the field backreacts on the FLRW background (leading to a dark energy component, as we have seen). Furthermore, working in the massless limit is sufficient for an interesting comparison between the stochastic formalism and field theory, which is the goal of this Section.
In this case, the mode function $U_\epsilon(\eta,k)$ which solves eqs. (\ref{eq: U eom}--\ref{eq: U Wronskian}) can be expressed as a linear combination of the Chernikov-Tagirov-Bunch-Davies (CTBD)  mode functions in the form (see \cite{Glavan:2015cut})
\begin{equation}
U_\epsilon(\eta,k) = \alpha_\epsilon(k) u_\epsilon(\eta,k)+\beta_\epsilon(k) u_\epsilon^*(\eta,k) \,,
\label{eq: full mode}
\end{equation}
where the CTBD mode function $u_\epsilon(\eta,k)$ is
\begin{eqnarray}
&&u_\epsilon(\eta,k) = \sqrt{\frac{\pi}{4|1-\epsilon|\mathcal{H}(\eta)}} H_{\nu_\epsilon}^{(1)} \left( \frac{k}{|1-\epsilon| \mathcal{H}(\eta)} \right) \,,\nonumber\\
&&{\rm with}~~~\nu_\epsilon \equiv \sqrt{\frac{1}{4}+(1-6\xi)\frac{2-\epsilon}{(1-\epsilon)^2}} \,.
\label{eq: CTBD mode}
\end{eqnarray}
$H_{\nu_\epsilon}^{(1)}$ is the Hankel function of the first kind with order $\nu_\epsilon$.

The coefficients $\alpha_\epsilon(k)$ and $\beta_\epsilon(k)$ are known as Bogolyubov coefficients. Since the CTBD mode function $u_\epsilon(\eta,k)$ in (\ref{eq: CTBD mode}) satisfies the Wronskian condition $u_\epsilon~u_\epsilon^{\prime *}-u_\epsilon^*~u_\epsilon^{\prime}=i$, then the same condition (\ref{eq: U Wronskian}) for the more general mode functions $U_\epsilon(\eta,k)$ requires
\begin{equation}
|\alpha_\epsilon(k)|^2 - |\beta_\epsilon(k)|^2 = 1\,.
\label{eq: Bogol}
\end{equation}
The form of the Bogolyubov coefficients in different epochs depends on the initial conditions at the the beginning of inflation, as well as on the precise way one matches periods with different $\epsilon$, i.e. reheating and matter-radiation transition. Following \cite{Belgacem:2021ieb},\cite{Glavan:2017jye}, \cite{Glavan:2015cut}, we assume that during inflation the full mode function is simply the CTBD mode function specialized to de Sitter\footnote{See also subsection III.C of \cite{Glavan:2015cut} for interesting comments on the initial state.} ($\epsilon=0$),
\begin{equation}
\label{eq: full mode inflation}
U_I(\eta,k)=\sqrt{\frac{\pi}{4{\mathcal H}(\eta)}}H_{\nu_I}^{(1)} \left(\frac{k}{{\mathcal H}(\eta)}\right)\,,
\end{equation}
where ${\mathcal H}(\eta)=a(\eta) H_I$, with $H_I$ the constant Hubble parameter in de Sitter. We remind that, in the case of de Sitter inflation ($\epsilon=0$), the order of the Hankel function (\ref{eq: CTBD mode}) is $\nu_0=\sqrt{\frac94-12\xi}$ and it can be shown that a mass $m$ can be simply included by promoting this index to
\begin{equation}
\label{eq: nu inflation}
\nu_I\equiv\sqrt{\frac94-12\xi-\left(\frac{m}{H_I}\right)^2}\,.
\end{equation}
As for the subsequent phases of radiation and matter domination, the Bogolyubov coefficients were determined in \cite{Glavan:2015cut} in the approximation of a sudden transition between different $\epsilon$ epochs\footnote{Limitations of this approximation were discussed in refs.~\cite{Glavan:2013mra, Glavan:2014uga}.}. Adopting a similar procedure as in \cite{Glavan:2015cut} (where the focus was on the computation of energy density and pressure), in the massless limit\footnote{The authors of ref. \cite{Glavan:2015cut} also discussed the exact massive CTBD mode function in radiation era (and an approximation for it) as well as the approximation at $\mathcal{O}\Bigl( \frac{ma}{\mathcal{H}} \Bigr)^{\!2}$ of the mode function in matter era. Here we only work in the massless limit.} we arrive, by imposing continuity of the mode function and its first derivative with respect to conformal time, at the following expressions for the full mode functions in radiation and matter epochs\footnote{The result in radiation era~(\ref{eq: full mode radiation}) looks simpler than in matter era~(\ref{eq: full mode matter}) because in radiation era the non-minimal coupling $\xi$ plays no role. Indeed, the order $\nu_\epsilon$ of the massless CTBD mode function (\ref{eq: CTBD mode}) reduces to 1/2 when $\epsilon=2$ and the Hankel function of order 1/2 is $H_{1/2}^{(1)}(z)=-i\sqrt{\frac{2}{\pi z}}~{\rm e}^{iz}$.}:

\begin{eqnarray}
U_R(\eta,k)&\approx& U_I(\eta_{\rm end},k)\left[\left(\nu_I-\frac12\right)\frac{{\mathcal H}_{\rm end}}{k}\sin{\left(\frac{k}{{\mathcal H}(\eta)}\right)}-\left(\nu_I-\frac32\right)\cos{\left(\frac{k}{{\mathcal H}(\eta)}\right)}\right]\,,\label{eq: full mode radiation}\\
U_M(\eta,k)&\approx&U_I(\eta_{\rm end},k)\left(\nu_I-\frac12\right)\frac{{\mathcal H}_{\rm end}}{{\mathcal H}_{\rm eq}}\left(\frac{{\mathcal H}_{\rm eq}}{{\mathcal H}(\eta)}\right)^{\frac12}\,\nonumber\\
&&\times\frac12\left[\left(\frac{3}{2}+\nu_M\right)\Gamma(\nu_M)\left(\frac{{\mathcal H}_{\rm eq}}{k}\right)^{\nu_M}J_{\nu_M}\left(\frac{2k}{{\mathcal H}(\eta)}\right)+\left(\nu_M\rightarrow-\nu_M\right)\right]\,,
\label{eq: full mode matter}
\end{eqnarray}
where $\nu_M$ in eq. (\ref{eq: full mode matter}) is the Hankel index of eq.~(\ref{eq: CTBD mode}) specialized to matter era\footnote{In the massless case, where eq. (\ref{eq: CTBD mode}) is valid, matter era and de Sitter inflation give the same order $\nu_\epsilon$ of the Hankel function. This is true because $\epsilon$ affects $\nu_\epsilon$ only through the combination $\frac{2-\epsilon}{(1-\epsilon)^2}$, which evaluates to $2$ for both $\epsilon=0$ and $\epsilon=\frac32$. More generally $\epsilon$ and $\frac{3-2\epsilon}{2-\epsilon}$ give the same $\nu_\epsilon$. An important difference is, however, that while in inflation comoving modes exit the Hubble sphere, in matter era they enter it.}, namely
\begin{equation}
\label{eq: nuM massless}
\nu_M=\sqrt{\frac94-12\xi}\,
\end{equation}
and $J_{\nu_M}$ is the Bessel function of order $\nu_M$. In the last line we denoted by $\left(\nu_M\rightarrow-\nu_M\right)$ a contribution equal to previous expression, but with $\nu_M$ replaced by $-\nu_M$.

In eqs. (\ref{eq: full mode radiation}--\ref{eq: full mode matter}), ${\mathcal H}_{\rm end}$ and ${\mathcal H}_{\rm eq}$ are the reduced Hubble parameter at the end of inflation and at matter-radiation equality, respectively. The matching with the inflationary mode function at the end of inflation originates the factor $U_I(\eta_{\rm end},k)$ in eqs. (\ref{eq: full mode radiation}--\ref{eq: full mode matter}). For $k\ll{\mathcal H}_{\rm end}$, it can be approximated as
\begin{equation}
U_I(\eta_{\rm end},k) \approx -\frac{i}{\sqrt{\pi}} 2^{\nu_I-1} \Gamma(\nu_I)\mathcal{H}_{\rm end}^{\nu_I-1/2} k^{-\nu_I}\,.
\label{eq: U infl end}
\end{equation}

It is also useful to have at hand the mode functions approximations for long modes (low $k$) because they will be useful for the infrared (IR) part of the integrals in the next subsection. At leading order in $k/{\mathcal H}(\eta)$, the three mode functions (\ref{eq: full mode inflation}, \ref{eq: full mode radiation}, \ref{eq: full mode matter}) reduce to
\begin{eqnarray}
U_I^{\rm (IR)}(\eta,k) &\approx& -\frac{i}{\sqrt{\pi}} 2^{\nu_I-1} \Gamma(\nu_I)\mathcal{H}(\eta)^{\nu_I-1/2} k^{-\nu_I}\,,\label{eq: mode IR}\\
U_R^{\rm (IR)}(\eta,k)&\approx& U_I(\eta_{\rm end},k)\left[\left(\nu_I-\frac12\right)\frac{{\mathcal H}_{\rm end}}{{\mathcal H}(\eta)}-\left(\nu_I-\frac32\right)\right]\,,\nonumber\\
U_M^{\rm (IR)}(\eta,k)&\approx& U_I(\eta_{\rm end},k)\left(\nu_I-\frac12\right)\frac{{\mathcal H}_{\rm end}}{{\mathcal H}_{\rm eq}}\frac12\left[\left(1+\frac{3}{2\nu_M}\right)\left(\frac{{\mathcal H}_{\rm eq}}{{\mathcal H}(\eta)}\right)^{\frac12+\nu_M}+\left(\nu_M\rightarrow-\nu_M\right)\right]\,.\nonumber
\end{eqnarray}
From eq.~(\ref{eq: mode IR}), it is straightforward to check the continuity of the mode function and of its first derivative with respect to conformal time in the transition between two epochs.

\subsection{The field 2-point function}
Once the mode functions have been determined, we can move to the determination of $\Delta_{\Phi\Phi}$ using eq. (\ref{eq: Delta U relation}). Here we outline a strategy to compute the required regularized integral. First one can formally write it as
\begin{equation}
\label{eq: Delta finite}
\Delta_{\Phi\Phi}\left(\eta,r\right)=\Delta_{\Phi\Phi}^{(\infty)}-\Delta_{\Phi\Phi}^{(k_0)}\,,
\end{equation}
where we defined
\begin{eqnarray}
\Delta_{\Phi\Phi}^{(\infty)}\equiv\frac{1}{2\pi^2 a^2(\eta)}\int_{k_{\rm IR}}^{\infty}{\rm d}k~k^2 j_0(kr)|U(\eta,k)|^2\,,\nonumber\\
\Delta_{\Phi\Phi}^{(k_0)}\equiv\frac{1}{2\pi^2 a^2(\eta)}\int_{k_{\rm IR}}^{k_0}{\rm d}k~k^2 j_0(kr)|U(\eta,k)|^2\,.
\label{eq: Delta parts}
\end{eqnarray}
The quantity $k_{\rm IR}$, used as the lower extreme of integration in eq.~(\ref{eq: Delta parts}), is a deep IR cutoff, $k_{\rm IR}\ll k_0$. It regulates the IR divergences, which would otherwise affect $\Delta_{\Phi\Phi}^{(\infty)}$ and $\Delta_{\Phi\Phi}^{(k_0)}$ if one had used $k=0$ as a lower extreme. The divergence is there\footnote{This can be seen from eq.~(\ref{eq: Delta infrared 2}). Since $j_0(0)=1$, the integral~(\ref{eq: Delta infrared 2}) would diverge for $\nu_I\geq\frac32$ if one had used $k=0$ as a lower extreme instead of $k=k_{\rm IR}$. The condition $\nu_I\geq\frac32$ applies to our case $(m/H_I)^2\ll|\xi|\ll1$ with $\xi<0$, giving from eq. (\ref{eq: nu inflation}), $\nu_I\simeq\frac32+4|\xi|$.} in the case of a massless (or very light) scalar field with negative non-minimal coupling $\xi<0$, which we are interested in, reducing eq.~(\ref{eq: nu inflation}) to
\begin{equation}
\label{eq: massless negaive xi}
\nu_I\simeq\frac32+4|\xi|\,.
\end{equation}
\noindent One can compute expressions for the integrals $\Delta_{\Phi\Phi}^{(\infty)}$ and $\Delta_{\Phi\Phi}^{(k_0)}$, and subtract them to obtain the result of~(\ref{eq: Delta finite}), where the contributions from $k_{\rm IR}$ cancel exactly.

Let us see explicitly how this works during inflation. The expression of $\Delta_{\Phi\Phi}^{(\infty)}$, using the mode function $U_I(\eta,k)$ in (\ref{eq: full mode inflation}), is

\begin{eqnarray}
\Delta_{\Phi\Phi, I}^{(\infty)}\left(\eta,r\right)&=&\frac{1}{2\pi^2 a^2(\eta)}\int_{k_{\rm IR}}^{\infty}{\rm d}k~k^2 j_0(kr)|U_I(\eta,k)|^2\,\nonumber\\
&=&\frac{1}{8\pi a^2(\eta){\mathcal H}(\eta)}\sqrt{\frac{\pi}{2r}}\int_{k_{\rm IR}}^{\infty}{\rm d}k~k^{\frac32}J_{\frac12}(kr)\bigg|H_{\nu_I}\left(\frac{k}{{\mathcal H}(\eta)}\right)\bigg|^2\,,
\label{eq: Delta infinity}
\end{eqnarray}
where the subscript ``$I$" in the first line stands for quantities computed during inflation (thus ${\mathcal H}(\eta)=a(\eta) H_I$) and we used the relation $j_0(z)=\sqrt{\frac{\pi}{2z}}J_{\frac12}(z)$ between spherical and ordinary Bessel functions.

\noindent By analytic continuation of eq. (6.578.10) in \cite{Gradshteyn} or directly from eq. (27) of \cite{Janssen:2008px}, one finds for the integral in (\ref{eq: Delta infinity})
\begin{equation}
\label{eq: Delta infinity final}
\Delta_{\Phi\Phi,I}^{(\infty)}\left(\eta,r\right)=\frac{H_I^2}{16\pi^2}~\Gamma\left(\frac32+\nu_I\right)\Gamma\left(\frac32-\nu_I\right)~_2F_1\left(\frac32+\nu_I,\frac32-\nu_I;2;1-\frac{{\mathcal H}^2(\eta)r^2}{4}\right)-\Delta_{\rm IR}\left(\eta,r\right)\,,
\end{equation}
where $\Delta_{\rm IR}\left(\eta,r\right)$ is the contribution due to the lower extreme of integration $k_{\rm IR}$; it will be evaluated in eq.~(\ref{eq: Delta deep IR}).

As for the infrared integral $\Delta_{\Phi\Phi, I}^{(k_0)}$ in inflation, one can use the low-$k$ approximation for the mode function (\ref{eq: mode IR}) instead of the full result (\ref{eq: full mode inflation}) because $k_0\ll{\mathcal H(\eta)}$, to get\footnote{ More precisely, since $k_0={\mathcal O}(1){\mathcal H}_{\rm in}$ is of the order of the inverse comoving Hubble radius at the beginning of inflation, the relation $k_0\ll{\mathcal H(\eta)}$ is not valid in the very early stages of inflation, but after a few e-foldings $N(\eta)=\ln{\left(\frac{a(\eta)}{a_{\rm in}}\right)}$, one can say that $\frac{k_0}{{\mathcal H}(\eta)}={\mathcal O}(1)~{\rm e}^{-N(\eta)}\ll1$, thus decaying exponentially with $N(\eta)$, where we used that ${\mathcal H(\eta)}=a(\eta)H_I$ in inflation.}

\begin{equation}
\label{eq: Delta infrared}
\Delta_{\Phi\Phi,I}^{(k_0)}\left(\eta,r\right)=\frac{2^{2\nu_I-3}\Gamma^2(\nu_I){\mathcal H}(\eta)^{2\nu_I-1}}{\pi^3a^2(\eta)}\int_{k_{\rm IR}}^{k_0}{\rm d}k~k^{2-2\nu_I}j_0(kr)\,.
\end{equation}
Let us rewrite the integral appearing in (\ref{eq: Delta infrared}) in terms of a dimensionless integration variable $z=kr$ as
\begin{equation}
\label{eq: Delta infrared 2}
\int_{k_{\rm IR}}^{k_0}{\rm d}k~k^{2-2\nu_I}j_0(kr)=r^{2\nu_I-3}\int_{k_{\rm IR} r}^{k_0 r}{\rm d}z~z^{2-2\nu_I}j_{0}(z)\,.
\end{equation}
Now we use the formula\footnote{Eq. (\ref{eq: formula IR}) can be checked by the series expansions $j_0(z)=\mathlarger{\sum}_{n=0}^{\infty}\frac{\left(-z^2\right)^n}{(2n+1)!}$ and $~_1F_2\left(\frac32-\nu_I;\frac32,\frac52-\nu_I;-\frac{z^2}{4}\right)=\mathlarger{\sum}_{n=0}^{\infty}\frac{\left(\frac32-\nu_I\right)_n}{\left(\frac32\right)_n \left(\frac52-\nu_I\right)_n}\frac{1}{n!}\left(-\frac{z^2}{4}\right)^n$, where $(c)_n=\frac{\Gamma(c+n)}{\Gamma(c)}$ is the Pochhammer symbol.}
\begin{equation}
\label{eq: formula IR}
\int_{z_{\rm IR}}^{z_0}{\rm d}z~z^{2-2\nu_I}j_{0}(z)=\frac{z_0^{3-2\nu_I}}{3-2\nu_I}~_1F_2\left(\frac32-\nu_I;\frac32,\frac52-\nu_I;-\frac{z_0^2}{4}\right)-\frac{z_{\rm IR}^{3-2\nu_I}}{3-2\nu_I}~_1F_2\left(\frac32-\nu_I;\frac32,\frac52-\nu_I;-\frac{z_{\rm IR}^2}{4}\right)\,,
\end{equation}
and substitute $z_{\rm IR}=k_{\rm IR}r$ and $z_0=k_0r$, to obtain
\begin{eqnarray}
\Delta_{\Phi\Phi,I}^{(k_0)}\left(\eta,r\right)=\frac{H_I^2}{\pi^3}\frac{\Gamma^2(\nu_I)}{3-2\nu_I}\!\!\!\!\!\!&&\left[\left(\frac{2{\mathcal H}(\eta)}{k_0}\right)^{2\nu_I-3}~_1F_2\left(\frac32-\nu_I;\frac32,\frac52-\nu_I;-\frac{k_0^2 r^2}{4}\right)\right.\,\nonumber\\
&&~\left.-\left(\frac{2{\mathcal H}(\eta)}{k_{\rm IR}}\right)^{2\nu_I-3}~_1F_2\left(\frac32-\nu_I;\frac32,\frac52-\nu_I;-\frac{k_{\rm IR}^2 r^2}{4}\right)\right]\,.\label{eq: Delta infrared final}
\end{eqnarray}
The second line of eq.~(\ref{eq: Delta infrared final}) contains the deep IR cutoff $k_{\rm IR}$ and it corresponds exactly to the quantity $\Delta_{\rm IR}\left(\eta,r\right)$ appearing in eq.~(\ref{eq: Delta infinity final}), in the sense that
\begin{equation}
\label{eq: Delta deep IR}
\Delta_{\rm IR}\left(\eta,r\right)=\frac{H_I^2}{\pi^3}\frac{\Gamma^2(\nu_I)}{3-2\nu_I}\left(\frac{2{\mathcal H}(\eta)}{k_{\rm IR}}\right)^{2\nu_I-3}~_1F_2\left(\frac32-\nu_I;\frac32,\frac52-\nu_I;-\frac{k_{\rm IR}^2 r^2}{4}\right)\,.
\end{equation}

Subtracting eq. (\ref{eq: Delta infrared final}) from eq. (\ref{eq: Delta infinity final}), as specified in (\ref{eq: Delta finite}), the deep IR term $\Delta_{\rm IR}\left(\eta,r\right)$ cancels out and one finally finds

\begin{eqnarray}
\Delta_{\Phi\Phi,I}\left(\eta,r\right)&=&-\Delta_{\Phi\Phi,I}^{(k_0)}\left(\eta,r\right)+\Delta_{\Phi\Phi,I}^{(\infty)}\left(\eta,r\right)\,\label{eq: Delta reg final}\\
&=&\frac{H_I^2}{\pi^2}\left[\frac{\Gamma^2(\nu_I)}{(2\nu_I-3)\pi}\left(\frac{2{\mathcal H}(\eta)}{k_0}\right)^{2\nu_I-3}~_1F_2\left(\frac32-\nu_I;\frac32,\frac52-\nu_I;-\frac{k_0^2 r^2}{4}\right)\right.\,\nonumber\\
&&\left.+\frac{1}{16}~\Gamma\left(\frac32+\nu_I\right)\Gamma\left(\frac32-\nu_I\right)~_2F_1\left(\frac32+\nu_I,\frac32-\nu_I;2;1-\frac{{\mathcal H}^2(\eta)r^2}{4}\right)\right]\,.\,\nonumber
\end{eqnarray}
This is the result during inflation for the field 2-point function (\ref{eq: Delta U relation}).

\subsubsection{Coincident correlator}
\label{subsubsec: coinc limit}
Let us consider the coincident limit $r=0$ in inflation. The coincident correlator has been computed within the stochastic approximation in~\cite{Glavan:2017jye} and compared to the field theory result of~\cite{Glavan:2015cut}. Here we summarize the QFT result, referring the reader to~\cite{Glavan:2015cut} for more details. In the coincident limit, one cannot work in $D=4$ spacetime dimensions from the beginning, but it is required to work with general $D$ spacetime dimensions in order to regularize divergences (via dimensional regularization). As a starting point for understanding this, let us consider eq.~(\ref{eq: Delta reg final}), which was obtained in $D=4$, and specialize it to $r=0$. Using ${\mathcal H}(\eta)=a(\eta)H_I$, eq. (\ref{eq: Delta reg final}) reduces to
\begin{eqnarray}
\Delta_{\Phi\Phi,I}\left(\eta,0\right)&=&\frac{H_I^2}{\pi^2}\left[\frac{\Gamma^2(\nu_I)}{(2\nu_I-3)\pi}\left(\frac{2 a_{\rm in}H_I}{k_0}\right)^{2\nu_I-3}\left(\frac{a(\eta)}{a_{\rm in}}\right)^{2\nu_I-3}\right.\,\nonumber\\
&&\left.+\frac{1}{16}~\Gamma\left(\frac32+\nu_I\right)\Gamma\left(\frac32-\nu_I\right)~_2F_1\left(\frac32+\nu_I,\frac32-\nu_I;2;1\right)\right]\,,
\label{eq: Delta coinc}
\end{eqnarray}
where, in the first line of (\ref{eq: Delta coinc}), we multiplied and divided by the same power of the scale factor at the beginning of inflation $a_{\rm in}$.
The second line of eq.~(\ref{eq: Delta coinc}) originates from $\Delta_{\Phi\Phi,I}\left(\eta,0\right)$ and it is a divergent constant, as one can see from Gauss' identity $~_2F_1(a,b;c;1)=\frac{\Gamma(c)\Gamma(c-a-b)}{\Gamma(c-a)\Gamma(c-b)}$ because $c-a-b=2-\left(\frac32+\nu_I\right)-\left(\frac32-\nu_I\right)=-1$, which is a simple pole of the gamma function. This divergence can be cured by dimensional regularization, working in $D$ spacetime dimensions instead of $D=4$ and promoting $D$ to a complex variable. In general $D$ spacetime dimensions, one can prove\cite{Chernikov:1968zm} (see also eq. (30) of \cite{Janssen:2008px}) that the result (\ref{eq: Delta infinity final}) for $\Delta_{\Phi\Phi,I}^{(\infty)}\left(\eta,r\right)$ (omitting to write the deep IR contribution $\Delta_{\rm IR}(\eta,r)$, which cancels out in the final expression~(\ref{eq: Delta reg final}) for $\Delta_{\Phi\Phi,I}\left(\eta,r\right)$) generalizes to
\begin{eqnarray}
\Delta_{\Phi\Phi,I}^{(\infty)}\left(\eta,r\right)&=&\frac{H_I^{D-2}}{(4\pi)^{\frac{D}{2}}}\frac{\Gamma\left(\frac{D-1}{2}+\nu_{I,D}\right)\Gamma\left(\frac{D-1}{2}-\nu_{I,D}\right)}{\Gamma\left(\frac{D}{2}\right)}\,\nonumber\\
&&\times~_2F_1\left(\frac{D-1}{2}+\nu_{I,D},\frac{D-1}{2}-\nu_{I,D};\frac{D}{2};1-\frac{{\mathcal H}^2(\eta)r^2}{4}\right)\,,\label{eq: Delta infinity dim reg D}
\end{eqnarray}
with $\nu_{I,D}\equiv\sqrt{\left(\frac{D-1}{2}\right)^2-D(D-1)\xi}$ for a massless non-minimally coupled field.
In the coincident limit $r=0$, by applying Gauss' identity, one gets
\begin{equation}
\Delta_{\Phi\Phi,I}^{(\infty)}\left(\eta,0\right)=\frac{H_I^{D-2}}{(4\pi)^{\frac{D}{2}}}\frac{\Gamma\left(\frac{D-1}{2}+\nu_{I,D}\right)\Gamma\left(\frac{D-1}{2}-\nu_{I,D}\right)\Gamma\left(1-\frac{D}{2}\right)}{\Gamma\left(\frac12+\nu_{I,D}\right)\Gamma\left(\frac12-\nu_{I,D}\right)}\,,
\end{equation}
and the factor $\Gamma\left(1-\frac{D}{2}\right)$ is responsible for the divergence at $D=4$, having a simple pole there. Expanding around $D=4$ and introducing a renormalization energy scale $\bar{\mu}$, one gets
\begin{eqnarray}
\Delta_{\Phi\Phi,I}^{(\infty)}\left(\eta,0\right)=-\frac{H_I^2}{8\pi^2}\left(1-6\xi\right)\!\!\!\!\!\!\!\!\!\!&&\left[2\frac{\bar{\mu}^{D-4}}{D-4}+\ln\left(\frac{H_I^2}{4\pi\bar{\mu}^2}\right)+\gamma_E-\frac{\xi}{1-6\xi}+\psi\left(\frac{1}{2}+\nu_I\right)+\psi\left(\frac{1}{2}-\nu_I\right)\right.\,\nonumber\\
&&\left.+{\mathcal O}\left(D-4\right)\right]\,,
\label{eq: expand D-4}
\end{eqnarray}
where $\nu_I$ in eq.~(\ref{eq: expand D-4}) refers to the value in $D=4$ spacetime dimensions, $\nu_I\equiv\nu_{I,4}=\sqrt{\frac94-12\xi}$. The function $\psi(z)\equiv\Gamma'(z)/\Gamma(z)$ is the digamma function and $\gamma_E=-\psi(1)\approx 0.577$ is the Euler-Mascheroni constant.
The divergence at $D=4$ is removed by the addition of suitable counterterms so that the coincident correlator $\Delta_{\Phi\Phi,I}^{(\infty)}\left(\eta,0\right)$ in eq.~(\ref{eq: expand D-4}) reduces to a finite constant, which is then added to the contribution in the first line of~(\ref{eq: Delta coinc}), coming from $-\Delta_{\Phi\Phi,I}^{(k_0)}\left(\eta,0\right)$, to get a final result for the full renormalized\footnote{In fact one never renormalizes the propagator, but rather physical quantities. In general, the renormalized propagator cannot be used for full loop calculations, except for some specific cases. One such example is the expectation value of the energy-momentum tensor, where the coincident Feynman propagator appears: $\langle\hat{T}_{\mu\nu}\rangle\supset-\frac{m^2}{2}g_{\mu\nu}\left[i\Delta_{F}(x;x)\right]^{(\rm ren)}$. This quantity can be renormalized by a cosmological constant counterterm. Here, the renormalized coincident correlator is considered for the purpose of making comparison with the stochastic 2-point functions of Section~\ref{sect:model}.} coincident correlator $\Delta_{\Phi\Phi,I}^{(\rm ren)}\left(\eta,0\right)$. For comparison with the prediction of stochastic formalism, we choose the (non-minimal) subtraction scheme in such a way that $\Delta_{\Phi\Phi,I}^{(\rm ren)}\left(\eta_{\rm in},0\right)$ is zero at the beginning of inflation $(\eta=\eta_{\rm in})$. Then the time evolution of the renormalized correlator $\Delta_{\Phi\Phi,I}^{(\rm ren)}\left(\eta,0\right)$ is given by
\begin{equation}
\Delta_{\Phi\Phi,I}^{(\rm ren)}\left(\eta,0\right)=H_I^2\frac{\Gamma^2(\nu_I)}{(2\nu_I-3)\pi^3}\left(\frac{2 a_{\rm in}H_I}{k_0}\right)^{2\nu_I-3}\left[\left(\frac{a(\eta)}{a_{\rm in}}\right)^{2\nu_I-3}-1\right]\,.
\label{eq: Delta coinc renormalized}
\end{equation}
For $\nu_I>\frac32$, as in the case $\xi<0$ where (\ref{eq: massless negaive xi}) holds, the contribution $\left(\frac{a(\eta)}{a_{\rm in}}\right)^{2\nu_I-3}\approx\left(\frac{a(\eta)}{a_{\rm in}}\right)^{8|\xi|}$ originated from $-\Delta_{\Phi\Phi,I}^{(k_0)}\left(\eta,0\right)$ is amplified by the growth of the scale factor. Therefore, after a few e-foldings $N_I(\eta)=\ln{\left(\frac{a(\eta)}{a_{\rm in}}\right)}$ of inflation, a good approximation for eq.~(\ref{eq: Delta coinc renormalized}) is
\begin{equation}
\Delta_{\Phi\Phi,I}^{(\rm ren)}\left(\eta,0\right)\approx\frac{H_I^2}{32\pi^2|\xi|}\left(\frac{2 a_{\rm in}H_I}{k_0}\right)^{8|\xi|}{\rm e}^{8|\xi|N_I(\eta)}\,,
\label{eq: Delta coinc approx}
\end{equation}
where we used $\nu_I\simeq\frac32+4|\xi|$ and worked at leading order in $|\xi|\ll1$.
Comparing the quantum field theory result (\ref{eq: Delta coinc approx}) with the stochastic prediction~(\ref{sol_deSitter 2pt 5}) at coincidence (and with $c_s=1$), we find the same growth with $N_I$. We also find the same amplitude for $k_0=2 a_{\rm in}H_I$, which is compatible with the expectation that the IR cutoff $k_0$ should be of the order of the inverse comoving Hubble radius at the beginning of inflation $k_0={\mathcal O}(1)a_{\rm in}H_I$ (as discussed below eq. (\ref{eq: set corr QFT})).

By applying the same logic, the coincident correlators in radiation epoch, $\Delta_{\Phi\Phi,R}\left(\eta,0\right)$, and in matter epoch, $\Delta_{\Phi\Phi,M}\left(\eta,0\right)$, are also dominated by their infrared parts, namely $-\Delta_{\Phi\Phi,R}^{(k_0)}\left(\eta,0\right)$ (and $-\Delta_{\Phi\Phi,M}^{(k_0)}\left(\eta,0\right)$). They can be evaluated using the low-$k$ mode functions in the second and third line of eq.~(\ref{eq: mode IR}). 
The result is that, in radiation epoch, the coincident correlator stays roughly constant, as we find for the ratio with respect to its initial value (i. e. at the end of inflation, at conformal time $\eta_{\rm end}$),
\begin{equation}
\label{eq: coinc rad}
\frac{\Delta_{\Phi\Phi,R}\left(\eta,0\right)}{\Delta_{\Phi\Phi,I}\left(\eta_{\rm end},0\right)}=\left[\nu_I-\frac12-\left(\nu_I-\frac32\right){\rm e}^{-N_R(\eta)}\right]^2\,,
\end{equation}
where $N_R(\eta)$ is the number of e-foldings in radiation epoch measured with respect to the end of inflation, so that $N_R(\eta_{\rm end})=0$.
For $\nu_I\simeq\frac32+4|\xi|$ the ratio in eq.~(\ref{eq: coinc rad}) only grows from $1$ at the beginning  of radiation era ($N_R=0$) to $\left(\nu_I-\frac12\right)^2\simeq1+8|\xi|$ for large value of $N_R$ (radiation epoch lasts $\approx 50$ e-foldings), which means that $\Delta_{\Phi\Phi,R}\left(\eta,0\right)$ is approximately constant, in agreement with the stochastic result~(\ref{sol_rad 2pt 2}).

With the same strategy we find that in matter era, the coincident correlator grows with respect to its initial value at matter-radiation equality (which corresponds to conformal time $\eta_{\rm eq}$) by a factor
\begin{equation}
\label{eq: coinc matter}
\frac{\Delta_{\Phi\Phi,M}\left(\eta,0\right)}{\Delta_{\Phi\Phi,R}\left(\eta_{\rm eq},0\right)}=\frac14\left[1+\frac{3}{2\nu_M}+\left(1-\frac{3}{2\nu_M}\right){\rm e}^{-\nu_M N_M(\eta)}\right]^2~{\rm e}^{\left(\nu_M-\frac32\right)N_M(\eta)}\,,
\end{equation}
where $N_M(\eta)$ is the number of e-foldings in matter epoch measured starting from matter-radiation equality, meaning that $N_M(\eta_{\rm eq})=0$.
Using $\nu_I\simeq\frac32+4|\xi|$ (leading order in $|\xi|$ of~(\ref{eq: nuM massless}) for $\xi<0$), then eq.~(\ref{eq: coinc matter}) is well approximated by simply ${\rm e}^{4|\xi|N_M(\eta)}$, in agreement with the stochastic formalism prediction in eq.~(\ref{sol_mat 2pt 2}).

\noindent Thus quantum field theory confirms the one-loop prediction of the stochastic formalism in the coincident limit.

\subsubsection{Super-Hubble regime}
\label{subsubsec: super-Hubble separation}
Stochastic theory, following the dynamics of super-Hubble modes, is expected to be trustworthy for super-Hubble separations $r$, i.e. when ${\mathcal H}(\eta)r\gg1$. Here we want to check its validity in this regime against the quantum field theory prediction $\Delta_{\Phi\Phi,I}\left(\eta,r\right)=-\Delta_{\Phi\Phi,I}^{(k_0)}\left(\eta,r\right)+\Delta_{\Phi\Phi,I}^{(\infty)}\left(\eta,r\right)$ that we worked out in eq. (\ref{eq: Delta reg final}).

For this purpose, in inflation it is useful to rewrite the contribution from $\Delta_{\Phi\Phi,I}^{(\infty)}\left(\eta,r\right)$ in eq. (\ref{eq: Delta infinity final}), using the hypergeometric transformation formula (9.132.1) in~\cite{Gradshteyn} (and omitting to write the deep IR contribution $\Delta_{\rm IR}(\eta,r)$, which cancels out in the final expression for $\Delta_{\Phi\Phi,I}\left(\eta,r\right)$ in eq.~(\ref{eq: Delta reg final})), to get
\begin{align}
\Delta_{\Phi\Phi,I}^{(\infty)}\left(\eta,r\right)=\frac{H_I^2}{16\pi^2}\left[\frac{\Gamma\left(\frac32-\nu_I\right)\Gamma\left(2\nu_I\right)}{\Gamma\left(\frac12+\nu_I\right)}\left(\frac{{\mathcal H}(\eta)r}{2}\right)^{2\nu_I-3}~_2F_1\left(\frac32-\nu_I,\frac12-\nu_I;1-2\nu_I;\frac{4}{{\mathcal H}^2(\eta)r^2}\right)\right.\,\nonumber\\
\left.+\frac{\Gamma\left(\frac32+\nu_I\right)\Gamma\left(-2\nu_I\right)}{\Gamma\left(\frac12-\nu_I\right)}\left(\frac{{\mathcal H}(\eta)r}{2}\right)^{-2\nu_I-3}~_2F_1\left(\frac32+\nu_I,\frac12+\nu_I;1+2\nu_I;\frac{4}{{\mathcal H}^2(\eta)r^2}\right)\right]\,.
\label{eq: hypergeo transform}
\end{align}
For super-Hubble distances with ${\mathcal H}(\eta)r\gg1$ and for $\nu_I\simeq\frac32+4|\xi|$, the leading contribution to $\Delta_{\Phi,\Phi}^{(\infty)}$ comes from the power $\left(\frac{{\mathcal H}(\eta)r}{2}\right)^{2\nu_I-3}$ in the first line of (\ref{eq: hypergeo transform}) (the hypergeometric functions are close to $1$ for small arguments $\frac{4}{{\mathcal H}^2(\eta)r^2}$), resulting in
\begin{align}
\Delta_{\Phi\Phi,I}^{(\infty)}\left(\eta,r\right)\simeq-\frac{H_I^2}{32\pi^2|\xi|}\left(\frac{{\mathcal H}(\eta)r}{2}\right)^{8|\xi|}=-\frac{H_I^2}{32\pi^2|\xi|}\left(\frac{a_{\rm in}H_I r}{2}\right)^{8|\xi|}{\rm e}^{8|\xi|N_I(\eta)}\,.
\end{align}
Adding the contribution from $-\Delta_{\Phi\Phi,I}^{(k_0)}\left(\eta,r\right)$ (see eq. (\ref{eq: Delta infrared final})) and using $k_0=2a_{\rm in}H_I$ which we required from matching the coincident correlators from quantum field theory and stochastic formalism (see the discussion below eq.~(\ref{eq: Delta coinc approx})), one finds\footnote{The function $~_1F_2\left(\frac32-\nu_I;\frac32,\frac52-\nu_I;-\frac{k_0^2 r^2}{4}\right)$ in eq.~(\ref{eq: Delta infrared final}) can be approximated as $1$ even at super-Hubble scales (${\mathcal H}(\eta)r\gg1$) and not only in the coincident regime, because its argument $-\frac{k_0^2 r^2}{4}$ is still small. The IR cutoff $k_0$ sets an upper limit for comoving scales $r\lesssim1/k_0$. Another way to understand the smallness of $-\frac{k_0^2 r^2}{4}$ is to use $k_0=2a_{\rm in}H_I=2a(\eta)H_I{\rm e}^{-N(\eta)}=2{\mathcal H}(\eta){\rm e}^{-N(\eta)}$ and therefore $-\frac{k_0^2 r^2}{4}=-\left({\mathcal H}(\eta)r\right)^2{\rm e}^{-2N(\eta)}$. This is suppressed by ${\rm e}^{-2N(\eta)}$, so that for $N(\eta)$ large enough, $-\frac{k_0^2 r^2}{4}$ is small even for large values $1\ll{\mathcal H}(\eta)r\ll{\rm e}^{2N(\eta)}$.}
\begin{equation}
\label{eq: infl superHubble}
\Delta_{\Phi\Phi,I}\left(\eta,r\right)=-\Delta_{\Phi\Phi,I}^{(k_0)}\left(\eta,r\right)+\Delta_{\Phi\Phi,I}^{(\infty)}\left(\eta,r\right)\simeq\frac{H_I^2}{32\pi^2|\xi|}~{\rm e}^{8|\xi|N_I(\eta)}\left[1-\left(\frac{a_{\rm in}H_I r}{2}\right)^{8|\xi|}\right]\,.
\end{equation}
Eq.~(\ref{eq: infl superHubble}) agrees with the stochastic prediction in eqs.~(\ref{sol_deSitter 2pt 5})--(\ref{eq: profile 2-pt s_2(r)}), whose spatial profile is proportional to $1-\left(\mu a_{\rm in}H_I r\right)^{8|\xi|}$ where $\mu$ sets the UV cutoff of the stochastic formalism. The two results have the same amplitude (for $k_0=2a_{\rm in}H_I$) and a profile described by a power law, proportional to $1-\left(\frac{r}{r_0}\right)^{8|\xi|}$. In both cases $r_0$ is of the order of $\left(a_{\rm in}H_I\right)^{-1}$, which is the comoving Hubble horizon at the beginning of inflation.

\subsubsection{Full comparison}
\label{subsubsec: stoch vs field plot}
\begin{figure}[H]
\centering
\includegraphics[width=0.9\columnwidth]{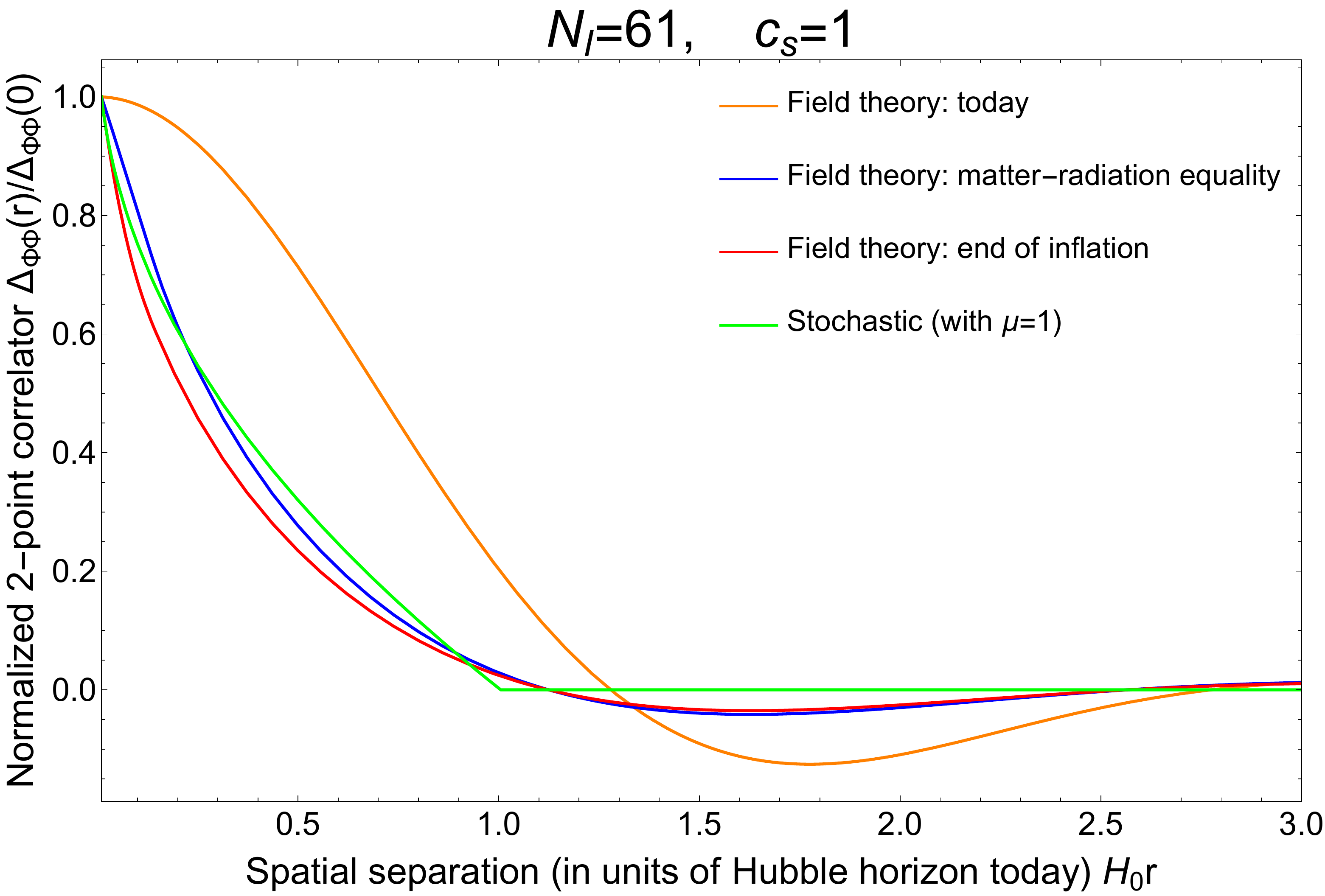}
\caption{
The field correlator $\Delta_{\Phi\Phi}$ computed with quantum field theory, as a function of distance. Distances are measured in units of the current Hubble horizon $H_0^{-1}$, thus a value of $1$ on the horizontal axis stands for a separation whose physical size today is the Hubble horizon $H_0^{-1}$. Correlators are normalized with respect to their respective value at coincidence. For simplicity and better readability of the plot, we considered three key moments in the history of the Universe: end of inflation (red), matter-radiation equality (blue) and today (orange). The stochastic formalism prediction (green) does not change relevantly from the end of inflation until today. We see that, at sub-Hubble scales, the stochastic theory only agrees with quantum field theory at early times (from the end of inflation until matter-radiation equality), but it fails at later times in matter-dominated era, whose result has been extrapolated until today. The plot assumes a sudden transition between different epochs and the values chosen for the cosmological parameters are: $\Omega_M=0.3$, $\Omega_R=9.1\times10^{-5}$ (matter and radiation energy density fractions today), $H_0=10^{-33}~{\rm eV}$ (Hubble parameter today) and $H_I=10^{13}~{\rm GeV}$ (Hubble parameter in inflation). 
They imply a length of radiation epoch of $N_R\simeq 57$ e-foldings and the matter era started $N_M\simeq 8.1$ e-foldings ago from now.
We worked in the massless limit for the field, with a negative non-minimal coupling $\xi=-0.06$. For simplicity, the duration of inflation assumed is $N_I=61$ e-foldings, such that, given the other parameters listed before, the comoving Hubble length at the beginning of inflation is equal to the Hubble horizon today.
Note that this is not necessarily the realistic number of e-foldings needed to reproduce the right amount of dark energy today: the correct value also depends on the mass $m$ of the scalar field and the speed of sound $c_s$, as in eq.~(\ref{inflation_efolds_relation}).
In this plot, we assumed a speed of sound $c_s=1$ for the field. The effect of $c_s<1$ is depicted in Fig.~\ref{fig: full comparison cs<1}.
\label{fig: full comparison cs=1}
}
\end{figure}

We have seen that stochastic formalism is able to reproduce QFT results in the coincident limit (subsection~\ref{subsubsec: coinc limit}) and for super-Hubble separations (subsection~\ref{subsubsec: super-Hubble separation}).
At intermediate non-coincident separations (sub-Hubble and near-Hubble scales) there is no reason to trust stochastic formalism. Hence it is interesting to compare its predictions and the field theory results. The stochastic formalism predicts that the spatial dependence (normalized with respect to the value at coincidence) of the field 2-point function inherited from inflation is not substantially altered in radiation and matter epochs and therefore it is the same today as it was at the end of inflation. This was a consequence of the negligibility of stochastic sources after the end of inflation, so that the time evolution could only modify the amplitude (in stochastic formalism, $\Delta_{\phi\phi}$ grows in matter era, while it is roughly constant in radiation era), but not the spatial dependence of correlators. In Fig.~\ref{fig: full comparison cs=1} we compare this stochastic prediction to the full quantum field theory result, highlighting how, according to field theory, the spatial profile gets modified today (assuming the matter domination approximation) with respect to the end of inflation.
The stochastic prediction for the spatial profile works much better at the end of inflation and in radiation epoch than in matter epoch. In other words the stochastic formalism fails to predict the right spatial dependence of field correlators in matter era and today. This can be largely understood as an effect of the non-negligibility of spatial gradients at late times in the life of the Universe, which were dropped by the stochastic formalism calculation in Section~\ref{sect:model}. For a given Fourier mode, with comoving momentum $k$, the  effect of spatial gradients on the equation of motion of the field is determined at conformal time $\eta$ by $\left(\frac{k}{{\mathcal H}(\eta)}\right)^2$. Since the comoving Hubble radius ${\mathcal H}^{-1}(\eta)$ was much smaller at the end of inflation than today, the gradients play a largely more relevant role today than in the early Universe. This helps us to understand the problem with stochastic formalism today (or in general at late times), because a steep decrease in the field correlator like the one predicted by stochastic theory (green curve in Fig.~\ref{fig: full comparison cs=1}) would imply a large energy of the field configuration associated to the spatial gradients and it is therefore energetically disfavored. On the contrary, quantum field theory predicts today (orange curve in Fig.~\ref{fig: full comparison cs=1}) a softer decrease of correlations with distance, thus diminishing the energy content of the field configuration. We also note that the coherent oscillations on super-Hubble scales predicted  by QFT are absent in the stochastic approximation.

\subsubsection{Effect of a reduced speed of sound}
\label{subsubsec: QFT cs<1}
\begin{figure}[H]
\centering
\includegraphics[width=0.49\columnwidth]{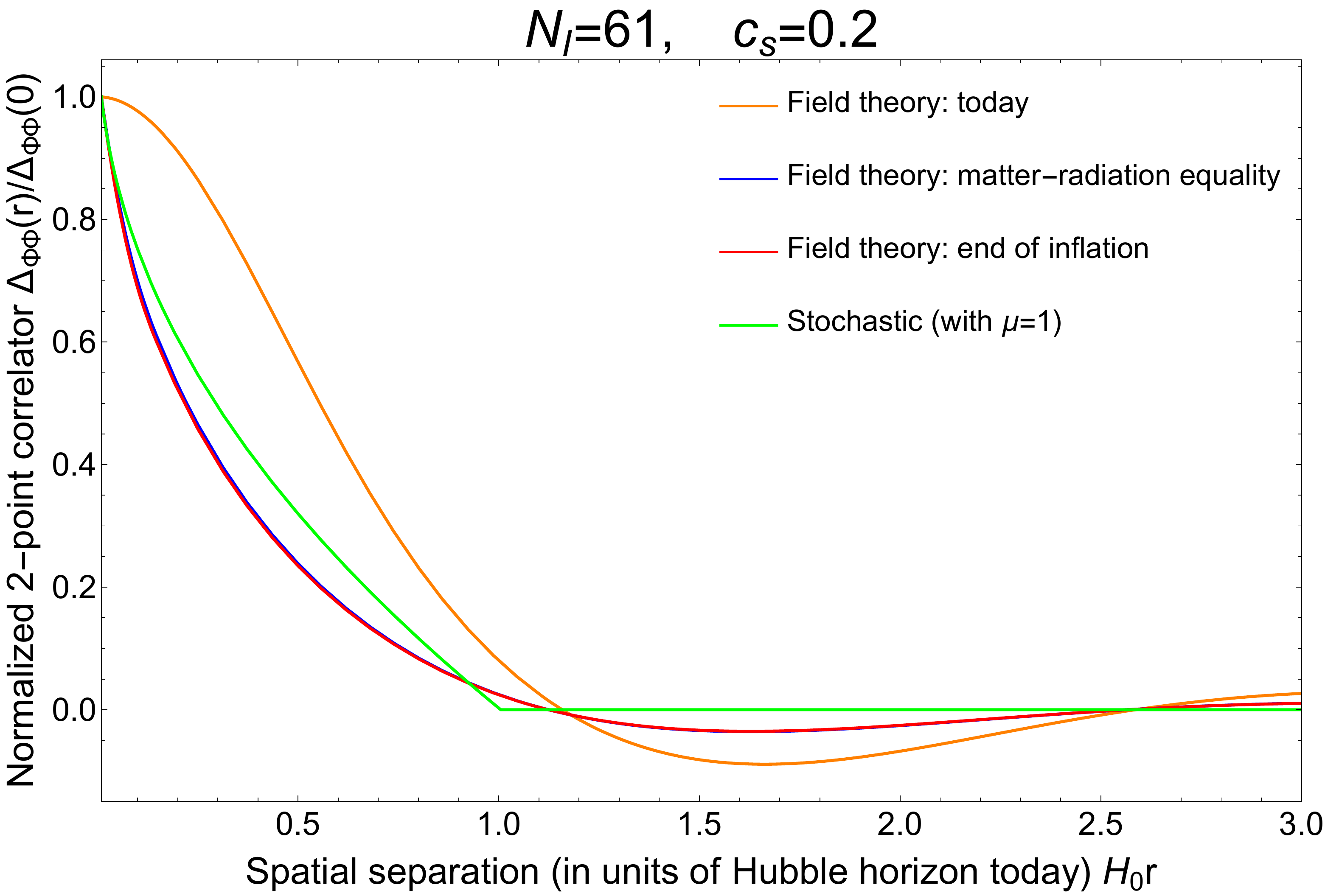}
\includegraphics[width=0.49\columnwidth]{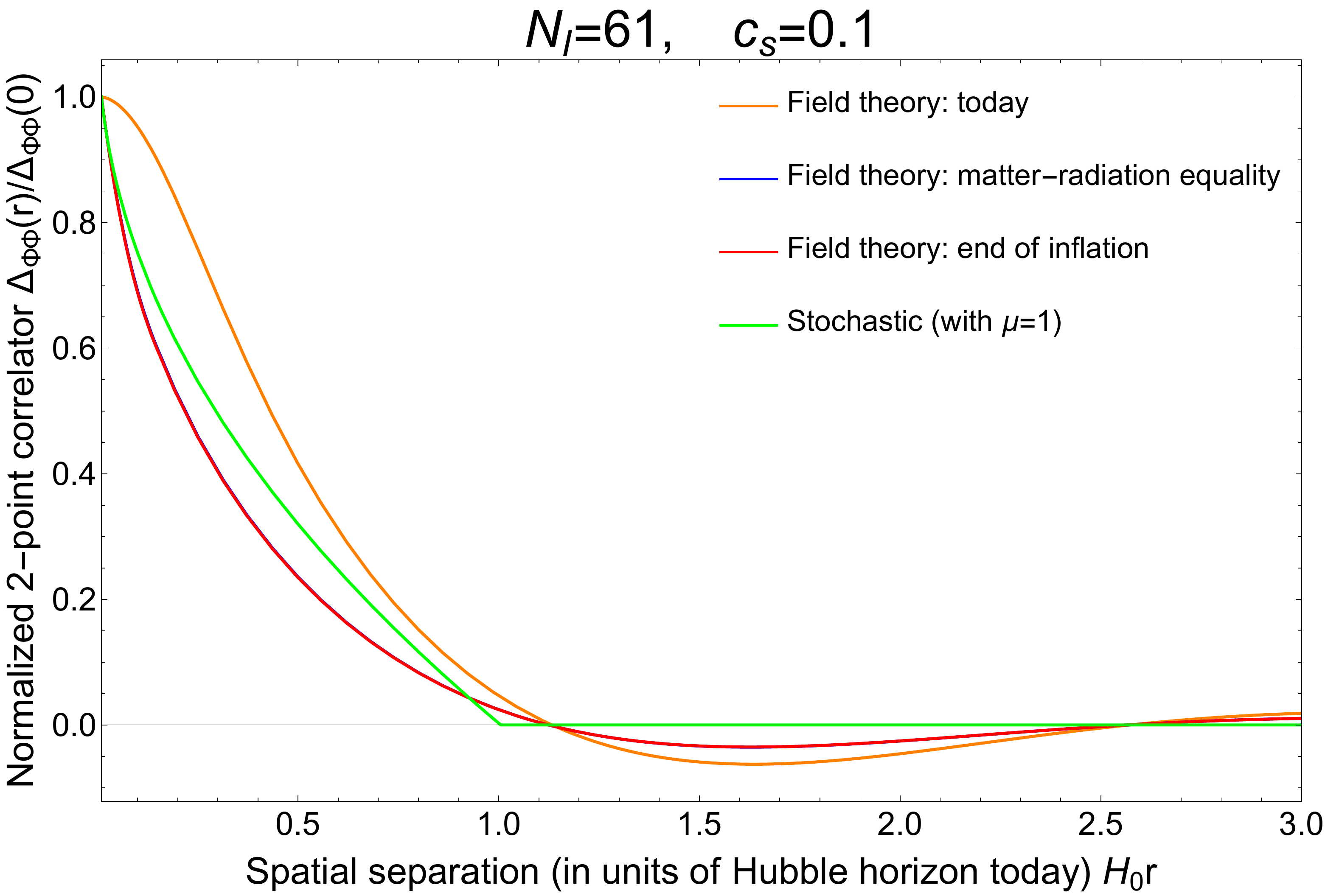}
\includegraphics[width=0.49\columnwidth]{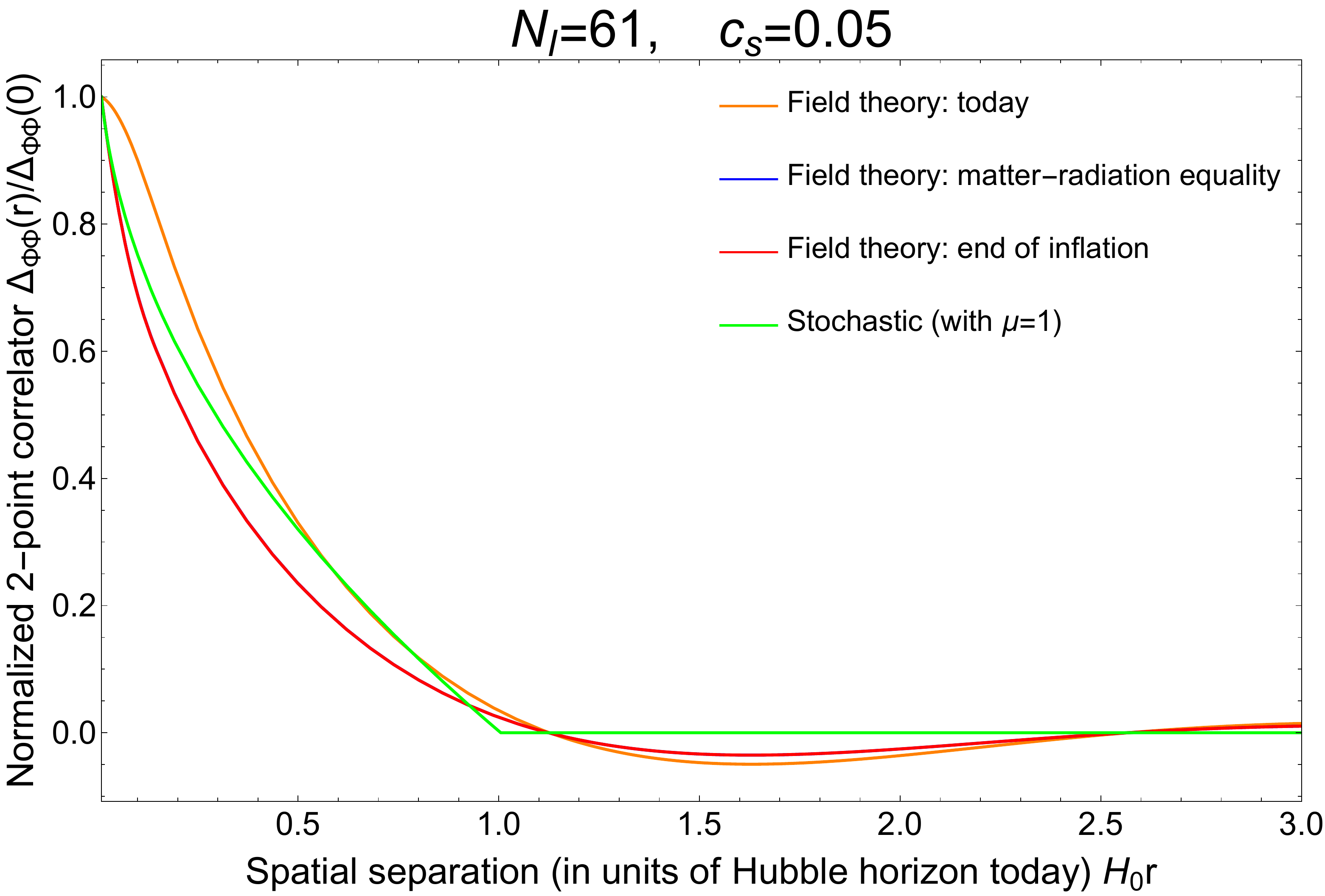}
\includegraphics[width=0.49\columnwidth]{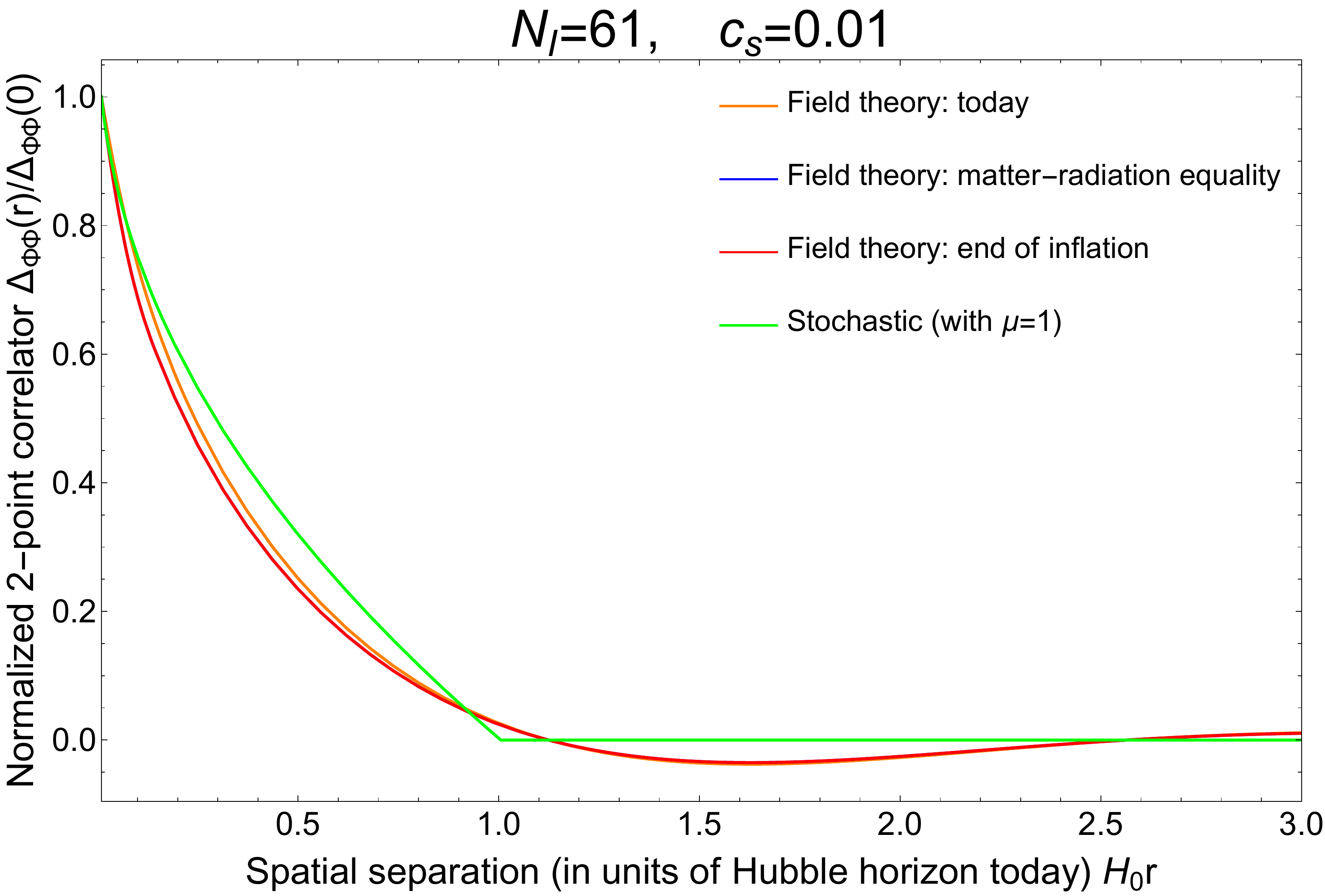}
\caption{
Effect of a  reduced speed of sound $c_s<1$ on the spatial dependence of the field correlator $\Delta_{\Phi,\Phi}$, computed with quantum field theory: $c_s=0.2$ (top left), $c_s=0.1$ (top right), $c_s=0.05$ (bottom left) and $c_s=0.01$ (bottom right). The plots should be compared to Fig.~\ref{fig: full comparison cs=1} (which uses $c_s=1$).
For $c_s=0.2$ the spatial dependence at the end of inflation and at matter radiation-equality (already well-fitted by the stochastic formalism for $c_s=1$) are indistinguishable and correlations at sub-Hubble scales today decrease more rapidly than in the case with $c_s=1$. By diminishing $c_s$, the spatial profiles at the three different times considered (end of inflation, matter-radiation equality and today) are barely distinguishable, in incrementally better agreement with the prediction from stochastic theory (constant spatial dependence after the end of inflation, up to the evolution in time of the overall amplitude). The cosmological parameters and the number of inflationary e-foldings have been chosen as for Fig.~\ref{fig: full comparison cs=1}.
\label{fig: full comparison cs<1}
}
\end{figure}

Stochastic theory fails to reproduce the correct spatial dependence of the field correlator at sub-Hubble scales in matter epoch, as shown by Fig.~\ref{fig: full comparison cs<1}. However, the discussion in subsection~\ref{subsubsec: stoch vs field plot} contains a hint on how agreement could be restored even at late times (today). Based on the role of spatial gradients, a simple strategy consists in decreasing their effect via the introduction of a reduced speed of sound. Such a situation is common in many studied and motivated modified gravity theories, e. g. scalar-tensor theories in the Horndeski class.

Let us introduce a constant speed of sound $c_s\leq1$. The equation of motion for the quantum field $\hat{\Phi}(t,\vec{x})$ is given in (\ref{eq: EOM Phi 2nd order cs<1}) and the equation of motion (\ref{eq: U eom}) for the mode function $U\left(\eta,k\right)$ gets simply modified into
\begin{equation}
\label{eq: U eom cs}
U^{\prime\prime}+\left[c_s^2k^2+m^2a^2-(1-6\xi)(2-\epsilon){\mathcal H}^2\right]U=0\,,
\end{equation}
i.e. $k$ is replaced by $c_s k$.
One can then repeat the calculations of this Section~\ref{sec: QFT} to predict the spatial profile of the field 2-point function in the presence of a constant reduced speed of sound.
The result is plotted in Fig.~(\ref{fig: full comparison cs<1}) and  it confirms the intuition that a $c_s<1$ enhances the stochastic formalism performance, reconciling it with quantum field theory.

\subsection{Free fields and Wick's theorem}
\label{subsec: QFT Wick}
We conclude this Section by discussing the QFT prediction for the higher-order correlators (computed on the vacuum state $|\Omega\rangle$ defined by the property~(\ref{eq: vacuum state})). We focus on
\begin{equation}
\Delta_{\Phi^2,\Phi^2}(t,r)  \equiv 
	\Bigl\langle \hat{\Phi}^2 (t,\vec{x}_1) \hat{\Phi}^2 (t,\vec{x}_2) \Bigr\rangle \,,
\end{equation}
which is the first of the 4-point functions considered in subsection~\ref{subsect:corr 4pt}, but with the {\it full} field instead of just its long-wavelength part.
Using the decomposition~(\ref{eq: canon Phi}--\ref{eq: canon Pi}) and the identities
\begin{eqnarray}
\Bigl\langle \hat{b}(\vec{k}_1)\hat{b}(\vec{k}^\prime_1)\hat{b}^{\dag}(\vec{k}_2)\hat{b}^{\dag}(\vec{k}^\prime_2) \Bigr\rangle&=&\delta^{3}\left(\vec{k}_1-\vec{k}_2\right)\delta^{3}\left(\vec{k}^\prime_1-\vec{k}^\prime_2\right)+\delta^{3}\left(\vec{k}_1-\vec{k}^\prime_2\right)\delta^{3}\left(\vec{k}^\prime_1-\vec{k}_2\right) \,,\nonumber\\
\Bigl\langle \hat{b}(\vec{k}_1)\hat{b}^{\dag}(\vec{k}^\prime_1)\hat{b}(\vec{k}_2)\hat{b}^{\dag}(\vec{k}^\prime_2) \Bigr\rangle&=&\delta^{3}\left(\vec{k}_1-\vec{k}^\prime_1\right)\delta^{3}\left(\vec{k}_2-\vec{k}^\prime_2\right)\,, \label{eq: 4 b bdag indentities}
\end{eqnarray}
which follow from eqs.~(\ref{eq: b bdag commutation rules}) and (\ref{eq: vacuum state}), one gets
\begin{equation}
\Delta_{\Phi^2,\Phi^2}(t,r)=\left[\frac{1}{2\pi^2}\int_{k_0}^{\infty}{\rm d}k~k^2 |\varphi(t,k)|^2\right]^2+2\left[\frac{1}{2\pi^2}\int_{k_0}^{\infty}{\rm d}k~k^2 j_0(kr)|\varphi(t,k)|^2\right]^2\,,
\end{equation}
where, again, we regularized the infrared sector with the IR cutoff $k_0$. Comparing to the first of eqs.~(\ref{eq: set corr QFT}), we immediately recognize that
\begin{equation}
\label{eq: Wick true}
\Delta_{\Phi^2,\Phi^2}(t,r)=\left[\Delta_{\Phi\Phi}(t,0)\right]^2+2\left[\Delta_{\Phi\Phi}(t,r)\right]^2\,.
\end{equation}
This result agrees with Wick's theorem, i.e. the statement that higher-order correlators of free fields can be expressed as the sum of all possible pair contraction products:
\begin{eqnarray}
\Delta_{\Phi^2,\Phi^2}(t,r)=\Bigl\langle \hat{\Phi}(t,\vec{x}_1)\hat{\Phi}(t,\vec{x}_1)\hat{\Phi}(t,\vec{x}_2)\hat{\Phi}(t,\vec{x}_2)\Bigr\rangle&=&\Bigl\langle\hat{\Phi}(t,\vec{x}_1)\hat{\Phi}(t,\vec{x}_1)\Bigr\rangle \Bigl\langle\hat{\Phi}(t,\vec{x}_2)\hat{\Phi}(t,\vec{x}_2)\Bigr\rangle\,\nonumber\\
&+&2\Bigl\langle\hat{\Phi}(t,\vec{x}_1)\hat{\Phi}(t,\vec{x}_2)\Bigr\rangle \Bigl\langle\hat{\Phi}(t,\vec{x}_1)\hat{\Phi}(t,\vec{x}_2)\Bigr\rangle\,\nonumber\\
&=&\left[\Delta_{\Phi,\Phi}(t,0)\right]^2+2\left[\Delta_{\Phi,\Phi}(t,r)\right]^2\,.\label{eq: Wick contractions}
\end{eqnarray}
This is a consequence of the one-loop approximation to QFT and therefore the use of free quantum fields in the correlators. Wick's theorem also applies more generally in QFT than just eq.~(\ref{eq: Wick contractions}), where some $n$-point function is given in terms of products of 2-point functions plus irreducible $n$-pt functions.
\noindent Canonical quantization via eqs.~(\ref{eq: canon Phi}), (\ref{eq: canon Pi}), (\ref{eq: b bdag commutation rules}) applies to free fields, which is the case for the action~(\ref{eq:action}) as long as one considers a fixed classical background metric.
In the most recent cosmological times, the backreaction of the scalar field becomes large (eventually dominating the expansion as dark energy) and therefore one should treat the metric as a quantum field to take into account the strong fluctuations induced by the energy-momentum tensor of the scalar field. A simple way to approximate these last stages of non-Gaussian evolution is to assume that the Friedmann equation is still valid, but the Hubble rate is a local quantum operator $\hat{H}(t,\vec{x})$ (therefore fluctuating):

\begin{equation}
\label{eq: Hubble quantum 1}
3M_P^2 \hat{H}^2(t,\vec{x})=\rho_{C}(t)\unit+\hat{\rho}_{Q}(t,\vec{x})\,,
\end{equation}
where $M_P$ is the reduced Planck mass $M_P=\left(8\pi G\right)^{-\frac12}$, $\rho_{C}(t)$ is the classical\footnote{$\rho_{C}$ is dominated by dark matter, which may also be of quantum origin and therefore with a fluctuating character~\cite{Friedrich:2017glg, Friedrich:2018qjv, Friedrich:2019zic, Vinke:2020}, which should be included in a complete treatment. For simplicity, here we opt against the inclusion of such effects.} energy density contribution which dominates the expansion until the backreaction of the scalar field is large, $\unit$ is the identity on the Hilbert space of quantum states and $\hat{\rho}_{Q}(t,\vec{x})$ is the energy density operator of the scalar field given by eq.~(\ref{eq: rho_Q operator}). The quantum version~(\ref{eq: Hubble quantum 1}) of the Friedmann equation will be the starting point of Section~\ref{sect: qbackr}.

Contrarily to the quantum field theory answer~(\ref{eq: Wick true}), even when fixing the classical background metric (as FLRW) so that the spectator scalar field is Gaussian, the result of stochastic formalism in eqs.~(\ref{4pt_inflation})--(\ref{spatial_inflation}) does not respect Wick's theorem. This is the main reason for studying an alternative stochastic approach in the next Section~\ref{sec: FP stochastic}.

\section{Stochastic solution from Fokker-Planck equation}
\label{sec: FP stochastic}

In this Section, we discuss an alternative solution for the time evolution of correlators during inflation within stochastic theory. Instead of deriving and solving a set of differential equations obeyed by the correlators as in eqs.~(\ref{eomPhiPhi})--(\ref{eomPiPi}), we will work with probability distributions for the field, extending the formalism developed by Starobinsky and Yokoyama in~\cite{Starobinsky:1994bd}. The result for the 2-point field correlator $\langle\hat{\phi}(t,\vec{x}) \hat{\phi}(t,\vec{y})\rangle$ will be the same as the one derived from the system of equations in subsection~\ref{subsect:corr}, but remarkably the two methods give a different prediction for the dependence of the 4-pt function $\langle\hat{\phi}^2(t,\vec{x}) \hat{\phi}^2(t,\vec{y})\rangle$ on the relative distance between the two points $\vec{x}$ and $\vec{y}$. Indeed, the method discussed in this Section will provides a result obeying Wick's theorem while the outcome obtained from the method described in subsection~\ref{subsect:corr 4pt} did not satisfy it.

For simplicity we focus on the evolution during de Sitter inflation, but the treatment could be extended to post-inflationary phases taking the inflationary results as initial conditions. Similarly to what we discussed in the previous approach of subsections~\ref{subsect:corr} and \ref{subsect:corr 4pt}, in post-inflationary epochs the contribution of stochastic sources is negligible with respect to the initial conditions inherited from inflation. 

If $H_I$ is the constant Hubble rate in de Sitter inflation, in the slow-roll regime the long-wavelength part (coarse-grained) of the scalar field obeys a Langevin equation with stochastic white noise, see eqs.~(9)-(11) of~\cite{Starobinsky:1994bd}. This observation is the basis for simplifying the quantum problem into a classical stochastic process.

\subsection{One-point probability distribution}
\label{subsec: 1-pt PDF}
Following~\cite{Starobinsky:1994bd}, let us start with the evolution of the one-point probability distribution function (PDF) $\rho_1\left[\varphi(t, \vec{x})\right]=\rho_1(\varphi,t)$ expressing the probability density function for the value of the coarse-grained field at comoving position $\vec{x}$ at the time $t$. Due to space homogeneity this cannot depend on the position $\vec{x}$, but only on time $t$ and the field value $\varphi$.
Coincident correlators and, more generally, expectation values of a function $F_1$ of the field at a single spacetime point $(t,\vec{x})$ can be computed as
\begin{equation}
\label{eq: 1pt func expect val}
\langle F_1\left(\varphi(t, \vec{x})\right)\rangle=\int {\rm d}\varphi~\rho_1(\varphi,t)F_1(\varphi)
\end{equation}
where we normalize $\rho_1(\varphi,t)$ to unity:
\begin{equation*}
\int {\rm d}\varphi~\rho_1(\varphi,t)=1\,,
\end{equation*}
the integral being evaluated on the support of $\rho_1(\varphi,t)$.

Following~\cite{Starobinsky:1994bd}, just like for Brownian motion, the Langevin equation for the coarse-grained field implies a Fokker-Planck equation for the evolution of the probability distribution (see eq.~(12) of~\cite{Starobinsky:1994bd}):
\begin{equation}
\label{eq: FP 1 pt}
\frac{\partial}{\partial t}\rho_1(\varphi,t)=\hat{\Gamma}_\varphi \rho_1(\varphi,t)\equiv\frac{1}{3H_I}\frac{\partial}{\partial \varphi}\left(V^\prime(\varphi)\rho_1(\varphi,t)\right)+\frac{H_I^3}{8\pi^2}\frac{\partial^2}{\partial \varphi^2}\rho_1(\varphi,t)
\end{equation}
where the second equality defines the action of the derivative operator $\hat{\Gamma}_\varphi$ on $\rho_1(\varphi,t)$ and $V(\varphi)$ is the potential of the field $\varphi$. The subscript ``1" in $\rho_1(\varphi,t)$ just reminds that this is a one-point PDF. Eq.~(\ref{eq: FP 1 pt}) has been used in~\cite{Starobinsky:1994bd} to study dynamical mass generation in de Sitter from a self-interacting potential $V(\varphi)$, e.g. quartic self-interaction.
In the present work what is relevant for us is just the case of~(\ref{eq:action}), which corresponds to a free field and therefore a non-interacting potential $V(\varphi)=\frac12 M^2\varphi^2$. Then $V^\prime(\varphi)=M^2\varphi$ where $M^2=m^2+\xi R=m^2+12\xi H_I^2$ is the effective squared mass, which is constant in de Sitter.
Thus we can reduce~(\ref{eq: FP 1 pt}) to
\begin{equation}
\frac{1}{H_I}\frac{\partial}{\partial t}\rho_1(\varphi,t)=\beta\frac{\partial}{\partial \varphi}\left(\varphi\rho_1(\varphi,t)\right)+\frac{H_I^2}{8\pi^2}\frac{\partial^2}{\partial \varphi^2}\rho_1(\varphi,t)\,,
\end{equation}
where we introduced the dimensionless parameter $\beta$ as
\begin{equation}
\label{eq: def beta}
\beta\equiv\frac13\frac{M^2}{H_I^2}=\frac13\frac{m^2}{H_I^2}+4\xi\,.
\end{equation}
At the beginning of inflation $t=t_{\rm in}$ we take the initial PDF to be
\begin{equation}
\label{eq: FP 1 pt ics}
\rho_1(\varphi,t_{\rm in})=\delta(\varphi)\,,
\end{equation}
so that the field has no condensate and all higher order self-correlations are zero, in agreement with the initial conditions that we used in subsections~\ref{subsect:corr} and \ref{subsect:corr 4pt}. 
The solution of the Fokker-Planck equation in the non-interacting case with Dirac delta initial condition is known to be a Gaussian distribution. It is straightforward to check the validity of the following solution (properly normalized to one):
\begin{equation}
\label{eq: 1pt prob sol}
\rho_1(\varphi,t)=\frac{1}{\sigma(t)\sqrt{2\pi}}\exp{\left[-\frac{\varphi^2}{2\sigma^2(t)}\right]}\,,\qquad\sigma^2(t)=\frac{H_I^2}{8\pi^2}\frac{1-e^{-2\beta H_I(t-t_{\rm in})}}{\beta}\,.
\end{equation}
The property $\sigma^2(t_{\rm in})=0$ ensures compliance with the initial condition $\rho_1(\varphi,t_{\rm in})=\delta(\varphi)$.
We are ultimately interested in $\beta\approx-4|\xi|<0$ corresponding to a negative non-minimal coupling dominating over the very light mass, but the solution~(\ref{eq: 1pt prob sol}) is valid for any sign of $\beta$. It is also interesting to observe that the free massless minimally coupled case $\beta=0$ can be obtained by taking the limit $\beta\to 0$ of~(\ref{eq: 1pt prob sol}), which then gives the random walk behavior with variance growing linearly in time $\sigma^2(t)=H_I^3(t-t_{\rm in})/(4\pi^2)$. The presence of a non-zero mass (i.e. $\beta$) is responsible for deviations from this pure random walk evolution. When $\beta>0$ the variance $\sigma^2(t)$ asymptotically approaches the finite value $H_I^2/(8\pi^2\beta)$ and the existence of a stationary solution is a consequence of the stability of the potential $V(\varphi)=(3/2) \beta H_I^2 \varphi^2$, which has $V''(\varphi)=M^2=3\beta H_I^2$ positive for $\beta>0$ (upward concavity of the parabola). However, in the most relevant case for us $\beta<0$ and the variance grows exponentially, which is in agreement with what one would expect from an unstable potential (downward concave parabola).

We can immediately read the coincident 2-point function $\Delta_{\phi\phi}(t)=\langle\hat{\phi}^2(t,\vec{x})\rangle$ from~(\ref{eq: 1pt prob sol}) because it is simply given by the variance $\sigma^2(t)$. In terms of the number of e-foldings $N=H_I(t-t_{\rm in})$ measured from the beginning of inflation one has
\begin{equation}
\Delta_{\phi\phi}(N)=\frac{H_I^2}{8\pi^2}\frac{1-e^{-2\beta N}}{\beta}\,,
\end{equation}
which, when $\beta\simeq-4|\xi|$, shows the same exponential growth with $N$ predicted before in~\cite{Glavan:2017jye} and written in eq.~(\ref{sol_deSitter 2pt 4}).

\subsection{Two-point probability distribution}
We now turn our attention to non-coincident correlators. Their study requires knowledge of the joint probability distribution at two different spacetime points. Of course, for our purposes it is sufficient to consider fields at the same time $t$ but different positions $\vec{x}_1$ and $\vec{x}_2$. Our starting point is eq. (73) in~\cite{Starobinsky:1994bd}, which is the Fokker-Planck equation for the joint two-point PDF $\rho_2(\varphi_1,t;\varphi_2,t)$:
\begin{equation}
\label{eq: FP joint 2 pt}
\frac{\partial}{\partial t}\rho_2(\varphi_1,t;\varphi_2,t)=\hat{\Gamma}_{\varphi_1} \rho_2+\hat{\Gamma}_{\varphi_2} \rho_2+j_0\left(\mu a(t)H_Ir\right)\frac{H_I^3}{4\pi^2}\frac{\partial^2}{\partial \varphi_1 \partial \varphi_2}\rho_2\,.
\end{equation}
The subscript ``2" in $\rho_2$ signals that this is a two-point joint PDF while the arguments $\varphi_1,t$ and $\varphi_2,t$ mean that we are considering the probability density for the field at $\vec{x}_1$ and time $t$ to have value $\varphi_1$ and for the field at $\vec{x}_2$ and same time $t$ to have value $\varphi_2$. The distance $r$ appearing in the argument of the $j_0$ spherical Bessel function is the comoving distance between $\vec{x}_1$ and $\vec{x}_2$ given by $r=\|\vec{x}_2-\vec{x}_1\|$. The derivative operators $\hat{\Gamma}_{\varphi_1}$ and $\hat{\Gamma}_{\varphi_2}$ have the same form as $\hat{\Gamma}_{\varphi}$ defined in~(\ref{eq: FP 1 pt}), but now with respect to ${\varphi_1}$ and ${\varphi_2}$ respectively.
Just like for~(\ref{eq: 1pt func expect val}), expectation values of a function $F_2$ of fields at the two spacetime points $(t, \vec{x}_1)$ and $(t, \vec{x}_2)$ can be computed using $\rho_2(\varphi_1,t;\varphi_2,t)$ as
\begin{equation}
\label{eq: 2pt func expect val}
\langle F_2\left(\varphi(t, \vec{x}_1);\varphi(t, \vec{x}_2)\right)\rangle=\int {\rm d}\varphi_1\int {\rm d}\varphi_2~\rho_2(\varphi_1,t;\varphi_2,t)F_2(\varphi_1;\varphi_2)\,,
\end{equation}
with the normalization condition
\begin{equation*}
\int {\rm d}\varphi_1\int {\rm d}\varphi_2~\rho_2(\varphi_1,t;\varphi_2,t)=1\,.
\end{equation*}

We now explain how to obtain $\rho_2(\varphi_1,t;\varphi_2,t)$ solving eq.~(\ref{eq: FP joint 2 pt}). The reader who is mainly interested in the results for correlators might want to skip this part and jump directly to the summary in subsection~\ref{subsubsec: essential}.

\subsubsection{Solution for the joint two-point PDF}

Following~\cite{Starobinsky:1994bd}, since we are dealing with the coarse-grained field (long-wavelength part), the spherical Bessel function $j_0\left(\mu a(t)H_I r\right)$ should be approximated by a Heaviside step function:
\begin{equation}
\label{eq: Bessel Heaviside approx}
j_0\left(\mu a(t)H_I r\right)\simeq\theta\left(1-\mu a(t)H_I r\right)\,.
\end{equation}
This corresponds exactly to the approximation that we use for noise sources in eq.~(\ref{eq: definition Fj}). Given the two comoving positions $\vec{x}_1$ and $\vec{x}_2$ separated by a comoving distance $r$, we can distinguish two periods in the time evolution of $\rho_2(\varphi_1,t;\varphi_2,t)$ depending on whether the Heaviside step function~(\ref{eq: Bessel Heaviside approx}) evaluates to unity or zero. The first phase starts from the beginning of inflation $t_{\rm in}$ and lasts until the time $t_r\equiv t_{\rm in}-\ln{(\mu a_{\rm in}H_Ir)}/H_I$. In this time interval the Heaviside function evaluates  to unity. The second phase is $t>t_r$, and the Heaviside function becomes zero.
In the first phase $t_{\rm in}<t<t_r$ one can check that the 2-point Fokker-Planck equation~(\ref{eq: FP joint 2 pt}) admits a solution of the form
\begin{equation}
\label{eq: 2pt PDF first}
\rho_2(\varphi_1,t;\varphi_2,t)=\rho_1(\varphi_1,t)\delta(\varphi_1-\varphi_2)\,,
\end{equation}
where $\rho_1(\varphi_1,t)$ is the one-point PDF studied in the previous subsection~\ref{subsec: 1-pt PDF} evaluated at time $t$ for the field value $\varphi_1$.
The existence of such a solution was pointed out in eq. (75) of~\cite{Starobinsky:1994bd} in the stationary late-times regime, but it is actually valid in full generality with time dependence, so that is applicable to study the time evolution of correlators.
As a check of the initial conditions that we are using, at the beginning of inflation $t_{\rm in}$ the solution~(\ref{eq: 2pt PDF first}) gives $\rho_2(\varphi_1,t_{\rm in};\varphi_2,t_{\rm in})=\delta(\varphi_1)\delta(\varphi_2)$, because of the initial condition used for the one-point PDF $\rho_1(\varphi_1,t_{\rm in})=\delta(\varphi_1)$. This initial value of $\rho_2(\varphi_1,t_{\rm in};\varphi_2,t_{\rm in})$ implies $\langle\varphi(t_{\rm in},\vec{x}_1)\varphi(t_{\rm in},\vec{x}_2)\rangle=0$ which is equivalent to the zero initial conditions for correlators used in Section~\ref{sect:model}.

It is important to notice that the first phase $t_{\rm in}<t<t_r$ really exists only as long as $t_r>t_{\rm in}$, i.e. when $r<r_0\equiv1/(\mu a_{\rm in}H_I)$. Indeed when $r>r_0$ the Heaviside function evaluates to zero already at the beginning of inflation and it keeps being null at later times. Therefore when $r>r_0$ we are immediately in the ``second" phase, already from the beginning of inflation $t_{\rm in}$ and we will get back to this case later.

For now let us assume $r<r_0$ and discuss the solution for the two-point joint PDF in the second phase $t>t_r$, when the last term in~(\ref{eq: FP joint 2 pt}) drops to zero. The initial conditions are those inherited from the first phase at time $t_r$, i.e.
\begin{equation}
\label{eq: 2pt PDF second ics}
\rho_2(\varphi_1,t_r;\varphi_2,t_r)=\rho_1(\varphi_1,t_r)\delta(\varphi_1-\varphi_2)\,.
\end{equation}
The relevant information for the solution at $t>t_r$ can be extracted from eq. (79) of~\cite{Starobinsky:1994bd} paying attention to the fact that we are interested in the full time evolution and not in the stationary regime. The solution at $t>t_r$ can be expressed as
\begin{equation}
\label{eq: 2pt PDF second}
\rho_2(\varphi_1,t;\varphi_2,t)=\int {\rm d}\varphi_r~\rho_1(\varphi_r,t_r)\Pi(\varphi_1,t|\varphi_r,t_r)\Pi(\varphi_2,t|\varphi_r,t_r)\,,
\end{equation}
where $\Pi(\varphi_1,t|\varphi_r,t_r)$ is the conditional probability density that the field at position $\vec{x}_1$ and time $t$ has value $\varphi_1$ given that at the same position $\vec{x}_1$ the field at time $t_r$ had value $\varphi_r$. A completely analogous meaning is attributed to $\Pi(\varphi_2,t|\varphi_r,t_r)$ but now the fields refer to position $\vec{x}_2$. The integral in~(\ref{eq: 2pt PDF second}) goes over all possible values of $\varphi_r$. The conditional probability $\Pi(\varphi_1,t|\varphi_r,t_r)$ satisfies a Fokker-Planck equation (see eq. (26) of~\cite{Starobinsky:1994bd}):
\begin{equation}
\label{eq: condprob1 eq}
\frac{\partial}{\partial t}\Pi(\varphi_1,t|\varphi_r,t_r)=\hat{\Gamma}_{\varphi_1}\Pi(\varphi_1,t|\varphi_r,t_r)\,,
\end{equation}
with initial conditions
\begin{equation}
\label{eq: condprob1 ics}
\Pi(\varphi_1,t_r|\varphi_r,t_r)=\delta(\varphi_1-\varphi_r)\,.
\end{equation}
A similar equation holds with respect to the second argument $\varphi_r,t_r$ (see eq. (27) of~\cite{Starobinsky:1994bd}), but it will not be needed.
In the same way $\Pi(\varphi_2,t|\varphi_r,t_r)$ follows the equation
\begin{equation}
\label{eq: condprob2 eq}
\frac{\partial}{\partial t}\Pi(\varphi_2,t|\varphi_r,t_r)=\hat{\Gamma}_{\varphi_2}\Pi(\varphi_2,t|\varphi_r,t_r)\,,
\end{equation}
with initial conditions
\begin{equation}
\label{eq: condprob2 ics}
\Pi(\varphi_2,t_r|\varphi_r,t_r)=\delta(\varphi_2-\varphi_r)\,.
\end{equation}
It is straightforward to check that if the conditional probabilities satisfy eqs.~(\ref{eq: condprob1 eq}) and~(\ref{eq: condprob2 eq}) with their initial conditions~(\ref{eq: condprob1 ics}) and~(\ref{eq: condprob2 ics}) respectively, then~(\ref{eq: 2pt PDF second}) is a solution of ~(\ref{eq: FP joint 2 pt}) (with the spherical Bessel function $j_0$ approximated by zero as it is appropriate for $t>t_r$) with initial conditions~(\ref{eq: 2pt PDF second ics}). Thus the problem completely reduces to solving eqs.~(\ref{eq: condprob1 eq}) and~(\ref{eq: condprob2 eq}).

The solution for conditional probabilities can be expressed in terms of the Mehler kernel\footnote{Defining $\Psi=\Pi \exp{\left[-\frac{\beta}{2}\left(H_It-4\pi^2\frac{\varphi_1^2}{H_I^2}\right)\right]}$ and changing variables from $t$ and $\varphi_1$ to $\tau\equiv\frac12 |\beta|H_I t$ and $\chi\equiv2\pi\sqrt{|\beta|}\frac{\varphi_1}{H_I}$, eq. (\ref{eq: condprob1 eq}) is equivalent to $\frac{\partial}{\partial\tau}\Psi=\left(\frac{\partial^2}{\partial\chi^2}-\chi^2\right)\Psi$. The Green's function of this equation is known as the Mehler kernel, see also Section V.A of~\cite{Prokopec:2019srf}. Its importance in physics stems from being the propagator of the quantum harmonic oscillator (upon changing $\tau$ into the imaginary $i\tau$), see eq. (2.5.18) of Sakurai's textbook~\cite{Sakurai:1167961} for its expression in the quantum mechanics context.}, which can be obtained by making a Gaussian ansatz, checking its validity and solving for the mean and variance as functions of time, thus reducing the problem to ordinary differential equations. The result is that eq.~(\ref{eq: condprob1 eq}) with initial condition~(\ref{eq: condprob1 ics}) is solved by a Gaussian distribution with mean and variance evolving with time as follows:
\begin{align}
&&\Pi(\varphi_1,t|\varphi_r,t_r)=\frac{1}{\sigma_r(t)\sqrt{2\pi}}\exp{\left[-\frac{\left(\varphi_1-\mu_r(t)\right)^2}{2\sigma_r^2(t)}\right]}\,,\nonumber\\
&&\mu_r(t)=\varphi_r~e^{-\beta H_I(t-t_r)}\,,\qquad\,\sigma_r^2(t)=\frac{H_I^2}{8\pi^2}\frac{1-e^{-2\beta H_I(t-t_r)}}{\beta}\,.
\label{eq: sol cond1}
\end{align}
The time evolution of the variance $\sigma_r^2(t)$ is perfectly analogous to the solution for the one-point PDF~(\ref{eq: 1pt prob sol}). The novel element is the drift of the distribution, namely the translation of its mean value $\mu_r(t)$. This effect was not present in~(\ref{eq: 1pt prob sol}) because of the zero initial conditions~(\ref{eq: FP 1 pt ics}), which kept the mean to zero also at later times. On the contrary, a generic $\varphi_r$ appears in the initial conditions~(\ref{eq: condprob1 ics}).
In the same way, the solution of~(\ref{eq: condprob2 eq}) with initial conditions~(\ref{eq: condprob2 ics}) is
\begin{equation}
\label{eq: sol cond2}
\Pi(\varphi_2,t|\varphi_r,t_r)=\frac{1}{\sigma_r(t)\sqrt{2\pi}}\exp{\left[-\frac{\left(\varphi_2-\mu_r(t)\right)^2}{2\sigma_r^2(t)}\right]}\,,
\end{equation}
with the same $\mu_r(t)$ and $\sigma_r(t)$ as in~(\ref{eq: sol cond1}).

The only other result needed for evaluating the joint two-point PDF according to eq.~(\ref{eq: 2pt PDF second}) is the one-point PDF $\rho_1(\varphi_r,t_r)$ which is readily obtained by substituting its arguments in the general formula~(\ref{eq: 1pt prob sol}):
\begin{equation}
\label{eq: 1pt prob phir tr}
\rho_1(\varphi_r,t_r)=\frac{1}{\sigma(t_r)\sqrt{2\pi}}\exp{\left[-\frac{\varphi_r^2}{2\sigma^2(t_r)}\right]}\,,\qquad\sigma^2(t_r)=\frac{H_I^2}{8\pi^2}\frac{1-e^{-2\beta H_I(t_r-t_{\rm in})}}{\beta}\,.
\end{equation}
One can then perform the integration~(\ref{eq: 2pt PDF second}) paying attention to the dependence on $\varphi_r$ in $\mu_r(t)$ given by~(\ref{eq: sol cond1}). This leads after a few steps to the following two-point joint PDF for $t>t_r$:
\begin{align}
\label{eq: sol rho2 t>tr}
\rho_2(\varphi_1,t;\varphi_2,t)=\frac{1}{2\pi\sqrt{{\rm det}(\Sigma(t,r))}}\exp\left[-\frac12(\varphi_1, \varphi_2)\Sigma^{-1}(t,r)
\begin{pmatrix}
\varphi_1 \\
\varphi_2 \\
\end{pmatrix}\,
\right]\,,\nonumber \\
\Sigma(t,r)=
\begin{pmatrix}
\sigma^2(t) & \sigma^2(t)-\sigma_r^2(t)\\
\sigma^2(t)-\sigma_r^2(t) & \sigma^2(t)
\end{pmatrix}
\,,
\end{align}
which is a two-dimensional Gaussian distribution with zero mean for both $\varphi_1$ and $\varphi_2$ and with covariance matrix $\Sigma(t,r)$ specified above. Notice that the diagonal terms of $\Sigma(t,r)$ are both equal to $\sigma^2(t)$ consistently with the one-point PDF in eq.~(\ref{eq: 1pt prob sol}) that we are supposed to obtain when marginalizing over one of the two variables.
\noindent It is worth reminding that, for a given comoving spatial separation $r$ between the two points, the result~(\ref{eq: sol rho2 t>tr}) is valid when $t>t_r>t_{\rm in}$, which using $t_r=t_{\rm in}-\frac{1}{H_I}\ln\left(\frac{r}{r_0}\right)$ translates into $N>\ln\left(\frac{r_0}{r}\right)>0$. The last inequality makes sense only when $r<r_0$.

\noindent We also remind that when $t_{\rm in}<t<t_r$, which translates into $0<N<\ln\left(\frac{r_0}{r}\right)$, then the solution for $\rho_2(\varphi_1,t;\varphi_2,t)$ has been written in eq.~(\ref{eq: 2pt PDF first}). Again this requires $r<r_0$.

Now we study $\rho_2(\varphi_1,t;\varphi_2,t)$ in the case $r>r_0$. For such distances the Bessel function in~(\ref{eq: FP joint 2 pt}) (and approximated by a Heaviside step function) is always zero from the very beginning of inflation. Then the solution is given by
\begin{equation}
\label{eq: 2pt PDF r>r0}
\rho_2(\varphi_1,t;\varphi_2,t)=\int {\rm d}\varphi_{\rm in}~\rho_1(\varphi_{\rm in},t_{\rm in})\Pi(\varphi_1,t|\varphi_{\rm in},t_{\rm in})\Pi(\varphi_2,t|\varphi_{\rm in},t_{\rm in})\,,
\end{equation}
with the same meaning of terms as in~(\ref{eq: 2pt PDF second}), except for the replacement of $\varphi_r,t_r$ by $\varphi_{\rm in},t_{\rm in}$ because when $r>r_0$ there are no longer two distinct phases in the evolution of $\rho_2(\varphi_1,t;\varphi_2,t)$ and the $t_r$ makes no sense. Therefore one has to integrate over values $\varphi_{\rm in}$ of the field at time $t_{\rm in}$, instead of $\varphi_r$ at $t_r$.
The conditional probabilities are now given by
\begin{align}
&&\Pi(\varphi_1,t|\varphi_{\rm in},t_{\rm in})=\frac{1}{\sigma(t)\sqrt{2\pi}}\exp{\left[-\frac{\left(\varphi_1-\mu_{\rm in}(t)\right)^2}{2\sigma^2(t)}\right]}\,,\nonumber\\
&&\mu_{\rm in}(t)=\varphi_{\rm in}~e^{-\beta H_I(t-t_{\rm in})}\,,\qquad\,\sigma^2(t)=\frac{H_I^2}{8\pi^2}\frac{1-e^{-2\beta H_I(t-t_{\rm in})}}{\beta}\,,
\label{eq: sol cond1 r>r0}
\end{align}
and similarly
\begin{equation}
\Pi(\varphi_2,t|\varphi_{\rm in},t_{\rm in})=\frac{1}{\sigma(t)\sqrt{2\pi}}\exp{\left[-\frac{\left(\varphi_2-\mu_{\rm in}(t)\right)^2}{2\sigma^2(t)}\right]}\,.
\label{eq: sol cond2 r>r0}
\end{equation}
Notice that $\sigma(t)$ in eqs.~(\ref{eq: sol cond1 r>r0}--\ref{eq: sol cond2 r>r0}) is exactly the same defined for the one-point PDF~(\ref{eq: 1pt prob sol}). This is again a consequence of considering $r>r_0$, in which case one has initial conditions set at $t_{\rm in}$, instead of the other time $t_r$ introduced when $r<r_0$. Finally, due to the initial conditions~(\ref{eq: FP 1 pt ics}), one has
\begin{equation}
\rho_1(\varphi_{\rm in},t_{\rm in})=\delta(\varphi_{\rm in})\,.
\end{equation}
The integration in~(\ref{eq: 2pt PDF r>r0}) immediately gives the solution for $r>r_0$:
\begin{align}
\label{eq: sol rho2 r>r0}
\rho_2(\varphi_1,t;\varphi_2,t)=\frac{1}{2\pi\sigma^2(t)}\exp{\left[-\frac{\varphi_1^2+\varphi_2^2}{2\sigma^2(t)}\right]}=\rho_1(\varphi_1,t)\rho_1(\varphi_2,t)\,,
\end{align}
which is a Gaussian with diagonal covariance matrix, from which we infer the absence of correlations between $\varphi_1$ and $\varphi_2$ when $r>r_0$.

\subsubsection{Summary on joint PDF and non-coincident correlators}
\label{subsubsec: essential}
Before moving to the calculation of non-coincident 2-point and 4-point functions, let us summarize the results on the joint two-point probability distribution $\rho_2(\varphi_1,t;\varphi_2,t)$ resulting from the solution of the Fokker-Planck equation~(\ref{eq: FP joint 2 pt}). We introduced the scale
\begin{equation}
r_0\equiv\frac{1}{\mu a_{\rm in} H_I}\,,
\end{equation}
which sets the separation between long and short modes of the stochastic theory at the beginning of inflation.
In~(\ref{eq: def beta}) we also defined for convenience a dimensionless parameter related to the effective mass of the scalar field as
$\beta\equiv\frac13\frac{M^2}{H_I^2}=\frac13\frac{m^2}{H_I^2}+4\xi$. We computed $\rho_2(\varphi_1,t;\varphi_2,t)$ depending on the number of e-foldings since the beginning of inflation $N=H_I(t-t_{\rm in})$ and on the separation $r=\|\vec{x}_2-\vec{x}_1\|$ between the comoving positions $\vec{x}_1$ and $\vec{x}_2$. The results from~(\ref{eq: 2pt PDF first}),~(\ref{eq: sol rho2 t>tr}) and~(\ref{eq: sol rho2 r>r0}) can be effectively summarized as follows:

\begin{equation}
\label{eq: joint PDF final}
\boxed{\rho_2(\varphi_1,N;\varphi_2,N) =
\begin{cases}
         \frac{1}{\sigma(N)\sqrt{2\pi}}\exp{\left[-\frac{\varphi_1^2}{2\sigma^2(N)}\right]}\delta(\varphi_1-\varphi_2) & \text{if $r<r_0~ e^{-N}$}\\
        \frac{1}{2\pi\sqrt{{\rm det}(\Sigma(N,r))}}\exp\left[-\frac12(\varphi_1, \varphi_2)\Sigma^{-1}(N,r)
\begin{pmatrix}
\varphi_1 \\
\varphi_2 \\
\end{pmatrix}\,
\right] & \text{if $r_0~e^{-N}<r<r_0$}\\
        \frac{1}{2\pi\sigma^2(N)}\exp{\left[-\frac{\varphi_1^2+\varphi_2^2}{2\sigma^2(N)}\right]} & \text{if $r>r_0$}\,,
\end{cases}
}
\end{equation}
where $\sigma^2(N)$ is the variance of the one-point PDF defined in~(\ref{eq: 1pt prob sol}), which we rewrite in terms of $N$ as
\begin{equation}
\label{eq: variance 1pt DPF}
\sigma^2(N)=\frac{H_I^2}{8\pi^2}\frac{1-e^{-2\beta N}}{\beta}\,,
\end{equation}
and $\Sigma(N)$ is the matrix defined in~(\ref{eq: sol rho2 t>tr}), which, upon using $t_r=t_{\rm in}-\frac{1}{H_I}\ln\left(\frac{r}{r_0}\right)$, can be equivalently expressed as
\begin{equation}
\label{eq: Covariance t>tr}
\Sigma(N,r)=\frac{H_I^2}{8\pi^2\beta}
\begin{pmatrix}
1-e^{-2\beta N} & e^{-2\beta N}\left[\left(\frac{r}{r_0}\right)^{-2\beta}-1\right]\\
e^{-2\beta N}\left[\left(\frac{r}{r_0}\right)^{-2\beta}-1\right] & 1-e^{-2\beta N}
\end{pmatrix}
\,.
\end{equation}
One can immediately recognize from eq.~(\ref{eq: joint PDF final}) that $\rho_2(\varphi_1,N;\varphi_2,N)$ exhibits perfect correlation between $\varphi_1$ and $\varphi_2$ for separations $r<r_0~e^{-N}$ (due to the Dirac delta function $\delta(\varphi_1-\varphi_2)$) and it is completely uncorrelated for separations $r>r_0$. For intermediate separations between $\vec{x}_1$ and $\vec{x}_2$, correlations go from their maximum value to the minimum one, interpolating between the two extremes according to the off-diagonal element(s) of $\Sigma(N,r)$ in eq.~(\ref{eq: Covariance t>tr}).

\begin{figure}[H]
\centering
\includegraphics[width=0.9\columnwidth]{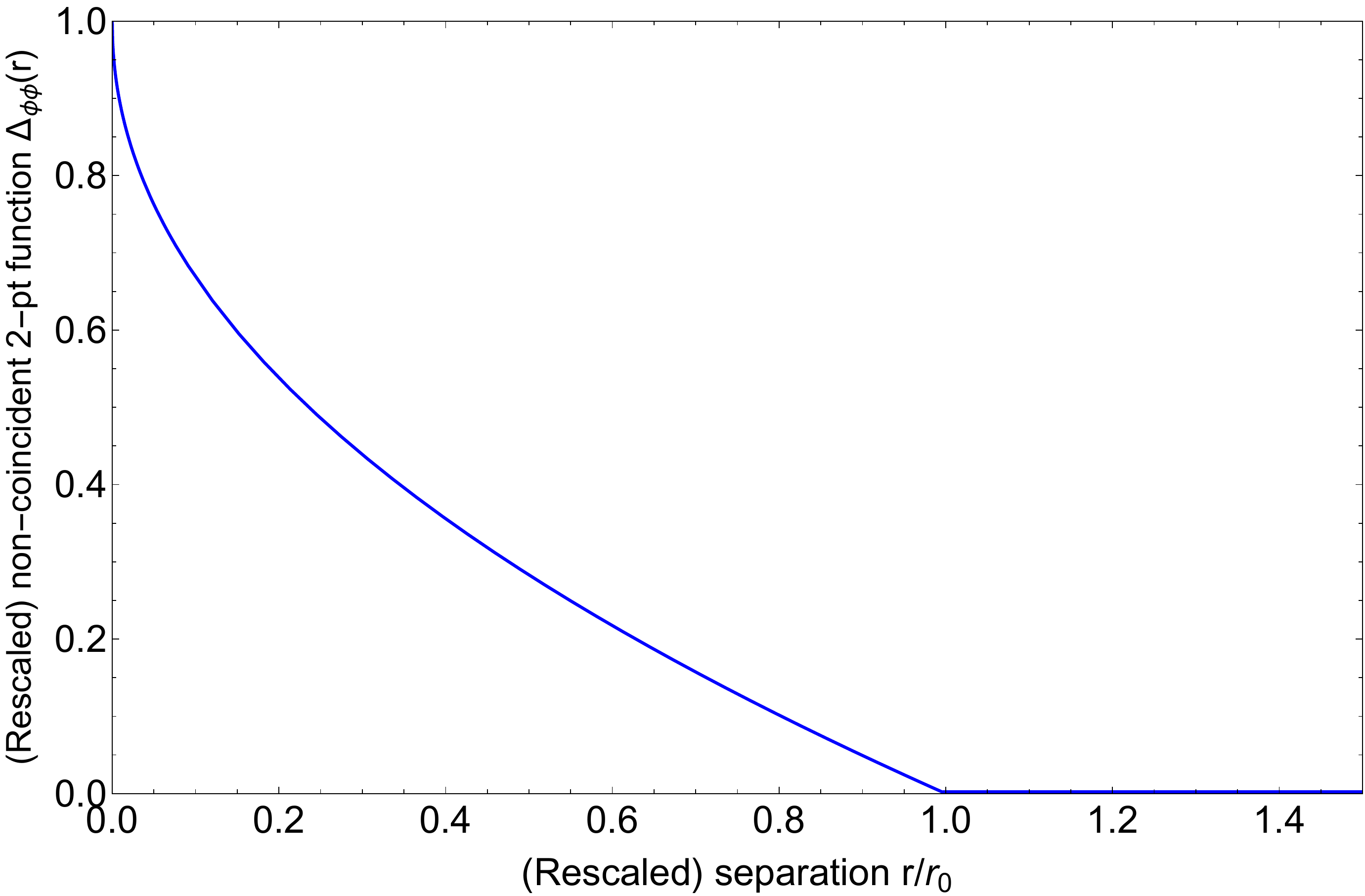}
\caption{The function $C(N,r)$ in~(\ref{eq: 2pt FP final simple}), equal to the non-coincident 2-point function $\Delta_{\phi\phi}(N,r)$  rescaled by its value at coincidence, plotted as a function of the separation $r$ between the points (rescaled by $r_0=1/(\mu a_{\rm in}H_I)$) at late times in inflation, $e^{8|\xi|N}\gg 1$. We used a negative non-minimal coupling $\xi=-0.06$. The function stays constant at zero for $r/r_0>1$. The other scale $r/r_0=e^{-N}$ appearing in eq.~(\ref{eq: 2pt FP final simple}) and setting the upper limit for the coincident regime is too small to be visible in the plot, $e^{-N}\ll1$.
\label{fig: 2pt noncoinc inflation}
}
\end{figure}

\begin{figure}[H]
\centering
\includegraphics[width=0.9\columnwidth]{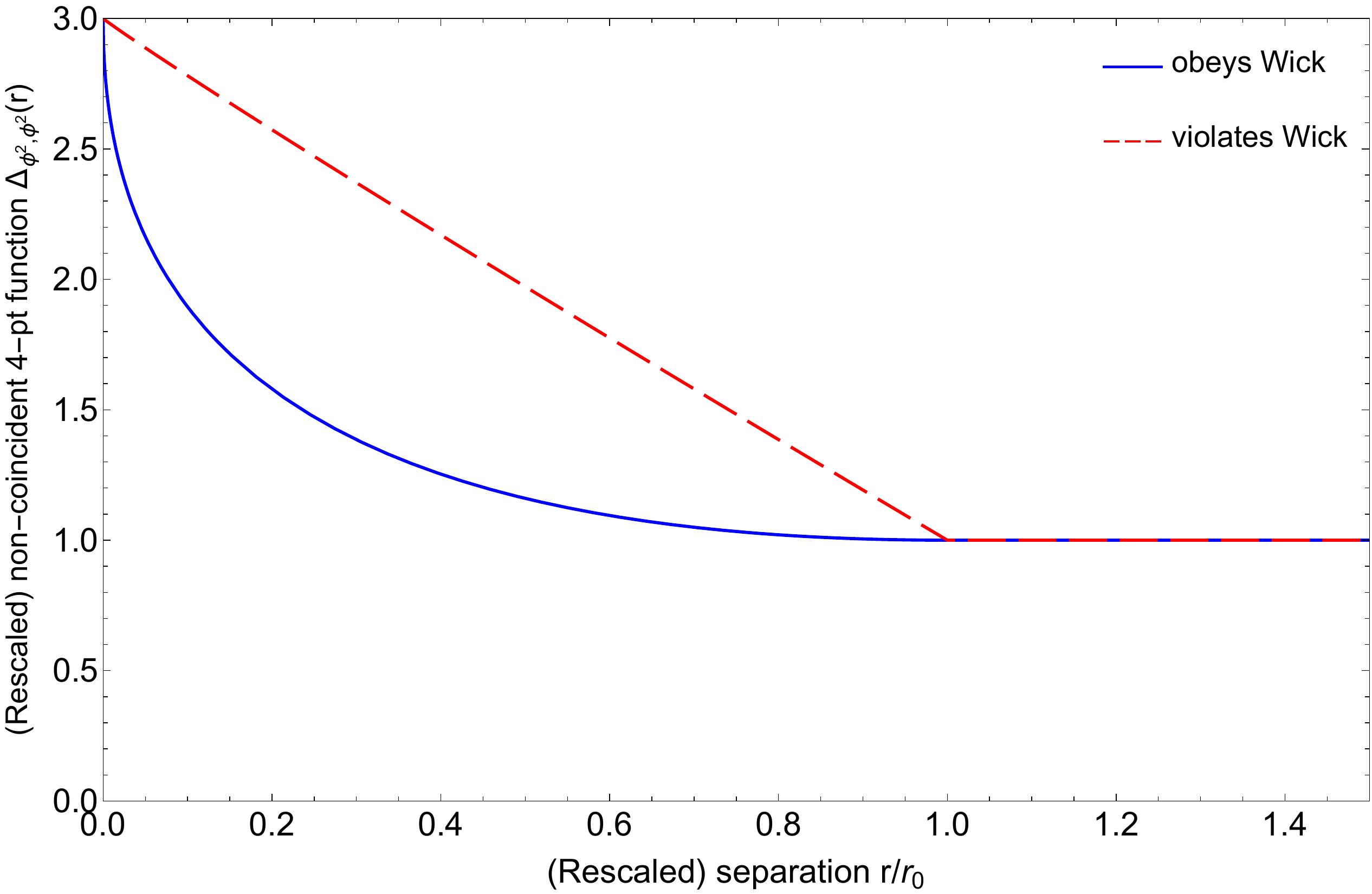}
\caption{Comparison between the spatial profile predicted by eq.~(\ref{eq: 4pt FP rescaled}) (blue solid curve) and by eq.~(\ref{spatial_inflation}) (red dashed curve). The blue solid curve, computed with the approach to stochastic formalism based on Fokker-Planck equations discussed in this Section~\ref{sec: FP stochastic}, obeys Wick's theorem. On the contrary, the red dashed curve, computed with the other approach to stochastic formalism adopted in Section~\ref{sect:model}, based on linear systems of equations for the time evolution of correlators, does not respect Wick's theorem.
\noindent We used $\xi=-0.06$ for the non-minimal coupling, as in Fig.~\ref{fig: 2pt noncoinc inflation}.
\label{fig: 4pt comparison}
}
\end{figure}

Quantitatively, the computation of the 2-point function $\langle\varphi_1\varphi_2\rangle$ follows from~(\ref{eq: 2pt func expect val}) giving, as a function of the separation $r$ and the number of e-foldings $N$,
\begin{equation}
\label{eq: 2pt FP final}
\langle\varphi_1\varphi_2\rangle(N,r)=
\begin{cases}
      \frac{H_I^2}{8\pi^2}\frac{1-e^{-2\beta N}}{\beta} & \text{if $r<r_0~e^{-N}$} \\
      \frac{H_I^2}{8\pi^2}e^{-2\beta N}\frac{\left(\frac{r}{r_0}\right)^{-2\beta}-1}{\beta} & \text{if $r_0~e^{-N}<r<r_0$} \\
      0 & \text{if $r>r_0$}\,.
\end{cases} 
\end{equation}
\noindent The result can also be expressed in terms of the correlation coefficient of the joint distribution~(\ref{eq: joint PDF final}), which we will denote by $C(N,r)$. For a distribution like $\rho_2(\varphi_1,N;\varphi_2,N)$ having $\langle\varphi_1\rangle=\langle\varphi_2\rangle=0$, the correlation coefficient is defined as customary as
\begin{equation}
\label{eq: C definition}
C(N,r)\equiv\frac{\langle\varphi_1\varphi_2\rangle(N,r)}{\sqrt{\langle\varphi_1\rangle^2(N)}\sqrt{\langle\varphi_2\rangle^2(N)}}=\frac{\langle\varphi_1\varphi_2\rangle(N,r)}{\sigma^2(N)}\,,
\end{equation}
that is by rescaling the 2-point function $\langle\varphi_1\varphi_2\rangle$ with respect to its value at coincidence. From~(\ref{eq: 2pt FP final}) it follows that
\begin{equation}
\label{eq: 2pt FP final rescaled}
C(N,r)=
\begin{cases}
      1 & \text{if $r<r_0~e^{-N}$} \\
      \frac{\left(\frac{r}{r_0}\right)^{-2\beta}-1}{e^{2\beta N}-1} & \text{if $r_0~e^{-N}<r<r_0$} \\
      0 & \text{if $r>r_0$}\,,
\end{cases} 
\end{equation}
which interpolates continuously from $1$ to $0$ with a power law.
For $\beta\simeq-4|\xi|$ and $N$ large enough so that $e^{-8|\xi|N}\ll 1$, this simplifies into
\begin{equation}
\label{eq: 2pt FP final simple}
\boxed{
C(N,r)\simeq
\begin{cases}
      1 & \text{if $r<r_0~e^{-N}$} \\
      1-\left(\frac{r}{r_0}\right)^{8|\xi|} & \text{if $r_0~e^{-N}<r<r_0$} \\
      0 & \text{if $r>r_0$}\,.
\end{cases}
}
\end{equation}
It agrees with the result found with another method in eq.~(\ref{eq: profile 2-pt s_2(r)}), thus providing further confirmation for its validity. 
The function $C(N,r)$ is plotted against $r/r_0$ in Fig.~\ref{fig: 2pt noncoinc inflation}. The precise choice of $N$ is irrelevant as long as it satisfies $e^{-8|\xi|N}\ll 1$ . We use $\xi=-0.06$ for the negative non-minimal coupling.

\begin{figure}[H]
\centering
\includegraphics[width=0.7\columnwidth]{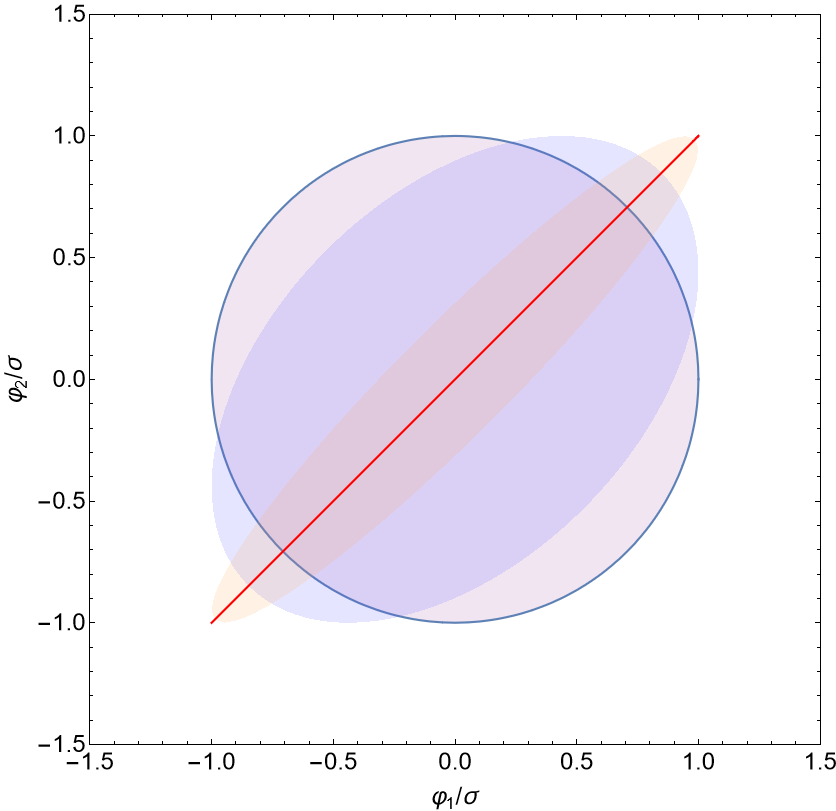}
\caption{The $1\sigma$ contour plot of the joint distribution $\rho_2(\varphi_1,N;\varphi_2,N)$ in~(\ref{eq: joint PDF final}) at a given large N (with $e^{8|\xi|N}\gg 1$), for different values of the comoving separation $r=\|\vec{x}_2-\vec{x}_1\|$ between the points $\vec{x}_1$ and $\vec{x}_2$, which the field values $\varphi_1$ and $\varphi_2$ refer to.
The variables $\varphi_1$ and $\varphi_2$ have been rescaled by their (equal) standard deviation $\sigma(N)$ written in~(\ref{eq: variance 1pt DPF}).
The $1\sigma$ contour is an ellipse and it goes from being degenerate as the red segment for $r<r_0 e^{-N}$ to becoming a circle (with the blue contour in the figure) when $r>r_0$. At intermediate distances, $r_0 e^{-N}<r<r_0$, the shape of the ellipse is determined by the corresponding correlation coefficient $C(N,r)$ according to eq.~(\ref{eq: 1sigma contour}).
\label{fig: 1sigma contour}
}
\end{figure}

The correlation coefficient $C(N,r)$ determines the shape of the contour levels of $\rho_2(\varphi_1,N;\varphi_2,N)$. Indeed it is straightforward to show from~(\ref{eq: joint PDF final}) that the $1\sigma$ contour of $\rho_2(\varphi_1,N;\varphi_2,N)$ can be written as
\begin{equation}
\label{eq: 1sigma contour}
\frac{\left(\frac{\varphi_1}{\sigma(N)}\right)^2-2~C(N,r)\frac{\varphi_1}{\sigma(N)}\frac{\varphi_2}{\sigma(N)}+\left(\frac{\varphi_2}{\sigma(N)}\right)^2}{1-C^2(N,r)}=1\,,
\end{equation}
where $\sigma(N)$ is the standard deviation of the one-point PDF evolving with time as in eq.~(\ref{eq: variance 1pt DPF}).
The $1\sigma$ regions corresponding to~(\ref{eq: 1sigma contour}) are plotted in Fig.~\ref{fig: 1sigma contour} at late times for different separations $r$.

Let us now study the correlator $\langle\varphi_1^2\varphi_2^2\rangle$ (which is the 4-point function $\Delta_{\phi^2, \phi^2}$ in the notation of Section~\ref{sect:model}) predicted by the Fokker-Planck solution~(\ref{eq: joint PDF final}). Again the evaluation is done by using~(\ref{eq: 2pt func expect val}) with the PDF in~(\ref{eq: joint PDF final}) to get from direct integration or by applying Isserlis' theorem~\cite{Isserlis:1918} (valid for Gaussian distributions)
\begin{equation}
\label{eq: 4pt FP final}
\langle\varphi_1^2\varphi_2^2\rangle(N,r)=
\begin{cases}
      3\left(\frac{H_I^2}{8\pi^2}\frac{1-e^{-2\beta N}}{\beta}\right)^2 & \text{if $r<r_0~e^{-N}$} \\
      \left(\frac{H_I^2}{8\pi^2}\frac{1-e^{-2\beta N}}{\beta}\right)^2+2\left[\frac{H_I^2}{8\pi^2}e^{-2\beta N}\frac{\left(\frac{r}{r_0}\right)^{-2\beta}-1}{\beta}\right]^2 & \text{if $r_0~e^{-N}<r<r_0$} \\
      \left(\frac{H_I^2}{8\pi^2}\frac{1-e^{-2\beta N}}{\beta}\right)^2 & \text{if $r>r_0$}\,,
\end{cases} 
\end{equation}
or, dividing by $\sigma^4(N)=\left(\frac{H_I^2}{8\pi^2}\frac{1-e^{-2\beta N}}{\beta}\right)^2$,
\begin{equation}
\label{eq: 4pt FP rescaled}
\boxed{
\frac{\langle\varphi_1^2\varphi_2^2\rangle(N,r)}{\sigma^4(N)}=
\begin{cases}
      3 & \text{if $r<r_0~e^{-N}$} \\
      1+2~C^2(N,r) & \text{if $r_0~e^{-N}<r<r_0$} \\
      1 & \text{if $r>r_0$}\,.
\end{cases}
}
\end{equation}
As one can check this result obeys Wick's theorem. Indeed this is just a consequence of the Gaussianity of $\rho_2(\varphi_1,N;\varphi_2,N)$ in~(\ref{eq: joint PDF final}). This is in agreement with the QFT calculation in Section~\ref{sec: QFT} for super-Hubble scales. For this reason we think that it correctly captures the classical stochastic limit. On the contrary the result~(\ref{spatial_inflation}) produced by solving the coupled system of equations~(\ref{eomPhi2,Phi2})--(\ref{eomPi2,Pi2}) violates Wick's theorem and is in disagreement with the QFT treatment. We do not understand yet the origin of the disagreement, considering that in the slow-roll regime, where the contribution of correlators involving the canonical momentum $\hat{\pi}(t,\vec{x})$ is subdominant, the two approaches should produce the same results based on the same approximation used for the source $n_{\phi^2,\phi^2}$ and for the last term in the Fokker-Planck equation~(\ref{eq: FP joint 2 pt}), where $j_0(z)\rightarrow\theta(1-z)$.
A possibility is that the approach of Section~\ref{sect:model} does not fully catch the interaction between long and short modes for higher-order correlators, but only for 2-point correlators, so that the sources appearing on the right-hand sides of eqs.~(\ref{eomPhi2,Phi2})--(\ref{eomPi2,Pi2}) would need to be modified with respect to those used, given in eq.~(\ref{eq: noise def}).

Due to the Gaussian nature of the probability distribution~(\ref{eq: joint PDF final}), Wick's theorem (or direct integration via~(\ref{eq: 2pt func expect val})) can be used to compute higher order correlators, like $\langle\varphi_1^n\varphi_2^n\rangle$ with $n$ a natural number, in terms of the correlation function $C(N,r)$ given in~(\ref{eq: 2pt FP final rescaled}) and the one-point variance $\sigma^2(N)$ in~(\ref{eq: variance 1pt DPF}). For simplicity we will omit the space ($r$) and time ($N$) dependence of $C(N,r)$ and $\sigma(N)$ in the following expressions. By inspection of the combinatorics, it can be shown that
\begin{equation}
\label{eq: 2n pt function}
\frac{\langle\varphi_1^n\varphi_2^n\rangle}{\sigma^{2n}}=\sum_{k=0\atop n-k={\rm even}}^n k!\left[{n \choose k}(n-k-1)!!\right]^2 C^k=\sum_{k=0\atop n-k={\rm even}}^n \left[\frac{\Gamma(n+1)}{\Gamma\left(\frac{n-k+2}{2}\right)}\right]^2\frac{C^k}{2^{n-k}\Gamma(k+1)}\,,
\end{equation}
where the sum goes only on $k$ with the same parity as $n$, as specified above by the condition that $n-k$ must be an even number. The last equality uses the double factorial identity $(2m-1)!!=\frac{(2m)!}{2^m m!}=\frac{\Gamma(2m+1)}{2^m\Gamma(m+1)}$, applied to the natural number $m=(n-k)/2$.

\noindent Eq.~(\ref{eq: 2n pt function}) is a polynomial of degree $n$ in the variable $C$ with parity determined by $n$ (even function in $C$ if $n$ is even, or odd if $n$ is odd).
The result can also be expressed in an equivalent (but less transparent) way in terms of Gauss' hypergeometric function $~_2F_1$ as
\begin{equation}
\label{eq: 2n pt simple}
\frac{\langle\varphi_1^n\varphi_2^n\rangle}{\sigma^{2n}}=\Gamma(n+1)~C^n~_2F_1\left(\frac{1-n}{2},-\frac{n}{2};1;\frac{1}{C^2}\right)\,.
\end{equation}

The same calculation in the framework of Section~\ref{sect:model} would be much more difficult because it would require solving a coupled system of equations involving both the field and the canonical momentum. Furthermore, we have already seen for $n=2$ (corresponding to the 4-pt function $\Delta_{\phi^2,\phi^2}$) that the procedure in Section~\ref{sect:model} fails to resproduce the correct form of non-coincident correlators expected from Wick's theorem, but it only succeeds in giving the right values in the coincident (and very short distance) regime ($r<r_0~e^{-N}$) and in the large distance regime ($r>r_0$) corresponding respectively to $C=1$ (perfectly correlated) and $C=0$ (completely uncorrelated).

\noindent The result for $C=1$ (coincidence) is readily obtained from Wick's theorem to be $\frac{\langle\varphi^{2n}\rangle}{\sigma^{2n}}=(2n-1)!!$. In Section~\ref{sect: qbackr} we will see an application of this coincident value when evaluating the quantum backreaction of the field on the expansion rate at late times, and how that affects the energy balance between classical and quantum contributions to the total energy density of the Universe.

\noindent At large distances $r>r_0$ (where $C=0$), eq.~(\ref{eq: 2n pt function}) saturates to $0$ if $n$ is odd and $\left[(n-1)!!\right]^2$ if $n$ is even, because only the term $k=0$ contributes to the sum.

The most general 2-point correlator $\langle\varphi_1^{n_1}\varphi_2^{n_2}\rangle$, with $n_1$ and $n_2$ arbitrary natural numbers obeys
\begin{eqnarray}
\label{eq: n1 n2 pt function}
\frac{\langle\varphi_1^{n_1}\varphi_2^{n_2}\rangle}{\sigma^{n_1+n_2}}&=&\sum_{\substack{k=0
\\ n_1-k={\rm even}\\ n_2-k={\rm even}}}
^{{\rm min}(n_1,n_2)} k!{n_1 \choose k}{n_2 \choose k}(n_1-k-1)!!(n_2-k-1)!!~C^k\,\nonumber\\
&=&\sum_{\substack{k=0
\\ n_1-k={\rm even}\\ n_2-k={\rm even}}}
^{{\rm min}(n_1,n_2)}\frac{\Gamma(n_1+1)\Gamma(n_2+1)}{\Gamma\left(\frac{n_1-k+2}{2}\right)\Gamma\left(\frac{n_2-k+2}{2}\right)}\frac{C^k}{2^{\frac{n_1+n_2}{2}-k}\Gamma(k+1)}\,.
\end{eqnarray}
Note that the sums in eq. (\ref{eq: n1 n2 pt function}) are non-empty only if $n_1$ and $n_2$ have the same parity, because $k$ runs over numbers with the same parity as both $n_1$ and $n_2$.  This is in agreement with the fact that only when $n_1+n_2$ is even it is possible to fully decompose the correlator in a number of contracted Wick pairs. When $n_1$ and $n_2$ do not have the same parity the result (\ref{eq: n1 n2 pt function}) must be simply read as $0$.  Of course, due to the symmetry under exchange of the positions where the fields $\varphi_1$ and $\varphi_2$ are evaluated, one could take $n_1\leq n_2$, but that is not necessary for the validity of (\ref{eq: n1 n2 pt function}).

\noindent Just like for eq. (\ref{eq: 2n pt simple}), again an expression in terms of Gauss' hypergeometric function is possible:
\begin{eqnarray}
\frac{\langle\varphi_1^{n_1}\varphi_2^{n_2}\rangle}{\sigma^{n_1+n_2}}&=&\frac{1+(-1)^{n_1+n_2}}{2}~2^{-\frac{|n_1-n_2|}{2}}~\frac{\Gamma\left(\max(n_1,n_2)+1\right)}{\Gamma\left(\frac{|n_1-n_2|}{2}+1\right)}\,\nonumber\\
&\times&C^{\min(n_1,n_2)}~_2F_1\left(\frac{1-\min(n_1,n_2)}{2},-\frac{\min(n_1,n_2)}{2};\frac{|n_1-n_2|}{2}+1;\frac{1}{C^2}\right)
\,,
\label{eq: n1 n2 pt simple}
\end{eqnarray}
where the coefficient $\frac{1+(-1)^{n_1+n_2}}{2}$ is basically a Kronecker delta on the relative parity of $n_1$ and $n_2$, giving $1$ if $n_1$ and $n_2$ have the same parity, and $0$ otherwise.
Equivalence between the functions of $C$ in eqs. (\ref{eq: n1 n2 pt function}) and (\ref{eq: n1 n2 pt simple}) has been checked with a symbolic mathematical software up to large values of $n_1$ and $n_2$, confirming its correctness.
When $n_1=n_2\equiv n$, then eqs. (\ref{eq: n1 n2 pt function}--\ref{eq: n1 n2 pt simple}) reduce to eqs. (\ref{eq: 2n pt function}--\ref{eq: 2n pt simple}).

\newpage

\section{Quantum backreaction}\label{sect: qbackr}

In the most recent stages of evolution, it is not possible to describe the evolution of the scalar field $\hat{\Phi}$ as a free field in a given classical background metric. The latter is the approximation used throughout the previous sections and, as we already mentioned in the quantum field theory treatment (Section~\ref{sec: QFT}), it corresponds to the one-loop approximation for the energy momentum tensor and the density-density correlator. In recent cosmological epochs, one cannot neglect the quantum backreaction of the scalar field on the expansion of the Universe, because the energy-momentum tensor of the scalar field becomes more and more important and eventually takes the lead of the expansion. As a consequence, the quantum nature of the gravitational field has to be taken into account and the metric unavoidably exhibits quantum fluctuations. In particular, the Hubble expansion rate is a fluctuating quantity and it is properly represented as a quantum operator at each spacetime point $\hat{H}\left(t,\vec{x}\right)$, obeying the energy conservation constraint. Assuming zero spatial curvature, the energy constraint takes the form of a quantum Friedmann equation:
\begin{equation}
\label{eq: Hubble quantum}
3M_P^2 \hat{H}^2(t,\vec{x})=\rho_{C}(t)\unit+\hat{\rho}_{Q}(t,\vec{x})\,,
\end{equation}
where $\rho_{C}(t)$ is the contribution to energy density coming from classical matter (baryons and cold dark matter\footnote{More precisely, $\rho_{C}$ only accounts for a fraction of matter because, as we have seen, $\hat{\rho}_Q$ also contains a small part which scales like matter.} and $\hat{\rho}_{Q}(t,\vec{x})$ is the quantum energy density due to the scalar field (part of which yields dark energy).

\noindent The quantum scalar field in turn obeys an equation motion which can be read from eq.~(\ref{eq: EOM Phi 2nd order}), but the Hubble rate is now a quantum operator $\hat{H}$, as well as the Ricci scalar $\hat{R}$. Neglecting explicit spatial gradients, the equation of motion for $\hat{\Phi}$ reads
\begin{equation}
\label{eq: EOM Phi quantum backreaction}
\ddot{\hat{\Phi}}+3\hat{H}\dot{\hat{\Phi}}+ m^2\hat{\Phi}+\xi \hat{R}\hat{\Phi} = 0\,.
\end{equation}
Eqs.~(\ref{eq: Hubble quantum}) and (\ref{eq: EOM Phi quantum backreaction}) constitute the leading order approximation in spatial gradients to the dynamical field equations and the relevant gravitational constraint equation. This is a standard approximation scheme used for studying inflationary dynamics, known as the separate Universe approach~\cite{Salopek:1990jq, Salopek:1990re}.

Solving the full quantum equation of motion~(\ref{eq: EOM Phi quantum backreaction}) goes beyond the scope of this paper and we will limit ourselves to take into account the quantum character of the metric in eq.~(\ref{eq: Hubble quantum}) via $\hat{H}^2$, but we neglect it for the dynamics of the field $\hat{\Phi}$ in eq.~(\ref{eq: EOM Phi quantum backreaction}), which we simply approximate as
\begin{equation}
\label{eq: EOM Phi no backreaction}
\ddot{\hat{\Phi}}+3H\dot{\hat{\Phi}}+ m^2\hat{\Phi}+\xi R\hat{\Phi} = 0\,.
\end{equation}
In other words, we only include the quantum backreaction effect in the gravity (energy) constraint equation~(\ref{eq: Hubble quantum}), but not in the full dynamics of the field.
Note that eq.~(\ref{eq: EOM Phi no backreaction}), which is the leading order approximation of eq.~(\ref{eq: EOM Phi quantum backreaction}), is linear in $\hat{\Phi}$ and therefore it preserves Gaussianity of the initial state, while eq.~(\ref{eq: EOM Phi quantum backreaction}) necessarily generates non-Gaussianities.

\noindent Analogously to the expansion that we will develop in this Section, one could also write down an expansion for the operators $\hat{H}$ and $\hat{R}$ appearing in eq.~(\ref{eq: EOM Phi quantum backreaction}). By defining $\bar{H}\equiv\sqrt{\langle\hat{H}^2\rangle}$, then, at linear order in the (squared) Hubble rate fluctuation $\hat{\delta}_{H^2}\equiv\frac{\hat{H}^2-\bar{H}^2}{\bar{H}^2}$, one gets $\hat{H}\simeq\bar{H}\left(\unit+\frac12\hat{\delta}_{H^2}\right)$. Similarly the Ricci operator $\hat{R}=12\hat{H}^2+6\dot{\hat{H}}$ can also be expanded using the second Friedmann equation for $\dot{\hat{H}}$.

As we will see, the approximation made when limiting ourselves to eqs.~(\ref{eq: Hubble quantum}) and~(\ref{eq: EOM Phi no backreaction}) will still allow us to find some interesting results from the resummation of scalar loops, arising from the energy conservation equation~(\ref{eq: Hubble quantum}). 

\noindent As a first step, let us recall that, in the model with action~(\ref{eq:action}), the quantum energy density operator receives contributions quadratic in the field/canonical momentum. Neglecting the subdominant $\{\hat{\Phi},\hat{\Pi}\}$ and $\hat{\Pi}^2$, neglecting spatial gradients and considering a negative non-minimal coupling $\xi<0$, the quantum energy density, which has to be used in eq.~(\ref{eq: Hubble quantum}), can be approximated from eq.~(\ref{eq: rho_Q operator}) as
\begin{equation}
\label{eq: rhoQ}
\hat{\rho}_Q=\left(\frac{m^2}{2}\unit-3|\xi|\hat{H}^2\right)\hat{\Phi}^2\,,
\end{equation}
where $\unit$ is the identity operator acting on the Hilbert space of quantum states.
At leading order in $|\xi|$, the pressure in matter era\footnote{We are using eq.~(\ref{eq: p_Q}) without taking expectation values and with $\epsilon=3/2$ (matter-dominated epoch). Spatial gradients have been neglected.} is
\begin{equation}
\label{eq: pQ}
\hat{p}_Q=-\frac{m^2}{2}\hat{\Phi}^2\,.
\end{equation}
As we see by comparing energy density and pressure, there are two contributions to the energy density: one ($m^2/2~\hat{\Phi}^2$) behaves like a cosmological constant (CC) (i.e. equation of state parameter $-1$) and the other ($-3|\xi|\hat{H}^2\hat{\Phi}^2$) mimics a non-relativistic matter-like contribution (pressureless).

An important detail that should not be underestimated is that, in the right-hand side of (\ref{eq: rhoQ}), the (squared) Hubble parameter itself appears as an operator (as denoted by its hat). This is the proper way to take into account the fact that, due to the intrinsic quantum nature of the field $\hat{\Phi}$, the local expansion rate necessarily inherits a quantum behavior, making it a fluctuating quantity. Indeed, the squared Hubble parameter is properly represented by an operator $\hat{H}^2(t,\vec{x})$, which obeys eq.~(\ref{eq: Hubble quantum}).
But then, since $\hat{H}^2$ is a quantum operator in (\ref{eq: Hubble quantum}), it should also be treated in the same way when it appears in~(\ref{eq: rhoQ}).

The quantum nature of $\hat{H}^2$ in~(\ref{eq: rhoQ}) is the quantum backreaction effect that we want to discuss in this Section. A consequence of it, which is readily obtained by combining~(\ref{eq: rhoQ}) and~(\ref{eq: Hubble quantum}), is the local relation between expansion rate and field
\begin{equation}
\label{eq: quantum Friedmann}
\hat{H}^2=\frac{1}{3M_P^2}\left(\rho_{C}\unit+\frac{m^2}{2}\hat{\Phi}^2\right)\left(\unit+|\xi|\frac{\hat{\Phi}^2}{M_P^2}\right)^{-1}\,,
\end{equation}
where the last (inverse) operator is exactly due to the quantum backreaction, which therefore renormalizes the Planck mass $M_P$. For simplicity of notation we omitted space-time dependence.

We would like to match our quantum DE model to the Universe as it appears today $t=t_0$, which is (mostly) made up by non-relativistic matter (cold dark matter and baryons) and dark energy well described by a cosmological constant. Firstly, the expectation value of $\hat{H}^2(t_0,\vec{x}) $ should be a proxy for the global cosmological squared Hubble parameter today $H_0^2$,
\begin{equation}
\label{eq: finding rhoC}
\langle\hat{H}^2\rangle=H_0^2\,.
\end{equation}
This condition basically defines what we mean by the (squared) Hubble parameter today within the quantum model. For simplicity we do not write explicitly that quantities and expectation values are considered at the current time $t=t_0$. Furthermore spatial homogeneity implies that conditions do not depend on the comoving position $\vec{x}$.

As a second step, we require that the CC part of $\hat{\rho}_Q$ should account for the cosmological constant observed, therefore
\begin{equation}
\label{eq: match CC}
\frac{m^2}{2}\langle\hat{\Phi}^2\rangle=3M_P^2 H_0^2\Omega_\Lambda\,,
\end{equation}
where $M_P=(8\pi G)^{-1/2}$ is the reduced Planck mass and $\Omega_\Lambda$ is the fraction of energy density today due to a cosmological constant.
Recalling the definition of the positive dimensionless quantity $\alpha$ (see eq.~(\ref{eq: def alpha 1} and use $H_{\rm DE}=H_0\sqrt{\Omega_\Lambda}$)
\begin{equation}
\label{eq: def alpha}
\alpha\equiv\frac{1}{6|\xi|\Omega_\Lambda}\left(\frac{m}{H_0}\right)^2\,,
\end{equation}
then eq. (\ref{eq: match CC}) gives
\begin{equation}
\label{eq: variance today}
\langle\hat{\Phi}^2\rangle=\frac{M_P^2}{\alpha|\xi|}\equiv\sigma^2\,,
\end{equation}
which is also the variance of the field $\sigma^2\equiv\langle\hat{\Phi}^2\rangle-\langle\hat{\Phi}\rangle^2$ because we always assume $\langle\hat{\Phi}\rangle=0$.

The classical energy density today $\rho_C$ is then determined by eq. (\ref{eq: finding rhoC}) where $\hat{H}^2$ obeys the Friedmann equation~(\ref{eq: quantum Friedmann}).

We then use the formal geometric series expansion
\begin{equation}
\left(\unit+\frac{|\xi|}{M_P^2}\hat{\Phi}^2\right)^{-1}=\sum_{n=0}^{\infty}\left(-|\xi|\frac{\hat{\Phi}^2}{M_P^2}\right)^n
\end{equation}
and the results from Wick's theorem at coincidence in the previous Section, $\langle\hat{\Phi}^{2n}\rangle=\sigma^{2n}(2n-1)!!$ and similarly $\langle\hat{\Phi}^{2n+2}\rangle=\sigma^{2n+2}(2n+1)!!$. Introducing the classical energy density fraction today
\begin{equation}
\Omega_C\equiv\frac{\rho_C}{3M_P^2 H_0^2}\,,
\end{equation}
one gets after a few steps
\begin{equation}
\label{eq: rhoC Wick prelim}
\langle\hat{H}^2\rangle=H_0^2\sum_{n=0}^{\infty}\left(-\frac{1}{\alpha}\right)^n \left[\Omega_C(2n-1)!!+\Omega_\Lambda(2n+1)!!\right]\,,
\end{equation}
and reindexing the second part of the series, the result can be written as
\begin{equation}
\label{eq: rhoC Wick}
\langle \hat{H}^2\rangle=H_0^2\left[\alpha\Omega_\Lambda+\left(\Omega_C-\alpha\Omega_\Lambda\right)\sum_{n=0}^{\infty}\left(-\frac{1}{\alpha}\right)^n (2n-1)!!\right]\,.
\end{equation}
This contains a formal power series in $1/\alpha$, or equivalently an asymptotic expansion at infinity in the variable $\alpha$. The convergence radius in the variable $1/\alpha$ is actually zero, meaning that, written like that, the series diverges for every finite $\alpha$. Nevertheless, if we can find a resummation for it, we would then be able to fix $\Omega_C$ (and therefore $\rho_C$) in terms of the other parameters of the problem by imposing the condition~(\ref{eq: finding rhoC}). A clever way to do this resummation consists in taking an alternative route which does not rely on any series expansion. Of course, at the end the two methods should agree, in the sense that the asymptotic expansion at $\alpha\to\infty$ of the result that we are going to find, despite not convergent, should match exactly~(\ref{eq: rhoC Wick}).

Note that Wick's theorem is nothing but a way to say that $\hat{\Phi}$ follows a Gaussian distribution. More precisely we can consider a classical stochastic variable with the same statistical properties as the quantum field $\hat{\Phi}$. Indeed that is the picture that one should have in mind for what we called $\varphi$ throughout the previous Section~\ref{sec: FP stochastic}, where we used Starobinsky's stochastic formalism. We remind that we are assuming zero VEV for $\hat{\Phi}$ and therefore, like in eq.~(\ref{eq: 1pt prob sol}), we can say that the equivalent classical stochastic variable $\varphi$ follows a distribution $P(\varphi)$ which is Gaussian (see the discussion after eq.~(\ref{eq: EOM Phi no backreaction}) about preservation of Gaussianity in the approximation adopted here), given by
\begin{equation}
\label{eq: P phi}
P(\varphi)=\frac{1}{\sigma\sqrt{2\pi}}\exp\left(-\frac{\varphi^2}{2\sigma^2}\right)\,,
\end{equation}
where $\sigma^2$ is given in eq.~(\ref{eq: variance today}) and $\varphi$ can assume any real value.
This could be applied at any time as long as $\hat{\Phi}$ is a free field, so that $\varphi$ is Gaussian. We proved explicitly the validity of this requirement for inflation in the previous Section and it keeps being true also in radiation and matter era if one treats $\hat{\Phi}$ as a free field, in the spirit of perturbation theory.

We are interested in the current cosmological time $t_0$, as we want to determine the value $\rho_C$ (or $\Omega_C$) today in terms of the model parameters $\xi$ and $m$ (or equivalently $\xi$ and $\alpha$) and the cosmological parameters $H_0$ and $\Omega_\Lambda=1-\Omega_M$.

Expectation values can then be evaluated using $P(\varphi)$ and mapping quantum operators to their corresponding classical stochastic quantities, which means that $3M_P^2\langle H^2\rangle$ is given by
\begin{equation}
3M_P^2\langle \hat{H}^2\rangle=\int_{-\infty}^{+\infty}{\rm d}\varphi~P(\varphi)\frac{\rho_C+\frac{m^2}{2}\varphi^2}{1+|\xi|\frac{\varphi^2}{M_P^2}}\,.
\end{equation}
Using eq.~(\ref{eq: P phi}) and changing the integration variable to the dimensionless $z\equiv\varphi/\sigma$ the expectation value reduces to
\begin{equation}
\label{eq: int z}
\langle \hat{H}^2\rangle=H_0^2\int_{-\infty}^{+\infty}{\rm d}z~\frac{1}{\sqrt{2\pi}}\exp\left(-\frac{z^2}{2}\right)\frac{\Omega_C+\Omega_\Lambda z^2}{1+\frac{z^2}{\alpha}}\,,
\end{equation}
whose result can be written in terms of the complementary error function\footnote{The complementary error function erfc(z) can be defined as ${\rm erfc}(z)\equiv\frac{2}{\sqrt{\pi}}\int_z^{+\infty}{\rm d}t~{\rm e}^{-t^2}$.} as
\begin{equation}
\label{eq: rhoC resummation}
\langle \hat{H}^2\rangle=H_0^2\left[\alpha\Omega_\Lambda+\left(\Omega_C-\alpha\Omega_\Lambda\right){\rm e}^{\alpha/2}\sqrt{\frac{\pi\alpha}{2}}~{\rm erfc}\left(\sqrt{\frac{\alpha}{2}}\right)\right]\,.
\end{equation}
This is the resummation that we are looking for. In order to check agreement of eq.~(\ref{eq: rhoC resummation}) with the result from application of Wick's theorem in eq.~(\ref{eq: rhoC Wick}), the following asymptotic expansion for the complementary error function ${\rm erfc}(x)$ is needed:
\begin{equation}
{\rm erfc}(x)=\frac{{\rm e}^{-x^2}}{x\sqrt{\pi}}\sum_{n=0}^{\infty}\left(-\frac{1}{2x^2}\right)^n (2n-1)!!\,.
\end{equation}
Using $x=\sqrt{\alpha/2}$, one immediately finds agreement with eq.~(\ref{eq: rhoC Wick}), confirming the correctness of the derivation.
We are now ready to impose the condition (\ref{eq: finding rhoC}), which enables us to express $\Omega_C$ today as
\begin{equation}
\label{eq: OmegaC backreaction}
\Omega_C(t_0)=\alpha\Omega_\Lambda+\left(1-\alpha\Omega_\Lambda\right)\left[{\rm e}^{-\alpha/2}\sqrt{\frac{2}{\alpha\pi}}\frac{1}{{\rm erfc}\left(\sqrt{\frac{\alpha}{2}}\right)}\right]\,,
\end{equation}
where we explicitly wrote that it refers to a value today.

As a consequence, the quantum energy density fraction today $\Omega_Q(t_0)\equiv\frac{\langle\hat{\rho}_Q\rangle(t_0)}{3M_P^2 H_0^2}=1-\Omega_C(t_0)$ is
\begin{equation}
\label{eq: OmegaQ backreaction}
\Omega_Q^{(\rm br)}(t_0)=\left(\Omega_\Lambda-\frac{1}{\alpha}\right)\left[{\rm e}^{-\alpha/2}\sqrt{\frac{2\alpha}{\pi}}\frac{1}{{\rm erfc}\left(\sqrt{\frac{\alpha}{2}}\right)}-\alpha\right]\,,
\end{equation}
where the superscript ``(br)", standing for ``backreaction", is there to stress that this is the correct expression taking into account the backreaction due to the quantum nature of the field $\hat{\Phi}$ which makes the Hubble parameter itself a stochastic quantity.
This result should be compared with what one gets when neglecting the quantum backreaction effect. Without quantum backreaction one would get the simplified formula (see eq.~(\ref{eq:rhoquantum}))
\begin{equation}
\label{eq: OmegaQ no backreaction}
\Omega_Q^{(\rm no-br)}(t_0)=\Omega_\Lambda-\frac{1}{\alpha}\,.
\end{equation}
This is true because, when neglecting the quantum nature of $\hat{H}^2$ in~(\ref{eq: rhoQ}), taking expectation values today gives
\begin{equation}
\label{eq: rhoQ no backreaction}
\langle\hat{\rho}_Q^{(\rm no-br)}\rangle=\left(\frac{m^2}{2}\unit-3|\xi|H_0^2\right)\langle\hat{\Phi}^2\rangle\,,
\end{equation}
which, upon  substituting ~(\ref{eq: variance today}), leads to~(\ref{eq: OmegaQ no backreaction}).

\begin{figure}[H]
\centering
\includegraphics[width=0.9\columnwidth]{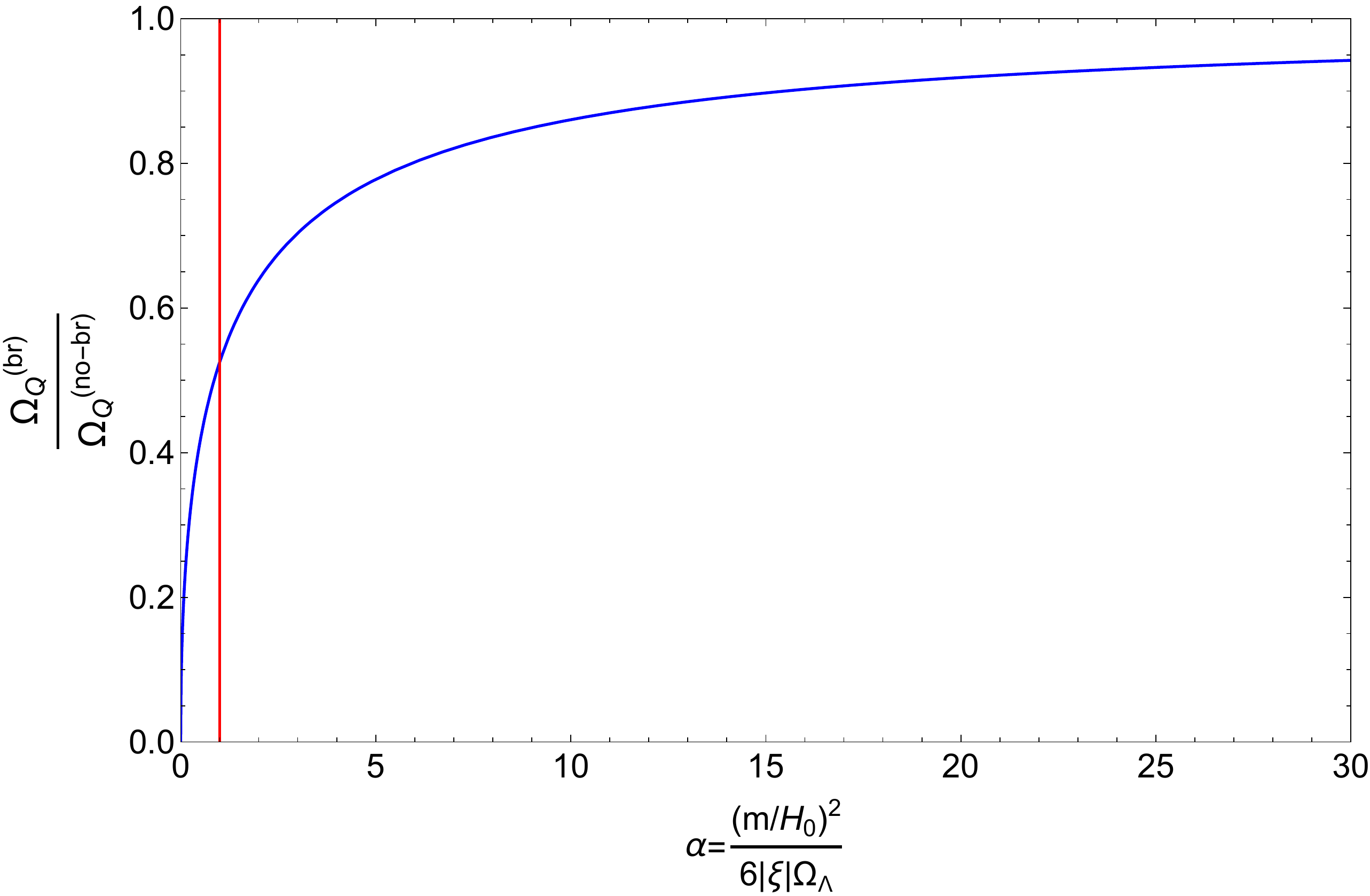}
\caption {The ratio between the average quantum energy densities evaluated with ($\Omega_Q^{(\rm br)}$) or without ($\Omega_Q^{(\rm no-br)}$) accounting for the backreaction due to the quantum nature of the field $\hat{\Phi}$ (which makes the Hubble parameter itself a stochastic quantity), as a function of the dimensionless combination of parameters $\alpha\equiv\frac{(m/H_0)^2}{6|\xi|\Omega_\Lambda}$. We denoted by $H_0$ and $\Omega_\Lambda$ the Hubble parameter and the cosmological constant energy density fraction today, while $m$ and $\xi<0$ are the mass and negative non-minimal coupling of the quantum spectator field $\hat{\Phi}$, as they appear in eq.~(\ref{eq:action}). Only values of $\alpha>1$, starting from the red vertical line, are physically relevant so that dark energy eventually completely dominates over matter in the expansion of the Universe. The plotted ratio, given by the right-hand side of eq.~(\ref{eq: rhoQ br ratio}), saturates to $1$ for large $\alpha$ values and it is roughly $0.53$ when $\alpha=1$, which is its minimum value. This means that neglecting the quantum backreaction leads to overestimating the fraction of energy density due to the quantum field $\hat{\Phi}$ by a factor of $1.9$ for $\alpha=1$ (and less for larger $\alpha$).
\label{fig: backreaction}
}
\end{figure}

Comparing eqs.~(\ref{eq: OmegaQ backreaction}) and (\ref{eq: OmegaQ no backreaction}) we see that the effect of quantum backreaction is encoded in a multiplicative function of the dimensionless parameter $\alpha$ that we remind to be defined as $\alpha\equiv\frac{m^2}{6|\xi|\Omega_\Lambda H_0^2}$. The ratio between $\Omega_Q^{(\rm br)}(t_0)$ and $\Omega_Q^{(\rm no-br)}(t_0)$ is
\begin{equation}
\label{eq: rhoQ br ratio}
\boxed{\frac{\Omega_Q^{(\rm br)}(t_0)}{\Omega_Q^{(\rm no-br)}(t_0)}={\rm e}^{-\alpha/2}\sqrt{\frac{2\alpha}{\pi}}\frac{1}{{\rm erfc}\left(\sqrt{\frac{\alpha}{2}}\right)}-\alpha}\,.
\end{equation}
In the limit of large $\alpha$ the ratio goes to unity\footnote{The first few terms of the expansion of~(\ref{eq: rhoQ br ratio}) in $\frac{1}{\alpha}$ are $1-\frac{2}{\alpha}+\frac{10}{\alpha^2}+{\mathcal O}\left(\frac{1}{\alpha^3}\right)$.}, which makes sense if one thinks about the limit $|\xi|\to0$ with $m/H_0$ finite. In this case the full quantum energy density behaves like dark energy (in the form of a cosmological constant), so both (\ref{eq: OmegaQ backreaction}) and (\ref{eq: OmegaQ no backreaction}) reduce to $\Omega_\Lambda$. For finite $\alpha$, the ratio between the quantum energy densities today with or without accounting for the backreaction can be seen in Fig.~\ref{fig: backreaction}, where we plot and discuss as a function of $\alpha$. There is a physical lower limit for $\alpha$, which is obtained by requiring that the dark energy contribution to the quantum energy density in eq.~(\ref{eq: rhoQ}) eventually dominates over its matter-like part. If this is true then at late times in the future $t\to+\infty$, one should get $\langle\hat{\rho}_Q(t)\rangle>0$, or equivalently,
\begin{equation}
\label{eq: condition large t}
\lim_{t\to+\infty}\Omega_Q^{(\rm br)}(t)>0\,.
\end{equation}
It is quite straightforward\footnote{Let us imagine to live in another epoch, say at time $t$, and repeat the matching procedure that we discussed starting from eq.~(\ref{eq: match CC}) at that time $t$, instead of $t_0$. What changes is just that we should refer to the energy density fraction made by the cosmological constant at the new time $\Omega_\Lambda(t)$ and to the Hubble parameter at that same time $H(t)$. Note that, for a cosmological constant, $\Omega_\Lambda(t)H^2(t)=\rho_\Lambda/(3M_P^2)$ is constant because the energy density $\rho_\Lambda$ for a cosmological constant is indeed constant. Thus there is no time dependence left in the dimensionless $\alpha(t)\equiv m^2/(6|\xi|H^2(t)\Omega_\Lambda(t))$, which is then the same as in eq.~(\ref{eq: def alpha}), $\alpha(t)=\alpha=(mM_P)^2/(2|\xi|\rho_\Lambda)={\rm constant}$. This explains why the only modification needed in eq.~(\ref{eq: OmegaQ backreaction}) to make it valid at any time $t$ is the replacement of $\Omega_\Lambda$ (which referred to $t_0$) with $\Omega_\Lambda(t)$. Incidentally, the constancy of $\alpha$ also implies that the result for the ratio in eq.~(\ref{eq: rhoQ br ratio}) is valid at any time $t$ and not only today (time $t_0$).} to extend eq.~(\ref{eq: OmegaQ backreaction}) to any time $t$ instead of today ($t_0$) just by replacing $\Omega_\Lambda$ measured today with the same quantity at time $t$, which is the fraction of energy density made up by the cosmological constant at time $t$. When $t\to+\infty$ that fraction approaches $1$, because the entire energy budget of the Universe will be completely made by the cosmological constant, therefore
\begin{equation}
\label{eq: OmegaQ backreaction large t}
\lim_{t\to+\infty}\Omega_Q^{(\rm br)}(t)=\left(1-\frac{1}{\alpha}\right)\left[{\rm e}^{-\alpha/2}\sqrt{\frac{2\alpha}{\pi}}\frac{1}{{\rm erfc}\left(\sqrt{\frac{\alpha}{2}}\right)}-\alpha\right]\,.
\end{equation}
Since the factor in square brackets in~(\ref{eq: OmegaQ backreaction large t}) is always positive (it is between $0$ and $1$, as plotted in Fig.~\ref{fig: backreaction}), then the condition (\ref{eq: condition large t}) imposes
\begin{equation}
\label{eq: condition alpha}
\alpha>1\,.
\end{equation}

\section{Theoretical aspects of the Hubble tension}\label{sect: Hubble tension}
This section contains the main theoretical steps needed to properly address the Hubble tension problem in the quantum dark energy model studied throughout the paper or other generalizations of it. In particular, we will refine some of the assumptions about probability distributions used in the previous work~\cite{Belgacem:2021ieb}, where it was shown that the same model considered here can relieve the Hubble tension. 
It is known that the $\Lambda$CDM standard cosmological model the value of the Hubble parameter inferred from early-Universe probes is in tension (at the level of more than $4\sigma$) with the expansion rate measured in the local Universe. We refer to them as global and local measurements of the Hubble parameter, respectively.
A brief overview on the Hubble tension problem and on different measures of the Hubble rate is contained in~\cite{Belgacem:2021ieb} and we refer the reader to that work (and references therein) for a more complete picture. Here we limit ourselves to recall only those concepts strictly necessary to put the Hubble tension in the context of quantum dark energy models and turn it into a probability statement. In that respect, the fundamental idea is that in a Universe where the expansion rate is a fluctuating quantity like in eq.~(\ref{eq: quantum Friedmann}), due to the quantum nature of the underlying fundamental field(s), it is natural that different probes of the expansion produce a different outcome depending on which length/time scales they are sensible to. The real matter is understanding how much their results can be apart and whether their difference is sufficient to explain the Hubble tension. Following~\cite{Belgacem:2021ieb} and due to its appearance in the Friedmann equation~(\ref{eq: quantum Friedmann}), we focus on the square of the Hubble parameter. Let us suppose that local measurements of the Hubble parameter (squared) can be seen as a spatial average over some volume $V_1$ at the current cosmological time $t_0$, and similarly we consider the global measurements as a spatial average are over a larger volume $V_2$ (including $V_1$) at the same time $t_0$\footnote{Real measurements are always done on the past light cone. However here we do not aim at describing a precise global or local measurement procedure, but we just want to encompass the general idea that some kind of averaging is necessarily part of the answer since a probe explores different regions of the sky. This physical ingredient is already contained in the simple procedure of spatial averaging adopted here. For a study of the effect of quantum dark energy fluctuations (and therefore a fluctuating Hubble rate) on specific cosmological observables, e. g. supernovae luminosity distances, see~\cite{inprep:2022lumdist}.}. As a proxy to quantify the Hubble tension, we consider the following conditional probability (see~\cite{Belgacem:2021ieb}), with ``P" standing for ``probability":

\begin{equation}
\label{eq: prob first}
{\rm P}\left(\left[H^2\right]_{V_1}>H_1^2\bigg|\left[H^2\right]_{V_2}<H_2^2\right)\,,
\end{equation}
where $\left[H^2\right]_{V_i},~(i=1,2)$ denotes the spatial average of the squared Hubble rate today over the local ($i=1$) or global ($i=2$) volumes, i.e. 
\begin{equation}
\left[H^2\right]_{V_i}\equiv(1/V_i)\int_{V_i} d^3x \, H^2(t_0,\vec{x})\,,\qquad(i=1,2)\,.
\end{equation}
\noindent The quantity $H^2(t_0,\vec{x})$ is meant as a classical stochastic variable with the same statistical properties as the quantum operator $\hat{H}^2(t_0,\vec{x})$ in eq.~(\ref{eq: quantum Friedmann}), just like we already introduced the classical stochastic variable $\varphi$ following the same statistics as the quantum field $\hat{\Phi}$.

As for $H_1$ and $H_2$, for definiteness we choose the local value $H_1$ as the one measured by the {\it SH0ES} collaboration \cite{Riess:2019cxk} from luminosity distances of supernovae Ia, and the global value $H_2$ as the one inferred by the {\it Planck} mission \cite{Planck:2018vyg} from CMB temperature fluctuations assuming the $\Lambda$CDM cosmology:

\begin{eqnarray}
H_1&=&74.03~{\rm km/s~Mpc^{-1}}\,,\nonumber \\
H_2&=&67.4~{\rm km/s~Mpc^{-1}}\,.
\label{eq: Hubble values}
\end{eqnarray}

\noindent Keeping in mind that the local value $H_1$ is larger than the global value $H_2$ (see eq.~(\ref{eq: Hubble values})), we can explain why the probability in eq.~(\ref{eq: prob first}) is a measure of the Hubble tension: it expresses how likely it is that local measurements give a result at least as large as the one actually measured, given that global measurements give a value at least as low as the one they really provide.
If the probability~(\ref{eq: prob first}) is large enough, then the Hubble tension is relieved by the quantum nature of dark energy. Indeed, showing that this actually happens was the main result of the letter~\cite{Belgacem:2021ieb}.
The volumes $V_1$ and $V_2$ are taken as spheres centered around the observer with radii $R_1$ and $R_2$. In~\cite{Belgacem:2021ieb}, we considered a global radius given by the Hubble horizon $R_2=H_2^{-1}\simeq4400$  Mpc and a local radius $R_1=100$ Mpc (we also studied another choice $R_1=1$ Gpc for the local radius to mimic other kinds of measurements from the {\it H0LiCOW} collaboration instead of {\it SH0ES}). The precise numerical values are not relevant for what we will discuss, which only deals with the theoretical approach for evaluating the probability~(\ref{eq: prob first}).

By making use of Bayes' theorem, we can rewrite the conditional probability in eq.~(\ref{eq: prob first}) in terms of a joint probability as
\begin{equation}
\label{eq: prob}
{\rm P}\left(\left[H^2\right]_{V_1}>H_1^2\bigg|\left[H^2\right]_{V_2}<H_2^2\right)=\frac{{\rm P}\left(\left[H^2\right]_{V_1}>H_1^2\bigcap\left[H^2\right]_{V_2}<H_2^2\right)}{{\rm P}\left(\left[H^2\right]_{V_2}<H_2^2\right)}\,.
\end{equation}
Evaluating the denominator requires knowing the distribution of the global variable $\left[H^2\right]_{V_2}$, while the numerator requires the joint probability distribution of the two variables $\left[H^2\right]_{V_1}$ (local) and $\left[H^2\right]_{V_2}$ (global). We shall study aspects of both of them, starting with the denominator, focusing on the methodology to approach the problem and not on numerical predictions. The latter were already obtained in ref.\cite{Belgacem:2021ieb} thanks to considerable simplifications stemming from the assumption that the squares of the fields $\varphi^2(t_0,\vec{x})$ can be approximately treated as Gaussian variables (with correlation between fields at different positions). Such a simplified approach takes into account the main physical ingredient for a possible relief of the Hubble tension in the context of a quantum dark energy model, namely the spatial correlations, and the main lesson from~\cite{Belgacem:2021ieb} is that spatially correlated dark energy can actually help with the Hubble tension issue. Nevertheless we know, e.g. from the results of Section~\ref{sec: FP stochastic}, that the field $\varphi(t_0,\vec{x})$ it is which follows a Gaussian distribution, not its square. The distribution of $\varphi^2$ at a given position is a chi-squared distribution, and the spatially averaged variables have the complication of spatial correlations, which create dependence between fields at different positions. The major purpose of the rest of this Section is to explore the possible path to a more rigorous statistical treatment of the problem.
As a general rule for notation, capital letter ``P" will be reserved for finite probabilities (i.e. integrated over a region), while we will use the small ``$p$" for probability distribution functions (PDF).

\subsection{Distribution of the global squared Hubble rate}
As a first step towards calculating the denominator ${\rm P}\left(\left[H^2\right]_{V_2}<H_2^2\right)$ of eq.~(\ref{eq: prob}), we discretize space in the global volume $V_2$ by a large set of $N_2$ points $\left\{\vec{x}_i\right\}_{i=1,2,\dots,N_2}$ evenly distributed in the volume $V_2$. This will make some manipulations more easy to understand, in the same spirit in which, for example, in the path integral approach to quantum mechanics it is useful to deal with an integral over a finite number of variables and only later take the continuum limit to get a functional integration. We will discuss some formal aspects of the continuum limit, however in practice a discretized approach is more suitable for numerical computations.
Since we always refer to quantities at the current cosmological time $t_0$ we will sometimes omit it from notation, for simplicity.

\noindent Due to discretization, the definition of the global squared Hubble parameter is now,
\begin{equation}
\left[H^2\right]_{2}\equiv\frac{1}{N_2}\sum_{i=1}^{N_2}H^2_i\,,
\end{equation}
where $H^2_i\equiv H^2\left(t_0,\vec{x}_i\right)$ is the Hubble rate at the point $\vec{x}_i$. $H_i^2$ can be considered as a local expansion parameter (squared), defined by the geodesic deviation equation or the Raychaudhuri equation.
Using eq.~(\ref{eq: quantum Friedmann}), one can relate the squared Hubble rate $H_i^2$ and the classical field variable $\varphi_i$ at $\vec{x}_i$. It is convenient to work with dimensionless variables to simplify as much as possible the notation and the following calculations. Thus we introduce the dimensionless field variable
\begin{equation}
\label{eq: z rescaled phi}
z_i\equiv\varphi_i/\sigma\,,
\end{equation}
where $\sigma^2=\langle\varphi_i^2\rangle$ is the variance of the field already introduced in eq.~(\ref{eq: variance today})(independent from the position because of statistical spatial homogeneity). Note that a similar variable $z$ was already introduced for the integration in eq.~(\ref{eq: int z}). The Friedmann equation~(\ref{eq: quantum Friedmann}) at $\vec{x}_i$ becomes
\begin{equation}
\label{eq: Hubble i first}
\frac{H_i^2}{H_0^2}=\frac{\Omega_C+\Omega_\Lambda z_i^2}{1+\frac{z_i^2}{\alpha}}\,,
\end{equation}
where $\alpha$ was defined in eq. (\ref{eq: def alpha}) and $H_0^2$ is the expectation value of the squared Hubble rate today (again space-independent). As we discussed in Section~\ref{sect: qbackr}, just before eq. (\ref{eq: OmegaC backreaction}), the classical energy density fraction $\Omega_C$ is determined by requiring that $\langle H_i^2\rangle=H_0^2$ (where the ensemble average $\langle\dots\rangle$ coincides with the quantum state average), resulting in the expression~(\ref{eq: OmegaC backreaction}).
As for the value of $H_0$ (which is again marginal for our mostly theoretical discussion), we can suppose that $H_2$ in~(\ref{eq: Hubble values}), being an average result over the large volume $V_2$, is a good approximation for $H_0$, so that
\begin{equation}
\label{eq: global H2 as H0}
H_0\simeq H_2\,.
\end{equation}

Introducing the dimensionless variable
\begin{equation}
h_i\equiv\frac{H_i^2}{H_0^2}\,,
\end{equation}
eq.~(\ref{eq: Hubble i first}) is equivalent to
\begin{equation}
\label{eq: Hubble i}
h_i=\frac{\Omega_C+\Omega_\Lambda z_i^2}{1+\frac{z_i^2}{\alpha}}\,.
\end{equation}
It follows from $\langle H_i^2\rangle=H_0^2$ that $\langle h_i\rangle=1$, which is guaranteed for $\Omega_C$ given by eq.~(\ref{eq: OmegaC backreaction}). After introducing the average variable
\begin{equation}
\label{eq: average h}
\left[h\right]_{2}\equiv\frac{1}{N_2}\sum_{i=1}^{N_2}h_i\,,
\end{equation}
where the subscript ``2" is meant to remind that it refers to an average over $N_2$ points (i.e. the global average), the probability ${\rm P}\left(\left[H^2\right]_{V_2}<H_2^2\right)$ is, in the discretized picture and with $H_0\simeq H_2$, equal to ${\rm P}\left(\left[h\right]_{2}<1\right)$.

Evaluating such a probability is possible if one manages to find how $\left[h\right]_{2}$ is distributed.
Note that $\langle h_i\rangle=1$ implies
\begin{equation}
\label{eq: h mean 1}
\langle\left[h\right]_{2}\rangle=1\,.
\end{equation}
However higher momenta (determining variance, skewness, kurtosis, etc...) are not so trivial to determine due to the correlations between different $h_i$, whose origin is in the non-zero off-coincident quantum correlations.
\noindent We know that the variables $\left\{z_i\right\}_{i=1,2,\dots,N_2}$ are simply obtained by rescaling the fields $\left\{\varphi_i\right\}_{i=1,2,\dots,N_2}$ by the constant $\sigma$. Thus their distribution is readily determined by the the distribution of fields. In Section \ref{sec: FP stochastic} we confirmed, by explicitly solving the Fokker-Planck equation (\ref{eq: FP joint 2 pt}), that the fields at a given couple of points separated by comoving distance $r$ (and at equal times) follow a Gaussian distribution with correlation coefficient $C(r)$ imprinted by inflation as in eq. (\ref{eq: 2pt FP final simple}). This is a consequence of the fields being free.
Due to the rescaling (\ref{eq: z rescaled phi}), it is clear from eq. (\ref{eq: C definition}) that the correlator $\langle z_i z_j\rangle$ is given by
\begin{equation}
\label{eq: corr z}
C_{ij}\equiv\langle z_i z_j\rangle=C(\|\vec{x}_i-\vec{x}_j\|)\,.
\end{equation}
In particular, when $i=j$ it correctly gives a variance $C_{ii}\equiv\langle z_i^2\rangle=C(0)=1$ in agreement with $z_i=\varphi_i/\sigma$ and $\langle\varphi_i^2\rangle=\sigma^2$.
Since we are now dealing with fields at $N_2$ points it is straightforward to write down a joint multivariate Gaussian distribution such that each pair of points $\{z_i, z_j\}$ has a correlator $C_{ij}$ given by eq.~(\ref{eq: corr z}). This is reached by

\begin{equation}
\label{eq: multi Gaussian}
p\left(z_1, z_2, \dots, z_{N_2}\right)=\frac{1}{\sqrt{(2\pi)^{N_2}{\rm det}(C)}}\exp{\left(-\frac12z^{T}\cdot C^{-1}\cdot z\right)}\,,
\end{equation}
where $z^T$ collects all the variables $z_i$ in a row vector as $z^T\equiv\left(z_1, z_2, \dots, z_{N_2}\right)$ and $z$ is the corresponding column vector. We also introduced the symmetric and positive definite $N_2\times N_2$ matrix $C$ with elements $C_{ij}$, its inverse $C^{-1}$ and used its determinant ${\rm det}(C)$ in the normalization coefficient to ensure that

\begin{equation}
\label{eq: norm multi Gaussian}
\int_{-\infty}^{+\infty}{\rm d}z_1\int_{-\infty}^{+\infty}{\rm d}z_2~\dots\int_{-\infty}^{+\infty}{\rm d}z_{N_2}~p\left(z_1, z_2, \dots, z_{N_2}\right)=1\,.
\end{equation}

We can now formulate in simple terms the temporary goal problem as follows, making reference to eqs.~(\ref{eq: Hubble i}), (\ref{eq: average h}) and (\ref{eq: multi Gaussian}): we are interested in the probability distribution function of the average variable $\left[h\right]_{2}$ defined in eq.~(\ref{eq: average h}), where each $h_i$ is given in terms of $z_i$ by eq.~(\ref{eq: Hubble i}), knowing that the variables $\left\{z_i\right\}_{i=1,2,\dots,N_2}$ follow the Gaussian distribution in eq.~(\ref{eq: multi Gaussian}).

A useful tool to find the distribution of $\left[h\right]_{2}$ is its moment generating function (MGF), defined as
\begin{equation}
\label{eq: def MGF}
M_{\left[h\right]_{2}}(s)\equiv\langle{\rm e}^{s\left[h\right]_{2}}\rangle\,,
\end{equation}
where $s$ is an auxiliary variable. As a reminder of the main features of the MGF, all the moments of the distribution for $\left[h\right]_{2}$ are encoded in $M_{\left[h\right]_{2}}(s)$ because
\begin{equation}
\label{eq: moments MGF}
M_{\left[h\right]_{2}}(s)=\sum_{n=0}^{\infty}\langle\left[h\right]_{2}^n\rangle\frac{s^n}{n!}\,,
\end{equation}
which means that
\begin{equation}
\label{eq: moments}
\langle\left[h\right]_{2}^n\rangle=\frac{{\rm d}^n}{{\rm d}s^n}M_{\left[h\right]_{2}}(s)\bigg|_{s=0}\,.
\end{equation}
A good strategy to determine the PDF of $\left[h\right]_{2}$ is to first compute its MGF as
\begin{eqnarray}
M_{\left[h\right]_{2}}(s)&=&\int_{-\infty}^{+\infty}{\rm d}z_1\int_{-\infty}^{+\infty}{\rm d}z_2~\dots\int_{-\infty}^{+\infty}{\rm d}z_{N_2}~p\left(z_1, z_2, \dots, z_{N_2}\right){\rm e}^{s\left[h\right]_{2}\left(z_1, z_2, \dots, z_{N_2}\right)}\,\nonumber\\
&=&\int\frac{{\rm d}z_1{\rm d}z_2\dots{\rm d}z_{N_2}}{\sqrt{(2\pi)^{N_2}{\rm det}(C)}}~\exp{\left[-\frac12z^{T}\cdot C^{-1}\cdot z+s\frac{1}{N_2}\sum_{i=1}^{N_2}\frac{\Omega_C+\Omega_\Lambda z_i^2}{1+\frac{z_i^2}{\alpha}}\right]}\,,
\label{eq: MGF 1}
\end{eqnarray}
where in the second line a shorthand notation for the integration over all variables $z_i$ was used.
However, the integral above is too complicated to be computed exactly due to the denominators of the form $1+\frac{z_i^2}{\alpha}$ inside the exponential, which are in turn originated by the quantum backreaction discussed in the previous Section. As we have seen explicitly in the computation of $\Omega_Q=1-\Omega_C$ leading to Fig.~\ref{fig: backreaction}, the backreaction effect decreases with larger $\alpha$ and gives a correction by a factor of roughly $0.5$ when $\alpha$ is close to its minimum allowed value of $1$. Since we have seen a case where the quantum backreaction effect does not change the results by orders of magnitude, but only by a factor between 0.5 and 1, it is worth to use also here some approximations to reduce the complication of integration, so that one can extract some analytical understanding.
Therefore let us approximate\footnote{Note that $z_i$ follows a Gaussian distribution centered at zero and with unitary variance while $\alpha$ is at least 1. Therefore it is not so likely that $\frac{z_i^2}{\alpha}$ is large, which justifies using the truncated expansion in eq. (\ref{eq: hi approx}). Clearly the approximation works at its best for large $\alpha$ values.},
\begin{equation}
\label{eq: hi approx}
h_i=\frac{\Omega_C+\Omega_\Lambda z_i^2}{1+\frac{z_i^2}{\alpha}}=\Omega_C+\left(\Omega_\Lambda-\frac{\Omega_C}{\alpha}\right)z_i^2+{\mathcal O}\left(z_i^4\right)\,,
\end{equation}
which gives
\begin{equation}
\label{eq: average h approx}
\left[h\right]_{2}\simeq\Omega_C+\left(\Omega_\Lambda-\frac{\Omega_C}{\alpha}\right)\frac{1}{N_2}\sum_{i=1}^{N_2}z_i^2\,.
\end{equation}
We remark that this simplification does not obey exactly the result $\langle\left[h\right]_{2}\rangle$ in eq.(\ref{eq: h mean 1}), indeed because of its approximate nature. From eq. (\ref{eq: average h approx}) and $\langle z_i^2\rangle=1$ one gets $\langle\left[h\right]_{2}\rangle\simeq\Omega_\Lambda+\Omega_C\left(1-\frac{1}{\alpha}\right)$, which only goes to $1$ in the limit of large $\alpha$, since then $\Omega_C$ approaches $\Omega_M=1-\Omega_\Lambda$.

\noindent Using the approximation (\ref{eq: average h approx}), the MGF in eq.~(\ref{eq: MGF 1}) simplifies to
\begin{eqnarray}
M_{\left[h\right]_{2}}(s)&\simeq&\int\frac{{\rm d}z_1{\rm d}z_2\dots{\rm d}z_{N_2}}{\sqrt{(2\pi)^{N_2}{\rm det}(C)}}~\exp{\left\{-\frac12z^{T}\cdot C^{-1}\cdot z+s\left[\Omega_C+\left(\Omega_\Lambda-\frac{\Omega_C}{\alpha}\right)\frac{1}{N_2}\sum_{i=1}^{N_2}z_i^2\right]\right\}}\,\nonumber\\
&=&\frac{{\rm e}^{s\Omega_C}}{\sqrt{{\rm det}\left[{\rm Id}_2-2s\left(\Omega_\Lambda-\frac{\Omega_C}{\alpha}\right)\frac{1}{N_2}C\right]}}\,,
\label{eq: MGF 2}
\end{eqnarray}
where the integral is solved in the last line and ${\rm Id}_2$ is the $N_2\times N_2$ identity matrix.

This approximate result, which basically depends on the characteristic polynomial of the covariance matrix $C$, can be used to extract some insight on the continuum limit. An expansion of eq. (\ref{eq: MGF 2}) is in order for this purpose\footnote{It comes from the matrix identities ${\rm det}(A)=\exp{\left[{\rm Tr}\log(A)\right]}$ and $-\log\left({\rm Id}_2-X\right)=\sum_{n=1}^\infty \frac{1}{n}X^n$ applied to $A\equiv {\rm Id}_2-X$ and $X\equiv2s\left(\Omega_\Lambda-\frac{\Omega_C}{\alpha}\right)\frac{1}{N_2}C$.}:
\begin{equation}
\label{eq: MGF series}
M_{\left[h\right]_{2}}(s)\simeq\exp{\left\{\Omega_C s+\frac12\sum_{n=1}^\infty \frac{1}{n}\left[2\left(\Omega_\Lambda-\frac{\Omega_C}{\alpha}\right)s\right]^n\frac{1}{N_2^n}{\rm Tr}\left(C^n\right)\right\}}\,,
\end{equation}
where ${\rm Tr}\left(C^n\right)$ is the trace of the n-th power of the matrix $C$.
The continuum limit is obtained from the following observation on the coefficients $f_n\equiv\frac{1}{N_2^n}{\rm Tr}\left(C^n\right)$:
\begin{eqnarray}
f_n\equiv\frac{1}{N_2^n}{\rm Tr}\left(C^n\right)=\frac{1}{N_2^n}\sum_{i_1=1}^{N_2}\sum_{i_2=1}^{N_2}\dots\sum_{i_n=1}^{N_2}C_{i_1 i_2}C_{i_2 i_3}\dots C_{i_n i_1}~~~~~~~~~~~~~~&&\,\nonumber\\
\rightarrow\frac{1}{V_2^n}\int_{V_2}{\rm d}^3\vec{x}_1\int_{V_2}{\rm d}^3\vec{x}_2~\dots\int_{V_2}{\rm d}^3\vec{x}_n~C(\|\vec{x}_1-\vec{x}_2\|)C(\|\vec{x}_2-\vec{x}_3\|)\dots C(\|\vec{x}_n-\vec{x}_1\|)&&\,.
\label{eq: traces continuum}
\end{eqnarray}
Note that $f_n\leq1$ (with equality holding for $n=1$) because the function $C$ lies between $0$ and $1$, see eq.~(\ref{eq: 2pt FP final simple}) and Fig.~\ref{fig: 2pt noncoinc inflation}.
Despite the elegant obtention of the continuum limit via the procedure outlined in eq. (\ref{eq: traces continuum}), for numerical calculations it is still preferable to use the discretized version (\ref{eq: MGF 2}) and check convergence for large $N_2$, instead of trying to resum the series (\ref{eq: MGF series}). Furthermore, the analytical computation of the integrals in the last line of eq. (\ref{eq: traces continuum}) is trivial when $n=1$ and still doable for $n=2$, but it becomes already too complicated for $n=3$, so that Monte Carlo integration has to be used.

For $n=1$ one gets $f_1=1$, while for $n=2$, the coefficient $f_2=\frac{1}{V_2^2}\int_{V_2}{\rm d}^3\vec{x}_1\int_{V_2}{\rm d}^3\vec{x}_2~C^2(\|\vec{x}_1-\vec{x}_2\|)$, where $V_2$ is the spherical global volume $V_2=\frac43\pi R_2^3$, can be reduced to a single integration on the relative distance $r\equiv\|\vec{x}_2-\vec{x}_1\|$. More precisely, the computation can be done\footnote{As a summary of the main steps involved, one goes from the coordinates $\vec{x}_1$ and $\vec{x}_2$, measured with respect to the origin located at the center of the sphere, to the relative and center-of-mass coordinates, given by $\vec{r}\equiv\vec{x}_2-\vec{x}_1$ and $\vec{R}_{CM}\equiv\left(\vec{x}_1+\vec{x}_2\right)/2$, respectively. Then the integration over the center-of-mass coordinate $\vec{R}_{CM}$ is done for a given relative distance $\vec{r}$. This must take into account the geometrical limits $\|\vec{R}_{CM}+\frac12\vec{r}\|\leq R_2$ and $\|\vec{R}_{CM}-\frac12\vec{r}\|\leq R_2$, which can be seen as the intersection of two spheres, so the problem is essentially reduced to Euclidean geometry. The relative distance $r=\|\vec{r}\|$ between two points inside a sphere of radius $R_2$ can take values between $0$ and $2R_2$ and its distribution function gives eq. (\ref{eq: rel dist prob}). This is properly normalized as a one-dimensional PDF with $\int_0^{2R_2}{\rm d}r~p(r)=1$ and we also checked its expression numerically by randomly generating pairs of points inside the sphere.} in terms of the probability distribution $p(r)$ for the relative distance between two points randomly chosen with uniform probability inside the sphere of radius $R_2$:
\begin{equation}
\label{eq: global int}
f_2=\int_0^{2R_2}{\rm d}r~p(r)C^2(r)\,,
\end{equation}
where
\begin{equation}
\label{eq: rel dist prob}
p(r)=\frac{12}{R_2}\left[\left(\frac{r}{2R_2}\right)^2-\frac32\left(\frac{r}{2R_2}\right)^3+\frac12\left(\frac{r}{2R_2}\right)^5\right]\,,\qquad {\rm defined~for}~r\in [0,2R_2]\,.
\end{equation}

The moments of the distribution for $\left[h\right]_{2}$ can be extracted from the series~(\ref{eq: MGF series}) and computed in terms of $f_n$'s. It is actually more convenient to work with cumulants so that the PDF of $\left[h\right]_{2}$ can be approximated by series expansions like the Gram-Charlier type A series~\cite{Gram:1883}--\cite{Charlier:1914}, which however is not always guaranteed to converge. The cumulant generating function (CGF) $K_{\left[h\right]_{2}}(s)$ is defined as the logarithm of the MGF and the cumulants $\kappa_n$ are defined from the coefficients of the expansion in $s$, as\footnote{In the language of quantum field theory (and up to factors of $i$), the MGF corresponds to the partition function $Z[J]$ (generating functional of Green's functions), while the CGF corresponds to generating functional of connected Green's functions $W[J]$ which obeys $Z[J]={\rm e}^{iW[J]}$, with the external current $J(x)$ playing the role of the source variable $s$. In this sense the cumulants are the connected parts of correlators.}
\begin{equation}
\log{\left[M_{\left[h\right]_{2}}(s)\right]}\equiv K_{\left[h\right]_{2}}(s)\equiv\sum_{n=1}^{\infty}\kappa_n\frac{s^n}{n!}\,.
\end{equation}

Therefore, the cumulants can be read immediately from the argument of the exponential in~(\ref{eq: MGF series}). The first one $\kappa_1$ is the mean value\footnote{See also the discussion on $\langle\left[h\right]_{2}\rangle$ right after eq.~(\ref{eq: average h approx}).}, $\kappa_1=\langle\left[h\right]_{2}\rangle\simeq\Omega_\Lambda+\Omega_C\left(1-\frac{1}{\alpha}\right)$. The second one $\kappa_2$ is the variance, $\kappa_2=\langle\left[h\right]_{2}^2\rangle-\langle\left[h\right]_{2}\rangle^2=2\left(\Omega_\Lambda-\frac{\Omega_C}{\alpha}\right)^2 f_2$. For $n\geq2$ the cumulants are
\begin{equation}
\label{eq: kappa f}
\kappa_n=\frac{\Gamma(n)}{2}\left[2\left(\Omega_\Lambda-\frac{\Omega_C}{\alpha}\right)\right]^n f_n\,, \qquad n\geq2\,.
\end{equation}
The Gram-Charlier series gives the distribution $p\left(\left[h\right]_{2}\right)$ as an expansion in cumulants around a reference distribution, usually taken to be Gaussian. Truncating for simplicity to the $4^{\rm th}$ cumulant, the PDF of $\left[h\right]_{2}$ is
\begin{equation}
\label{eq: Gram-Charlier}
p\left(\left[h\right]_{2}\right)\simeq\frac{1}{\sqrt{2\pi\kappa_2}}\exp{\left[-\frac12\frac{\left(\left[h\right]_{2}-\kappa_1\right)^2}{\kappa_2}\right]}\left[1+\frac{\kappa_3}{6\kappa_2^{3/2}}H_3\left(\frac{\left[h\right]_{2}-\kappa_1}{\sqrt{\kappa_2}}\right)+\frac{\kappa_4}{24\kappa_2^2}H_4\left(\frac{\left[h\right]_{2}-\kappa_1}{\sqrt{\kappa_2}}\right)+\dots\right]\,,
\end{equation}
where $H_3(x)=x^3-3x$ and $H_4(x)=x^4-6x^2+3$ are Hermite polynomials and the dots denote terms dependent on higher order cumulants with perfectly known structure~\cite{Gram:1883}--\cite{Charlier:1914}, but that we omit for brevity.
The physical information about dark energy spatial correlations is encoded in the cumulants $\kappa_n$ of eq.~(\ref{eq: kappa f}) via the integrals $f_n$ in eq.~(\ref{eq: traces continuum}).

Alternatively, a technique to recover the PDF, $p\left(\left[h\right]_{2}\right)$, which does not rely on the cumulant expansion, is via numerical evaluation of the inverse Laplace transform of the MGF $M_{\left[h\right]_{2}}(s)$. The starting point is the relation implied\footnote{The rigorous limits of integration, i. e. the possible values of $\left[h\right]_{2}$, can be deduced from eqs.~(\ref{eq: Hubble i}),~(\ref{eq: average h}) and are given by $\left[h\right]_{2}\in [{\rm min}\left(\Omega_C, \alpha\Omega_\Lambda\right), {\rm max}\left(\Omega_C, \alpha\Omega_\Lambda\right)]$ depending on the sign of $\Omega_\Lambda-\frac{\Omega_C}{\alpha}$. When $\alpha$ is not too close to $1$, so that the approximations (\ref{eq: hi approx}) and (\ref{eq: average h approx}) work well (better for large $\alpha$), $\Omega_\Lambda-\frac{\Omega_C}{\alpha}>0$ and the MGF in eq.~(\ref{eq: MGF 2}) can be fully trusted, the range of values for $\left[h\right]_{2}$ can be taken as $\left[h\right]_{2}\in [\Omega_C, +\infty)$.} by eq.~(\ref{eq: def MGF}),

\begin{equation}
\label{eq: MGF vs PDF 1}
M_{\left[h\right]_{2}}(s)\simeq\int_{\Omega_C}^{+\infty}{\rm d}\left[h\right]_{2}~{\rm e}^{s\left[h\right]_{2}}~p\left(\left[h\right]_{2}\right)\,.
\end{equation}
After a shift of variable and recalling the definition of Laplace transform
\begin{equation}
{\mathcal L}\left\{f(x)\right\}(s)\equiv\int_0^{+\infty}{\rm d}x~{\rm e}^{-sx}f(x)\,, 
\end{equation}
eq.~(\ref{eq: MGF vs PDF 1}) can be recast into
\begin{equation}
\label{eq: MGF vs PDF 2}
M_{\left[h\right]_{2}}(s)\simeq{\rm e}^{s\Omega_C}{\mathcal L}\left\{p(x+\Omega_C)\right\}(-s)\,,
\end{equation}
where $x$ is a dummy variable, and then eq.~(\ref{eq: MGF vs PDF 2}) is inverted as
\begin{equation}
\label{eq: inverse Laplace}
p\left(\left[h\right]_{2}\right)\simeq{\mathcal L}^{-1}\left\{{\rm e}^{s\Omega_C}M_{\left[h\right]_{2}}(-s)\right\}\left(\left[h\right]_{2}-\Omega_C\right)\,,
\end{equation}
where ${\mathcal L}^{-1}$ is the inverse Laplace transform, which can be evaluated numerically or expressed as a line integral (called Bromwich or Fourier-Mellin integral), see e.g. Section 15.12 of~\cite{Arfken}.
Once the PDF $p\left(\left[h\right]_{2}\right)$ is known, either from eq.~(\ref{eq: Gram-Charlier}) or via eq.~(\ref{eq: inverse Laplace}), the goal probability ${\rm P}\left(\left[H^2\right]_{V_2}<H_2^2\right)={\rm P}\left(\left[h\right]_{2}<1\right)$ can be found by integration.

\subsection{Joint distribution of the local and global squared Hubble rates}

Let us now discuss briefly the numerator ${\rm P}\left(\left[H^2\right]_{V_1}>H_1^2\bigcap\left[H^2\right]_{V_2}<H_2^2\right)$ of eq. (\ref{eq: prob}). As for the denominator, we discretize space and consider the same $N_2$ points in the global volume $V_2$ as before, with $N_1<N_2$ of them belonging to the local volume $V_1$ contained in $V_2$. Without loss of generality, we can suppose that points $\vec{x}_i$ with $i=1,2,\dots,N_1$ belong to $V_1$ and points $\vec{x}_i$ with $i=N_1+1,N_1+2,\dots,N_2$ lie in $V_2$ but outside of $V_1$.
Following eq. (\ref{eq: average h}) we consider the variables
\begin{equation}
\label{eq: average h joint}
\left[h\right]_{1}\equiv\frac{1}{N_1}\sum_{i=1}^{N_1}h_i\,,\qquad\qquad \left[h\right]_{2}\equiv\frac{1}{N_2}\sum_{i=1}^{N_2}h_i\,,
\end{equation}
and, using the identification in eq. (\ref{eq: global H2 as H0}), the probability ${\rm P}\left(\left[H^2\right]_{V_1}>H_1^2\bigcap\left[H^2\right]_{V_2}<H_2^2\right)$ is equivalent in the discretized picture to ${\rm P}\left(\left[h\right]_{1}>\frac{H_1^2}{H_2^2}\bigcap\left[h\right]_{2}<1\right)$, where $H_1$ and $H_2$ are the local and global measures of the Hubble parameter in (\ref{eq: Hubble values}).
Computing such a probability requires the joint PDF, $p\left(\left[h\right]_{1},\left[h\right]_{2}\right)$, of the local and global variables $\left[h\right]_{1}$ and $\left[h\right]_{2}$.

Following eq. (\ref{eq: def MGF}) we define the joint MGF for $\left[h\right]_{1}$ and $\left[h\right]_{2}$ as
\begin{equation}
\label{eq: def MGF joint}
M_{\left[h\right]_{1},\left[h\right]_{2}}(s_1,s_2)\equiv\langle{\rm e}^{s_1\left[h\right]_{1}+s_2\left[h\right]_{2}}\rangle\,,
\end{equation}
where we introduced two auxiliary variables $s_1$ and $s_2$.

Eq. (\ref{eq: moments MGF}) generalizes to
\begin{equation}
\label{eq: joint moments MGF}
M_{\left[h\right]_{1},\left[h\right]_{2}}(s_1,s_2)=\sum_{n_1=0}^{\infty}\sum_{n_2=0}^{\infty}\langle\left[h\right]_{1}^{n_1}\left[h\right]_{2}^{n_2}\rangle\frac{s_1^{n_1}}{n_1 !}\frac{s_2^{n_2}}{n_2 !}\,,
\end{equation}
and eq. (\ref{eq: moments}) generalizes to
\begin{equation}
\label{eq: moments joint}
\langle\left[h\right]_{1}^{n_1}\left[h\right]_{2}^{n_2}\rangle=\frac{{\rm d}^{n_1+n_2}}{{\rm d}s_1^{n_1}{\rm d}s_2^{n_2}}M_{\left[h\right]_{1},\left[h\right]_{2}}(s_1,s_2)\Big|_{s_1=0\atop s_2=0}\,.
\end{equation}
In the same approximations leading to eq. (\ref{eq: MGF 2}), it can be shown that the jont MGF is
\begin{equation}
M_{\left[h\right]_{1},\left[h\right]_{2}}(s_1,s_2)\simeq\frac{{\rm e}^{(s_1+s_2)\Omega_C}}{\sqrt{{\rm det}\left[{\rm Id}_2-2\left(\Omega_\Lambda-\frac{\Omega_C}{\alpha}\right)\left(\frac{s_2}{N_2}C+\frac{s_1}{N_1}{\rm Id}_1 C\right)\right]}}\,,
\label{eq: MGF joint 2}
\end{equation}
where ${\rm Id}_1$ is the $N_2\times N_2$ matrix given by the identity on the first $N_1$ rows and columns (the local subspace) and with zero entries otherwise. Hence all the three matrices ${\rm Id}_1$, ${\rm Id}_2$ and $C$ are $N_2\times N_2$ in size and their combination in eq. (\ref{eq: MGF joint 2}) is a legitimate operation.

A similar expansion to eq. (\ref{eq: MGF series}) applied to the result (\ref{eq: MGF joint 2}) and the introduction of the cumulant generating function $K_{\left[h\right]_{1},\left[h\right]_{2}}(s_1,s_2)\equiv\log{\left[M_{\left[h\right]_{1},\left[h\right]_{2}}(s_1,s_2)\right]}$, gives
\begin{equation}
\label{eq: CGF series joint}
K_{\left[h\right]_{1},\left[h\right]_{2}}(s_1,s_2)\simeq\Omega_C (s_1+s_2)+\frac12\sum_{n=1}^\infty \frac{1}{n}\left[2\left(\Omega_\Lambda-\frac{\Omega_C}{\alpha}\right)\right]^n{\rm Tr}\left[\left(\frac{s_2}{N_2}C+\frac{s_1}{N_1}{\rm Id}_1 C\right)^n\right]\,.
\end{equation}
The joint cumulants $\kappa_{n_1, n_2}$ are retrieved from the expansion
\begin{equation}
\label{eq: joint cumulants general}
K_{\left[h\right]_{1},\left[h\right]_{2}}(s_1,s_2)\equiv\sum_{n_1=0}^{\infty}\sum_{n_2=0}^{\infty}\kappa_{n_1, n_2}\frac{{s_1}^{n_1}}{n_1 !}\frac{{s_2}^{n_2}}{n_2 !}\,,
\end{equation}
with $\kappa_{0,0}=0$. Since the matrices $C$ and ${\rm Id}_1 C$ appearing in eq. (\ref{eq: CGF series joint}) do not commute, the general expression for $\kappa_{n_1, n_2}$ is not straightforward to obtain. A simple interesting case with a non-trivial result, which will also enable us to take its continuum limit, is $n_1=1, n_2=1$. Using the cyclic property of trace and the symmetry of $C$, one gets

\begin{eqnarray}
\label{eq: kappa11}
\kappa_{1, 1}&=&2\left(\Omega_\Lambda-\frac{\Omega_C}{\alpha}\right)^2\frac{1}{N_1 N_2}{\rm Tr}\left({\rm Id}_1 C^2\right)=2\left(\Omega_\Lambda-\frac{\Omega_C}{\alpha}\right)^2\frac{1}{N_1 N_2}\sum_{i=1}^{N_1}\sum_{j=1}^{N_2}C_{ij}^2~~~~~~~~~~~~~~\,\nonumber\\
&\rightarrow&2\left(\Omega_\Lambda-\frac{\Omega_C}{\alpha}\right)^2\frac{1}{V_1 V_2}\int_{V_1}{\rm d}^3\vec{x}_1\int_{V_2}{\rm d}^3\vec{x}_2~C^2(\|\vec{x}_1-\vec{x}_2\|)\,,
\end{eqnarray}
where in the last line the continuum limit involves integration of correlators between pairs of points, with the first point in the local volume $V_1=\frac{4}{3}\pi R_1^3$ and the second one in the global volume $V_2=\frac{4}{3}\pi R_2^3$.
Similarly to eq. (\ref{eq: global int}), also the integration in the last line of eq. (\ref{eq: kappa11}) can be recast as a single integral over the relative distance $r\equiv\|\vec{x}_1-\vec{x}_2\|$,
\begin{equation}
\label{eq: joint int}
\frac{1}{V_1 V_2}\int_{V_1}{\rm d}^3\vec{x}_1\int_{V_2}{\rm d}^3\vec{x}_2~C^2(\|\vec{x}_1-\vec{x}_2\|)=\int_{0}^{R_2+R_1}{\rm d}r~p_{12}(r)C^2(r)\,,
\end{equation}
where now $p_{12}(r)$ is the probability distribution for the relative distance between a point randomly chosen with uniform probability from the sphere of radius $R_1$ and a point randomly chosen with uniform probability from the sphere of radius $R_2>R_1$ with the same center. Its exact expression is\footnote{The strategy is the same as for $p(r)$ in eq. (\ref{eq: rel dist prob}), but with a slightly more complicated geometric setup. Note that when $R_1=R_2$, then $p_{12}(r)$ reduces to the aforementioned $p(r)$. Again we checked the correctness of the formula (\ref{eq: dist rel 12}) numerically (for different values of the radii ratio $R_1/R_2<1$) by randomly generating pairs of points inside the spheres. It is also correctly normalized as a one-dimensional distribution, namely $\int_0^{R_2+R_1}{\rm d}r~p_{12}(r)=1$.}
\begin{equation}
\label{eq: dist rel 12}
p_{12}(r)=  
\begin{cases}
      \frac{3r^2}{R_2^3}\,, & \text{if $0\leq r\leq R_2-R_1$}\,, \\
      \frac{3\left[r^5-6\left(R_1^2+R_2^2\right)r^3+8\left(R_1^3+R_2^3\right)r^2-3\left(R_2^2-R_1^2\right)^2 r\right]}{16 R_1^3 R_2^3}\,, & \text{if $R_2-R_1\leq r\leq R_2+R_1$}\,.
\end{cases}
\end{equation}
Other cumulants and their continuum limit can obtained case by case from eqs. (\ref{eq: CGF series joint}--\ref{eq: joint cumulants general}) and then one might attempt to approximate the joint PDF, $p\left(\left[h\right]_{1},\left[h\right]_{2}\right)$, by existing generalizations to two variables of the Gram-Charlier series (\ref{eq: Gram-Charlier}), see e.g.~\cite{Sauer:1979}. Another possibility is to compute numerically the joint MGF in eq. (\ref{eq: MGF joint 2}) and, also numerically, take its inverse Laplace transform in two variables, again a generalization of the approach used for eq. (\ref{eq: inverse Laplace}). We did not dig into the practical feasibility and challenges of such a program, but we expect the numerical problem to be computationally expensive, unless a fast routine is used.
Of course, if one succeeds in evaluating $p\left(\left[h\right]_{1},\left[h\right]_{2}\right)$, then the goal probability ${\rm P}\left(\left[H^2\right]_{V_1}>H_1^2\bigcap\left[H^2\right]_{V_2}<H_2^2\right)={\rm P}\left(\left[h\right]_{1}>\frac{H_1^2}{H_2^2}\bigcap\left[h\right]_{2}<1\right)$ is given by integration of $p\left(\left[h\right]_{1},\left[h\right]_{2}\right)$ over the proper two-dimensional region.

\section{Conclusions and outlook}\label{sect: concl}
In this paper we study the features of dark energy predicted by a simple model, namely a light scalar field non-minimally coupled to the metric, as in eq.~(\ref{eq:action}). Building on previous work in~\cite{Belgacem:2021ieb, Glavan:2013mra, Glavan:2014uga, Glavan:2017jye, Glavan:2015cut}, we see how this model is able to produce dark energy from the amplification of quantum fluctuations in inflation (and, less relevantly, in matter-dominated epoch). Due to its quantum nature, dark energy emerging at later times from the backreaction of the scalar field on the Universe expansion, is predicted to exhibit spatial correlations of a precise form.

A first technique used to get these results starts from the observation in~\cite{Glavan:2015cut, Glavan:2013mra, Glavan:2014uga} that the field backreaction is mostly made up of the energy-momentum tensor of infrared modes. For this reason, in Section~\ref{sect:model}, we apply Starobinsky's stochastic formalism to study how correlators of long-wavelength field/canonical momentum evolve from inflation until current times. The discussion in~Section~\ref{sect:model} completes the treatment in~\cite{Belgacem:2021ieb} and~\cite{Glavan:2017jye} and it presents the non-coincident 2-point and 4-point correlators at equal times. Even starting from zero initial conditions, the accelerated expansion in inflation amplifies quantum fluctuations and creates the spatial profile of correlators. We see how stochastic theory predicts that, up to the overall amplitude, the spatial shape of field correlators generated in inflation is transmitted, roughly unaltered, throughout the subsequent epochs of cosmic history. The scale entering the spatial profile of correlators is of the order of the comoving Hubble horizon at the beginning of inflation, see eqs.~(\ref{eq: profile 2-pt s_2(r)}),~(\ref{spatial_inflation}). The quantum fields at two points separated by a larger comoving distance lose any information about each other and fluctuate independently.

Interestingly, 2-point and 4-point correlators have profiles well described by simple power laws, but unexpectedly their relation fails to reproduce Wick's theorem. Apart from the few most recent e-foldings when the quantum field backreacts on the metric generating unavoidably non-Gaussianities, during all previous stages of evolution the backreaction is negligible and the field is Gaussian because it is well approximated by a free field in a given classical background metric. Therefore Wick's theorem should be respected in these conditions.
We checked numerically for possible contributions lost in approximations when solving the system of equations for the 4-point correlators, finding no evidence for that. For this reason, we think that the only explanation of the problem lies in the noise sources of 4-point functions, which would actually include some other contribution from short and long wavelength modes interaction, which is not taken into account by eq.~(\ref{eq: noise def}). Further research is needed to understand what exactly goes wrong.

The issue with Wick's theorem in the stochastic approach of Section~\ref{sect:model} is one of the main reasons for looking into another approach by which the dynamics of infrared modes could be studied. This is done in Section~\ref{sec: FP stochastic}, following the ideas introduced by Starobinsky and Yokoyama in~\cite{Starobinsky:1994bd}. The strategy consists in studying the evolution of the probability distribution of a classical field configuration subjected to stochastic noise reproducing the coupling to short-wavelength modes. This allows us to show, in an elegant way, that the joint distribution for the fields at two points is Gaussian and correlators computed from it obey Wick's theorem, differently from the previous results of Section~\ref{sect:model}, thus rescuing stochastic formalism. We also provide a concise form for the most general higher-order correlator between fields at two points.

Independently of its implementation, the only way to really assess whether (and in which conditions) stochastic formalism provides a good approximation for the evolution of the field and for dark energy generation, is to compare its predictions to the results from quantum field theory. This is the subject of Section~\ref{sec: QFT}, which presents a study for the non-coincident 2-point field correlator and compares its full result to the stochastic theory approximation. Even in its relative simplicity, this is to our knowledge the first test of stochastic formalism in the non-coincident regime beyond inflation, namely in radiation and matter epochs.
It confirms that stochastic theory gives the right amplitude of correlators and it also reproduces the right behavior for super-Hubble separations between the two points, but at intermediate scales the situation is more intriguing. Stochastic formalism, focusing on long modes, is not designed to catch the shape of correlations at sub-Hubble scales and we showed that in matter-dominated epoch, a full quantum field theory treatment predicts that correlations persist at deeper distances than the stochastic approximation suggests. The discrepancy is mostly attributable to the energy content of spatial gradients, which are ignored in stochastic formalism. We support quantitatively this argument by studying the effect of a reduced speed of sound, which suppresses the contribution of spatial gradients. The consequence is a better agreement between QFT and stochastic formalism when the speed of sound is smaller.

The exploration of the observational consequences of the model is crucial to test it using cosmological data and compare its performance to $\Lambda$CDM. At the background level, it was found in~\cite{Glavan:2017jye} that Euclid and LSST could be able to test the redshift-dependence of the dark energy equation of state predicted. Encouraging results in model comparison with $\Lambda$CDM also come from~\cite{Demianski:2019vmq}.

It is remarkable that the simple dark energy model considered in this paper shows a rich phenomenology of dark energy, which is not just a time-dependent and spatially homogeneous dark energy, but rather a spatially-correlated quantum field, whose fluctuations can be used to compute interesting effects on physical observables and ultimately test the model. An example in this direction is~\cite{inprep:2022lumdist}, where the consequences of spatial correlations on the luminosity distance of supernovae is quantified and compared to the signal expected from a perturbed $\Lambda$CDM Universe (null hypothesis). Furthermore, a reduced speed of sound would increase the signal of dark energy fluctuations imprinted on luminosity distance correlations, to the point that the LSST survey could detect it.

In~\cite{Belgacem:2021ieb} it was investigated the possibility that a dark energy of quantum origin could relieve the Hubble tension. The result is affirmative because, within the approximations used there, the tension goes from more than $4\sigma$ in $\Lambda$CDM down to $1\sigma$ in the quantum dark energy model. Independently of the details related to the model specifics, this shows that proposals of this kind have a great potential for the Hubble tension issue. In Section~\ref{sect: Hubble tension} of the present paper we review and further expand the theoretical aspects of the Hubble tension within our dark energy model, highlighting the role of spatial correlations in providing different answers for the average Hubble rate, depending on the scale probed. The path from spatial correlations of dark energy to the Hubble tension drawn in Section~\ref{sect: Hubble tension} can be used to extract a few general techniques. It could serve as a basis for numerical refinements and also be applied to other fluctuating dark energy models. 

Finally, the techniques developed here can be used for a more accurate modeling of not just dark energy, but also of dark matter, given that the ultimate origin of these mysterious components is quantum~\cite{Friedrich:2017glg, Friedrich:2018qjv, Friedrich:2019zic, Vinke:2020}.

\vspace{5mm}

\newpage

\noindent
{\bf Acknowledgments.} 
This work is part of the Delta ITP consortium, a program of the Netherlands Organisation for Scientific Research (NWO) that is funded by the Dutch Ministry of Education, Culture and Science (OCW) - NWO projectnumber 24.001.027.

\appendix
\section{Sources of 4-point function stochastic equations}
\label{app: 4-pt stoch sources}
The stochastic sources needed in eqs.~(\ref{eomPhi2,Phi2})--(\ref{eomPi2,Pi2}) are
\begin{eqnarray}
n_{\phi^2,\phi^2}(t,r)  &=&  \frac{1}{H(t)} \Bigl\langle 
	\left\{ \hat{f}_{\phi}(t,\vec{x}_1) ,\hat{\phi}(t,\vec{x}_1) \right\} \hat{\phi}^2(t,\vec{x}_2)+
          \hat{\phi}^2(t,\vec{x}_1) \left\{ \hat{f}_{\phi}(t,\vec{x}_2) ,\hat{\phi}(t,\vec{x}_2) \right\} 
		\Bigr\rangle \, ,\nonumber \\
n_{\phi^2,\phi\pi}(t,r)  &=&  \frac{1}{a^3(t)H^2(t)} \Bigl\langle
\hat{\phi}^2 (t,\vec{x}_1)\left\{ \hat{f}_{\phi}(t,\vec{x}_2) ,\hat{\pi}(t,\vec{x}_2) \right\}+\left\{ \hat{f}_{\phi}(t,\vec{x}_1) ,\hat{\pi}(t,\vec{x}_1) \right\}\hat{\phi}^2 (t,\vec{x}_2)   \Bigr. \nonumber \\
&&\Bigl. +\hat{\phi}^2 (t,\vec{x}_1)\left\{ \hat{\phi}(t,\vec{x}_2) , \hat{f}_{\pi}(t,\vec{x}_2) \right\}+\left\{ \hat{\phi}(t,\vec{x}_1) , \hat{f}_{\pi}(t,\vec{x}_1) \right\} \hat{\phi}^2 (t,\vec{x}_2)  \Bigr. \nonumber \\
	&&\Bigl.+\left\{ \hat{f}_{\phi}(t,\vec{x}_1) ,\hat{\phi}(t,\vec{x}_1) \right\} \left\{\hat{\phi} (t,\vec{x}_2),\hat{\pi} (t,\vec{x}_2)\right\}+\left\{\hat{\phi} (t,\vec{x}_1),\hat{\pi} (t,\vec{x}_1)\right\}\left\{ \hat{f}_{\phi}(t,\vec{x}_2) ,\hat{\phi}(t,\vec{x}_2) \right\} \Bigr\rangle \, ,\nonumber \\
n_{\phi\pi,\phi\pi}(t,r)  &=&  \frac{1}{a^6(t)H^3(t)} \Bigl\langle
\left\{\hat{\phi}(t,\vec{x}_1),\hat{\pi}(t,\vec{x}_1)\right\}\left\{ \hat{f}_{\phi}(t,\vec{x}_2) ,\hat{\pi}(t,\vec{x}_2) \right\}\Bigr. \nonumber \\
&& \Bigl.+\left\{ \hat{f}_{\phi}(t,\vec{x}_1) ,\hat{\pi}(t,\vec{x}_1) \right\}\left\{\hat{\phi}(t,\vec{x}_2),\hat{\pi}(t,\vec{x}_2)\right\}   \Bigr. \nonumber \\
&&\Bigl. +\left\{\hat{\phi}(t,\vec{x}_1),\hat{\pi}(t,\vec{x}_1)\right\}\left\{ \hat{\phi}(t,\vec{x}_2) , \hat{f}_{\pi}(t,\vec{x}_2) \right\}+\left\{ \hat{\phi}(t,\vec{x}_1) , \hat{f}_{\pi}(t,\vec{x}_1) \right\} \left\{\hat{\phi}(t,\vec{x}_2),\hat{\pi}(t,\vec{x}_2)\right\} \Bigr\rangle \, ,\nonumber \\
n_{\phi^2,\pi^2}(t,r)  &=&  \frac{1}{a^6(t)H^3(t)} \Bigl\langle 
	\left\{ \hat{f}_{\phi}(t,\vec{x}_1) ,\hat{\phi}(t,\vec{x}_1) \right\} \hat{\pi}^2(t,\vec{x}_2)+
          \hat{\pi}^2(t,\vec{x}_1) \left\{ \hat{f}_{\phi}(t,\vec{x}_2) ,\hat{\phi}(t,\vec{x}_2) \right\} \Bigr. \nonumber \\
    &&\Bigl.+\left\{ \hat{f}_{\pi}(t,\vec{x}_1) ,\hat{\pi}(t,\vec{x}_1) \right\} \hat{\phi}^2(t,\vec{x}_2)+
          \hat{\phi}^2(t,\vec{x}_1) \left\{ \hat{f}_{\pi}(t,\vec{x}_2) ,\hat{\pi}(t,\vec{x}_2) \right\}
		\Bigr\rangle \, ,\nonumber \\
n_{\phi\pi,\pi^2}(t,r)  &=&  \frac{1}{a^9(t)H^4(t)} \Bigl\langle
\hat{\pi}^2 (t,\vec{x}_1)\left\{ \hat{f}_{\pi}(t,\vec{x}_2) ,\hat{\phi}(t,\vec{x}_2) \right\}+\left\{ \hat{f}_{\pi}(t,\vec{x}_1) ,\hat{\phi}(t,\vec{x}_1) \right\}\hat{\pi}^2 (t,\vec{x}_2)   \Bigr. \nonumber \\
&&\Bigl. +\hat{\pi}^2 (t,\vec{x}_1)\left\{ \hat{\pi}(t,\vec{x}_2) , \hat{f}_{\phi}(t,\vec{x}_2) \right\}+\left\{ \hat{\pi}(t,\vec{x}_1) , \hat{f}_{\phi}(t,\vec{x}_1) \right\} \hat{\pi}^2 (t,\vec{x}_2)  \Bigr. \nonumber \\
	&&\Bigl.+\left\{ \hat{f}_{\pi}(t,\vec{x}_1) ,\hat{\pi}(t,\vec{x}_1) \right\} \left\{\hat{\phi} (t,\vec{x}_2),\hat{\pi} (t,\vec{x}_2)\right\}+\left\{\hat{\phi} (t,\vec{x}_1),\hat{\pi} (t,\vec{x}_1)\right\}\left\{ \hat{f}_{\pi}(t,\vec{x}_2) ,\hat{\pi}(t,\vec{x}_2) \right\}
\Bigr\rangle \, ,\nonumber \\
n_{\pi^2,\pi^2}(t,r)  &=&  \frac{1}{a^{12}(t)H^5(t)} \Bigl\langle 
	\left\{ \hat{f}_{\pi}(t,\vec{x}_1) ,\hat{\pi}(t,\vec{x}_1) \right\} \hat{\pi}^2(t,\vec{x}_2)+
          \hat{\pi}^2(t,\vec{x}_1) \left\{ \hat{f}_{\pi}(t,\vec{x}_2) ,\hat{\pi}(t,\vec{x}_2) \right\} \Bigr\rangle \,.
\label{eq: noise def}
\end{eqnarray}
The stochastic sources can be computed in terms of the mode function $\varphi(t,k)$. The results depend on integrals over Fourier modes, which we regulate at low momenta with an IR cutoff $k_0$ (while $\mu a H$ plays the role of an UV cutoff). Their expression is
\begin{eqnarray}
n_{\phi^2,\phi^2} &=& \frac{1}{2\pi^4} 
(\mu aH)^3(1-\epsilon)\left[|\varphi(t,k)|^2\right]_{k=\mu aH} \int_{k_0}^{\mu aH}dk~k^2 |\varphi(t,k)|^2\left[1+2j_0(\mu aHr)j_0(kr)\right] \,, \nonumber \\
n_{\phi^2,\phi\pi} &=& \frac{1}{2\pi^4} 
     \mu^3 a^3 H^2(1-\epsilon)\left\{\left[\frac{\partial}{\partial t}|\varphi(t,k)|^2\right]_{k=\mu aH} \int_{k_0}^{\mu aH}dk~k^2 |\varphi(t,k)|^2\left[1+2j_0(\mu aHr)j_0(kr)\right]\right. \nonumber \\
   & &  \left.+\left[|\varphi(t,k)|^2\right]_{k=\mu aH} \int_{k_0}^{\mu aH}dk~k^2 \frac{\partial}{\partial t}|\varphi(t,k)|^2\left[1+2j_0(\mu aHr)j_0(kr)\right]\right\} \,,\nonumber \\
 n_{\phi\pi,\phi\pi} &=& \frac{1}{\pi^4} 
     \mu^3 a^3 H(1-\epsilon)\left\{\frac{1}{2}\left[\frac{\partial}{\partial t}|\varphi(t,k)|^2\right]_{k=\mu aH} \int_{k_0}^{\mu aH}dk~k^2 \frac{\partial}{\partial t}|\varphi(t,k)|^2\left[1+j_0(\mu aHr)j_0(kr)\right]\right. \nonumber \\
 & & \left.+j_0(\mu aHr)\left[|\varphi(t,k)|^2\right]_{k=\mu aH} \int_{k_0}^{\mu aH}dk~k^2 |\dot{\varphi}(t,k)|^2\right.\nonumber \\
& &\left.+j_0(\mu aHr)\left[|\dot{\varphi}(t,k)|^2\right]_{k=\mu aH} \int_{k_0}^{\mu aH}dk~k^2 |\varphi(t,k)|^2+\frac{1}{2}\left(\frac{\mu H}{a}\right)^3 j_0(\mu aHr)\frac{j_1(\mu aHr)}{\mu aHr}\right\} \,,\nonumber \\
  n_{\phi^2,\pi^2} &=& \frac{1}{2\pi^4} 
     \mu^3 a^3 H(1-\epsilon)\left\{\left[|\varphi(t,k)|^2\right]_{k=\mu aH} \int_{k_0}^{\mu aH}dk~k^2 |\dot{\varphi}(t,k)|^2\right.\nonumber \\
& &\left.+\left[|\dot{\varphi}(t,k)|^2\right]_{k=\mu aH} \int_{k_0}^{\mu aH}dk~k^2 |\varphi(t,k)|^2\right. \nonumber \\
& &\left.+j_0(\mu aHr)\left[\frac{\partial}{\partial t}|\varphi(t,k)|^2\right]_{k=\mu aH} \int_{k_0}^{\mu aH}dk~k^2 \frac{\partial}{\partial t}|\varphi(t,k)|^2-\left(\frac{\mu H}{a}\right)^3 j_0(\mu aHr)\frac{j_1(\mu aHr)}{\mu aHr}\right\} \,,\nonumber \\
n_{\phi\pi,\pi^2} &=& \frac{1}{2\pi^4} 
     \mu^3 a^3(1-\epsilon)\left\{\left[\frac{\partial}{\partial t}|\varphi(t,k)|^2\right]_{k=\mu aH} \int_{k_0}^{\mu aH}dk~k^2 |\dot{\varphi}(t,k)|^2\left[1+2j_0(\mu aHr)j_0(kr)\right]\right. \nonumber \\
    & & \left.+\left[|\dot{\varphi}(t,k)|^2\right]_{k=\mu aH} \int_{k_0}^{\mu aH}dk~k^2 \frac{\partial}{\partial t}|\varphi(t,k)|^2\left[1+2j_0(\mu aHr)j_0(kr)\right]\right\} \,,\nonumber \\
n_{\pi^2,\pi^2} &=& \frac{1}{2\pi^4} 
     \frac{\mu^3 a^3}{H}(1-\epsilon)\left[|\dot{\varphi}(t,k)|^2\right]_{k=\mu aH} \int_{k_0}^{\mu aH}dk~k^2 |\dot{\varphi}(t,k)|^2\left[1+2j_0(\mu aHr)j_0(kr)\right]\,.
\label{eq: noise}
\end{eqnarray}
where $j_0(z)\equiv\frac{\sin z}{z}$ and $j_1(z)\equiv-\frac{d}{dz}j_0(z)$ are the spherical Bessel functions of order $0$ and $1$ respectively, which are evaluated at $z=kr$.

In de Sitter inflation with constant Hubble parameter $H_I$, assuming $\xi<0$ and $(m/H_I)^2\ll|\xi|\ll 1$, their leading behavior is given by eq.~(\ref{eq: noise_deSitter 4pt}).

\section{Correlators in matter + cosmological constant epoch}
\label{app: mat+CC}

This Appendix contains the evolution of the scalar field correlators after matter-radiation equality, including both matter and a cosmological constant (CC) as the fluids leading the expansion. This could serve as a possible refinement of the evolution in a pure matter-dominated Universe given in subsections~\ref{subsubsec: 2pt mat} and~\ref{subsubsec: 4pt mat}, because the energy-momentum tensor of the scalar field contains a part behaving like a cosmological constant that contributes to the expansion.
For simplicity, we focus on 2-pt functions, but it is straightforward to extend the treatment to 4-pt functions. Part of the work was already done in Appendix A.3 of~\cite{Belgacem:2021ieb}.
Our goal is to solve the system of equations~(\ref{eomPhiPhi}--\ref{eomPiPi}) in a matter+CC Universe to predict the 2-pt correlators today.

We set the number of e-foldings $N$ to $N=0$ at matter-radiation equality and denote by $N_0=\ln{\left(\frac{\Omega_M}{\Omega_R}\right)} \simeq8.1$ the current time (today). As usual, $\Omega_R\simeq9.1\times10^{-5}$, $\Omega_M\simeq0.3$ and $\Omega_\Lambda\simeq1-\Omega_M\simeq0.7$ are the radiation, matter and cosmological constant fractions of energy density today.
The parameter $\epsilon=-\dot{H}/H^2$ as a function of $N$ is
\begin{equation}
\label{eq: eps N}
\epsilon(N)=\frac32 \frac{1}{1+\frac{\Omega_\Lambda}{\Omega_M}e^{3(N-N_0)}}\,.
\end{equation}
Following Appendix A.3 of~\cite{Belgacem:2021ieb}, let us change variable from $N$ to
\begin{equation}
\label{eq: n N}
n(N)\equiv\frac{1}{1+\frac{\Omega_\Lambda}{\Omega_M}e^{3(N-N_0)}}\,.
\end{equation}
Then $\epsilon(n)=(3/2) n$ and the ratio between the effective mass and the Hubble rate $M(n)/H(n)$ is
\begin{equation}
\left(\frac{M(n)}{H(n)}\right)^2=\left(\frac{m}{H_{\rm DE}}\right)^2(1-n)-12|\xi|\left(1-\frac34 n\right)\,,
\end{equation}
where $H_{\rm DE}=H_0\sqrt{\Omega_\Lambda}$.
Matter-radiation equality $N=0$ corresponds to $n_{\rm eq}\equiv\left(1+\frac{\Omega_\Lambda}{\Omega_M}e^{-3N_0}\right)^{-1}$ while the current time $N_0$ gives $n_0\equiv\Omega_M$.
The derivatives with respect to $N$ are easily transformed to derivatives with respect to $n$ by using $dn/dN=3n(n-1)$.
Neglecting stochastic noise sources, as legitimate after inflation, eqs.~(\ref{eomPhiPhi}--\ref{eomPiPi}) are equivalent, in vector/matrix notation, to
\begin{equation}
\frac{{\mathrm d}}{{\mathrm d} n}\Delta_{(\rm 2)}(n,r)+A_{(\rm 2)}(n)~\Delta_{(\rm 2)}(n,r)=0\,,
\label{system_matter+CC}
\end{equation}
 where the matrix $A_{(\rm 2)}(n)$ can be decomposed as $A_{(\rm 2)}(n)=b_1(n)B_1+b_2(n)B_2+b_3(n)B_3$, with
\begin{equation}
b_1(n)\equiv\frac{2-n}{2n(n-1)}\,, \qquad b_2(n)\equiv-\frac{\left(\frac{m}{H_{\rm DE}}\right)^2}{3n}-|\xi|\frac{4-3n}{n(n-1)}\,, \qquad b_3(n)\equiv\frac{1}{3n(n-1)}\,,
\end{equation}
and
\begin{equation}
B_1=
\begin{pmatrix}
 0 & 0 & 0 \\
 0 & 1 & 0 \\
 0 & 0 & 2 \\
\end{pmatrix}\,,
\qquad
B_2=
\begin{pmatrix}
 0 & 0 & 0 \\
 2 & 0 & 0 \\
 0 & 1& 0 \\
\end{pmatrix}\,,
\qquad
B_3=
\begin{pmatrix}
 0 & -1 & 0 \\
 0 & 0 & -2 \\
 0 & 0 & 0 \\
\end{pmatrix}\,.
\end{equation}

The exact solution of~(\ref{system_matter+CC}) is

\begin{equation}
\label{eq: 2pt T-ordered}
\Delta_{(\rm 2)}(n,r)=T\exp\left[-\int_{n_{\rm eq}}^n dn'~A_{(\rm 2)}(n')\right]\Delta_{(\rm 2)}(n_{\rm eq},r)\,,
\end{equation}
where $T\exp$ denotes the time-ordered exponential, accounting for the non-commutativity of $B(n)$ matrices for different $n$.

As we will see at the end of this Appendix, in particular in Fig.~\ref{fig: 2pt coinc matter+CC}, neglecting the time ordering only induces errors at a few percent level.
Therefore, before embarking in a refined estimate using another method, let us mention what the result is when time-ordering is not considered.
In this case, as already done for 4-pt functions in~Appendix A.3 of~\cite{Belgacem:2021ieb}, one finds that at leading order in $|\xi|$ and $\left(\frac{m}{H_{\rm DE}}\right)^2$, the 2-pt correlators {\it today} have a relatively simple approximate expression\footnote{This result refers strictly to the current time $N_0=\ln{\left(\frac{\Omega_M}{\Omega_R}\right)}$. At intermediate times $0<N<N_0$ the correlators have more complicated expressions.}
\begin{equation}
\label{sol_mat+CC 2pt}
\begin{pmatrix}
\Delta_{\phi,\phi}(N_0,r)\\
\Delta_{\phi,\pi}(N_0,r)\\
\Delta_{\pi,\pi}(N_0,r)\\
\end{pmatrix}
\simeq 
\begin{pmatrix}
1\\
4\zeta\\
4\zeta^2\\
\end{pmatrix}
e^{4\zeta N_0}~\Delta_{\phi,\phi}(0,r)\,,
\end{equation}
where $\Delta_{\phi,\phi}(0,r)$ is the only 2-pt correlators inherited from radiation epoch (see eq.~(\ref{sol_rad 2pt 2})) and $\zeta$ is a combination of parameters (the same as in eq.~(15) of~\cite{Belgacem:2021ieb}) defined as
\begin{equation}
\label{eq: zeta def}
\zeta\equiv|\xi|\frac{\frac32\frac{\ln\left(\Omega_R\right)}{\ln\left(\Omega_M\right)}-\frac{1}{6|\xi|}\left(\frac{m}{H_{\rm DE}}\right)^2}{\frac32\frac{\ln\left(\Omega_R\right)}{\ln\left(\Omega_M\right)}-1}\,.
\end{equation}

We now propose another strategy to study the evolution of correlators, which circumvents the computation of time-ordered exponentials. We checked numerically that it reproduces the full time-ordered solution~(\ref{eq: 2pt T-ordered}). Finally, Fig.~\ref{fig: 2pt coinc matter+CC} will show that neglecting time-ordering in the exponential~(\ref{eq: 2pt T-ordered}) actually gives a surprisingly accurate result.
\subsection{Alternative solution}

The full quantum field $\hat{\Phi}$ obeys the second order equation of motion~(\ref{eq: EOM Phi 2nd order}). Trading cosmological time $t$ for the number of e-folding $N$ and omitting for simplicity of notation the spacetime dependence of $\hat{\Phi}$,
\begin{equation}
\label{eq: Phi vs N}
\partial_N^2 \hat{\Phi}+\left(3-\epsilon(N)\right)\partial_N \hat{\Phi}
-\frac{\nabla^2}{a^2(N) H^2(N)} \hat{\Phi}+ \frac{M^2(N)}{H^2(N)} \hat{\Phi} = 0 \,.
\end{equation}
As we already did before, for super-Hubble (infrared) modes $k\ll  aH$ we can neglect spatial gradients, which here means dropping the Laplacian term.
In matter+CC epoch we switch from $N$ to the variable $n=\frac23 \epsilon(n)$ as in eqs.~(\ref{eq: eps N})--(\ref{eq: n N}), and then eq.~(\ref{eq: Phi vs N}) becomes
\begin{equation}
\label{eq: Phi evolution matter+CC}
\left[n^2(1-n)^2\partial_n^2-\frac32 n^2(1-n)\partial_n+\left(\frac{m}{3H_{DE}}\right)^2(1-n)-\frac{|\xi|}{3}(4-3n)\right]\hat{\Phi} = 0 \,.
\end{equation}
Notice that, when $N$ increases, $n$ decreases, therefore the goal is to solve the equation~(\ref{eq: Phi evolution matter+CC}) from the initial $n_{\rm eq}=\left(1+\frac{\Omega_\Lambda}{\Omega_M}e^{-3N_0}\right)^{-1}$ close to 1 (but slightly smaller), down to smaller $n$ (our current time corresponding to $n_0=\Omega_M$ and the infinitely far future being $n\to 0$).

One can show that, with a suitable (time-dependent) rescaling of the field $\hat{\Phi}$, eq.~(\ref{eq: Phi evolution matter+CC}) reduces to a hypergeometric equation. The procedure consists in introducing a new variable $\hat{\tilde{\Phi}}$ related to the old $\hat{\Phi}$ by,
\begin{equation}
\label{eq: Phi to Phitilde}
\hat{\Phi}=n^\alpha (1-n)^\beta \hat{\tilde{\Phi}}\,,
\end{equation}
where $\alpha$ and $\beta$ are constants which are chosen in such a way that $\hat{\tilde{\Phi}}$ satisfies a hypergeometric equation. One can then check that, taking $\alpha$ and $\beta$ as solutions of the quadratic equations,
\begin{eqnarray}
\label{eq: sol alpha beta}
\alpha^2-\alpha+\left(\frac{m}{3H_{\rm DE}}\right)^2-\frac{4}{3}|\xi|&=&0\,, \nonumber \\
\beta^2+\frac12\beta-\frac13|\xi|&=&0\,,
\end{eqnarray}
then~(\ref{eq: Phi evolution matter+CC}) simplifies to
\begin{equation}
\label{eq: almost hypergeom}
n(1-n)\partial_n^2\hat{\tilde{\Phi}}+\left[\alpha-\left(\alpha+\beta+\frac32\right)n\right]\partial_n\hat{\tilde{\Phi}}-(\alpha+\beta)\left(\alpha+\beta+\frac12\right)\hat{\tilde{\Phi}}=0\,,\nonumber
\end{equation}
which is a hypergeometric equation.
Note that, under the assumption of light field $m/H_{\rm DE}<1$ that we are using in this work, we are guaranteed that the solutions $\alpha$ and $\beta$ of~(\ref{eq: sol alpha beta}) are real numbers. However we still have to make a choice for the solutions since each of those quadratic equations has two solutions. This will be done in a moment after requiring identification of~(\ref{eq: almost hypergeom}) with the standard form of hypergeometric equation (with variable $n$)
\begin{equation}
\label{eq: standard hypergeom}
n(1-n)\partial_n^2\hat{\tilde{\Phi}}+\left[c-\left(a+b+1\right)n\right]\partial_n\hat{\tilde{\Phi}}-ab~\hat{\tilde{\Phi}}=0\,,
\end{equation}
so that $c=\alpha$, $a+b+1=\alpha+\beta+\frac32$ and $ab=(\alpha+\beta)\left(\alpha+\beta+\frac12\right)$. The solution (up to the irrelevant symmetry between $a$ and $b$ built in~(\ref{eq: standard hypergeom}) itself) is, without loss of generality,
\begin{eqnarray}
\label{eq: abc vs alphabeta}
a&=&\frac12\left[\frac12+\alpha+\beta+\sqrt{\left(\frac12+\alpha+\beta\right)\left(\frac12-3\alpha-3\beta\right)}\right]\,,\nonumber \\ 
b&=&\frac12\left[\frac12+\alpha+\beta-\sqrt{\left(\frac12+\alpha+\beta\right)\left(\frac12-3\alpha-3\beta\right)}\right]\,,\nonumber \\ 
c&=&\alpha\,.
\end{eqnarray}
Note that $a$ and $b$ are real if the quantity under square root appearing in~(\ref{eq: abc vs alphabeta}) is non-negative, i.e. when $-\frac12\leq\alpha+\beta\leq\frac16$. This gives a way to select the solutions for $\alpha$ and $\beta$ in~(\ref{eq: sol alpha beta}). For our purposes, it is sufficient to work at linear order in the small quantities $(m/H_{\rm DE})^2$ and $|\xi|$. It is then straightforward to check that the only choice which complies with the condition $-\frac12\leq\alpha+\beta\leq\frac16$ for small but otherwise arbitrary values of $(m/H_{\rm DE})^2$ and $|\xi|$ is
\begin{eqnarray}
\label{eq: sol alpha beta final}
\alpha&=&\frac12\left(1-\sqrt{1-\left(\frac{2m}{3H_{\rm DE}}\right)^2+\frac{16}{3}|\xi|}\right)\,,\nonumber \\ 
\beta&=&\frac14\left(-1+\sqrt{1+\frac{16}{3}|\xi|}\right)\,.
\end{eqnarray}

\begin{figure}[H]
\centering
\includegraphics[width=0.9\columnwidth]{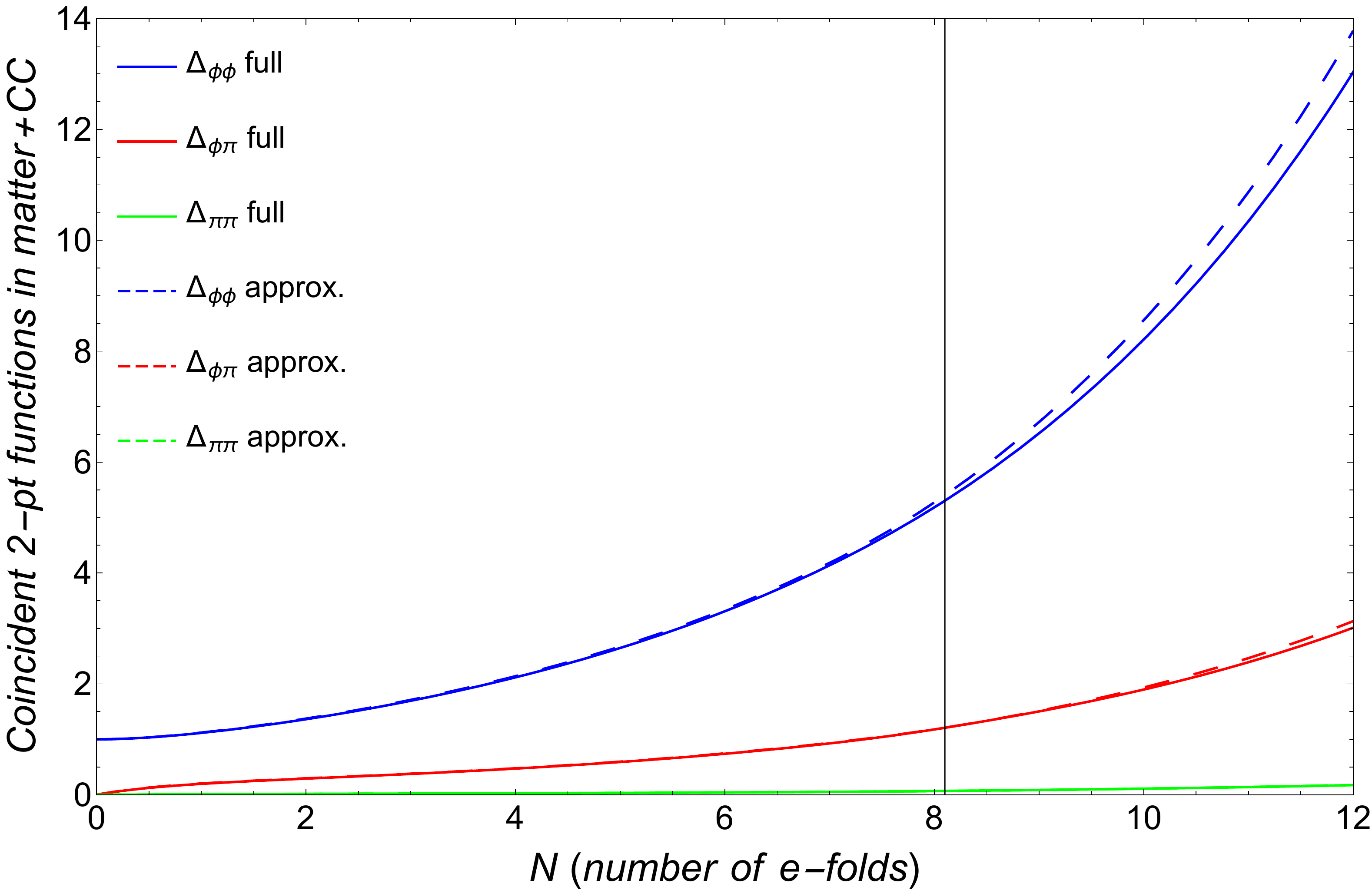}
\caption{The coincident 2-point functions $\Delta_{\phi\phi}$, $\Delta_{\phi\pi}$ and $\Delta_{\pi\pi}$ in the cosmological era dominated by non-relativistic matter and cosmological constant as functions of the number of e-foldings starting from matter-radiation equality. The full lines refer to the results in terms of hypergeometric functions given in~(\ref{eq: 2pt coincident matter+CC}), while the dashed curves are computed by neglecting time-ordering of matrix exponentials in~(\ref{eq: 2pt T-ordered}). The black vertical line marks the current epoch ($N_0\simeq8.1$). The curves are normalized with respect to the initial value of $\Delta_{\phi\phi}$ at matter-radiation equality. The correlators $\Delta_{\phi\pi}$ and $\Delta_{\pi\pi}$ have utterly negligible initial values because of their exponential suppression through radiation epoch.
The values used for the cosmological density fractions today are $\Omega_R=9.1\times10^{-5}$, $\Omega_M=0.3$, $\Omega_\Lambda\simeq1-\Omega_M\simeq0.7$, while for the model we adopt $\xi=-0.06$ and $m/H_{\rm DE}=0.6$. For the largest correlator $\Delta_{\phi\phi}$, the relative difference between the full result and the approximated is less than $2\%$ today and $6\%$ in the future when $N=12$. The difference is even smaller for the less relevant correlators $\Delta_{\phi\pi}$ and $\Delta_{\pi\pi}$.
\label{fig: 2pt coinc matter+CC}
}
\end{figure}

Due to the singular points at $n=0$, $n=1$ and $n\to\infty$ of the hypergeometric differential equation~(\ref{eq: standard hypergeom}), the choice of two linearly independent solutions for it depends on the interval where one has to solve it. For our problem, as already anticipated, the interval of $n$ values needed is $\left[\Omega_M, \left(1+\frac{\Omega_\Lambda}{\Omega_M}e^{-3N_0}\right)^{-1}\right]$ (if  we want to evolve from matter-radiation equality until today) or $\left(0, \left(1+\frac{\Omega_\Lambda}{\Omega_M}e^{-3N_0}\right)^{-1}\right]$ if we also want to extrapolate our predictions to the future (of course the singular point $n=0$ is never reached, since it corresponds to an infinite scale factor). In these intervals (and for non-integer values of $c-a-b$) two independent solutions are known to be $~_2F_1\left(a,b;1+a+b-c;1-n\right)$ and $(1-n)^{c-a-b}~_2F_1\left(c-a,c-b;1+c-a-b;1-n\right)$, where $~_2F_1$ is the Gauss' hypergeometric function. We can simplify some of the coefficients by using~(\ref{eq: abc vs alphabeta}) and then write the solution of~(\ref{eq: standard hypergeom}) as
\begin{equation}
\label{eq: sol phi tilde}
\hat{\tilde{\Phi}}(n,\vec{x})=\hat{A}(\vec{x})~_2F_1\left(a,b;\frac32+\beta;1-n\right)+\hat{B}(\vec{x})(1-n)^{-\frac12-\beta}~_2F_1\left(\alpha-a,\alpha-b;\frac12-\beta;1-n\right)\,.\nonumber
\end{equation}
In the equation above we wrote explicitly the dependence of the field on the spatial (comoving) position $\vec{x}$, which is reflected on the right-hand side by the quantum operators $\hat{A}(\vec{x})$ and $\hat{B}(\vec{x})$ (time-independent).
Thus, taking aso into account the factors in~(\ref{eq: Phi to Phitilde}) and the relation~(\ref{eq: abc vs alphabeta}), we find
\begin{eqnarray}
\label{eq: sol phi}
\hat{\Phi}(n,\vec{x})=F_A(n)\hat{A}(\vec{x})+F_B(n)\hat{B}(\vec{x})\,,
\end{eqnarray}
where we introduced a compact notation for the time-dependent functions
\begin{eqnarray}
\label{eq: hypergeomAB}
F_A(n)&=&n^\alpha(1-n)^\beta~_2F_1\left(a,b;\frac32+\beta;1-n\right)\,, \nonumber \\
F_B(n)&=&n^\alpha(1-n)^{-\frac12}~_2F_1\left(\alpha-a,\alpha-b;\frac12-\beta;1-n\right)\,,
\end{eqnarray}
with $a$ and $b$ determined by $\alpha$ and $\beta$ according to eq.~(\ref{eq: abc vs alphabeta}).
We can also write the corresponding solution for the canonical momentum operator $\hat{\Pi}(n,\vec{x})$ by looking at eq.~(\ref{FullEOM1}) and transforming from the cosmological time $t$ variable to the $x$ variable to get for the rescaled momentum $\hat{\Pi}/(a^3 H)$ relevant for our correlators
\begin{equation}
\frac{\hat{\Pi}(n,\vec{x})}{a^3(n) H(n)} = -3 n(1-n)~\partial_n \hat{\Phi}(n,\vec{x}) \,.
\label{eq: sol pi 1}
\end{equation}
Upon inserting~(\ref{eq: sol phi}) into~(\ref{eq: sol pi 1}) and after taking derivative of the hypergeometric functions, we obtain
\begin{equation}
\frac{\hat{\Pi}(n,\vec{x})}{a^3(n) H(n)} = G_A(n)\hat{A}(\vec{x})+G_B(n)\hat{B}(\vec{x}) \,,
\label{eq: sol pi 2}
\end{equation}
where the functions $G_A(n)$ and $G_B(n)$ are defined as
\begin{eqnarray}
\label{eq: hypergeomAB}
G_A(n)&=& -3 n^{\alpha+1}(1-n)^{\beta+1}\left[\left(\frac{\alpha}{n}-\frac{\beta}{1-n}\right)~_2F_1\left(a,b;\frac32+\beta;1-n\right)\right.\,\nonumber \\
&&\left.-\frac{(\alpha+\beta)(\alpha+\beta+\frac12)}{\frac32+\beta}~_2F_1\left(a+1,b+1;\frac52+\beta;1-n\right)\right]\,, \nonumber \\
G_B(n)&=&-3 n^{\alpha+1}(1-n)^{\frac12}\left[\left(\frac{\alpha}{n}+\frac{1/2}{1-n}\right)~_2F_1\left(\alpha-a,\alpha-b;\frac12-\beta;1-n\right)\right.\, \nonumber \\
&&\left.-\frac{\alpha^2+\beta(\alpha+\beta+\frac12)}{\frac12-\beta}~_2F_1\left(\alpha-a+1,\alpha-b+1;\frac32-\beta;1-n\right)\right]\,.
\end{eqnarray}
We want to use these results to evaluate the coincident 2-pt functions $\Delta_{\phi\phi}$, $\Delta_{\phi\pi}$ and $\Delta_{\pi\pi}$ defined in (\ref{DeltaPhiPhi}--\ref{DeltaPiPi}). We begin by observing that eqs.~(\ref{eq: sol phi}) and~(\ref{eq: sol pi 2}) imply
\begin{equation}
\begin{pmatrix}
\Delta_{\phi\phi}(n)\\
\Delta_{\phi\pi}(n)\\
\Delta_{\pi\pi}(n)\\
\end{pmatrix}
=\mathcal{M}(n)
\begin{pmatrix}
\langle\hat{A}^2\rangle \\
\langle\{\hat{A}, \hat{B}\}\rangle \\
\langle\hat{B}^2\rangle \\
\end{pmatrix}\,,
\end{equation}
where the matrix $\mathcal{M}(n)$ is defined as
\begin{equation}
\label{M matrix in matter+CC}
\mathcal{M}(n)=
\begin{pmatrix}
F_A^2(n) &  F_A(n)F_B(n) & F_B^2(n) \\
2F_A(n)G_A(n) &  F_A(n)G_B(n)+G_A(n)F_B(n) & 2F_B(n)G_B(n) \\
G_A^2(n) &  G_A(n)G_B(n) &G_B^2(n) \\
\end{pmatrix}\,.
\end{equation}
The constant correlators $\langle\hat{A}^2\rangle$, $\langle\{\hat{A}, \hat{B}\}\rangle$ and $\langle\hat{B}^2\rangle$ (which are also space-independent due to the homogeneity of the FLRW background) are determined by the values of $\Delta_{\phi\phi}(n_{\rm eq})$, $\Delta_{\phi\pi}(n_{\rm eq})$ and $\Delta_{\pi\pi}(n_{\rm eq})$ at matter-radiation equality inherited from radiation epoch, so that we conclude
\begin{equation}
\label{eq: 2pt coincident matter+CC}
\begin{pmatrix}
\Delta_{\phi\phi}(n)\\
\Delta_{\phi\pi}(n)\\
\Delta_{\pi\pi}(n)\\
\end{pmatrix}
=\mathcal{M}(n)\mathcal{M}^{-1}(n_{\rm eq})
\begin{pmatrix}
\Delta_{\phi\phi}(n_{\rm eq})\\
\Delta_{\phi\pi}(n_{\rm eq})\\
\Delta_{\pi\pi}(n_{\rm eq})\\
\end{pmatrix}\,.
\end{equation}
The same equation also holds for non-coincident correlators as one can see by considering fields at two comoving positions $\vec{x}$ and $\vec{y}$. As we know, homogeneity and isotropy of the FLRW background imply that the spatial dependence of these correlators only appears through the relative distance $r=\|\vec{x}-\vec{y}\|$. Then from eqs.~(\ref{eq: sol phi}) and~(\ref{eq: sol pi 2}) it follows that
\begin{equation}
\begin{pmatrix}
\Delta_{\phi\phi}(n, r)\\
\Delta_{\phi\pi}(n, r)\\
\Delta_{\pi\pi}(n, r)\\
\end{pmatrix}
=\mathcal{M}(n)
\begin{pmatrix}
\langle\hat{A}(\vec{x})\hat{A}(\vec{y})\rangle \\
\langle\hat{A}(\vec{x})\hat{B}(\vec{y})+\hat{B}(\vec{x})\hat{A}(\vec{y})\rangle \\
\langle\hat{B}(\vec{x})\hat{B}(\vec{y})\rangle \\
\end{pmatrix}\,,
\end{equation}
where $\mathcal{M}(n)$ is the same matrix defined in~(\ref{M matrix in matter+CC}).
In terms of initial conditions at matter-radiation equality
\begin{equation}
\begin{pmatrix}
\Delta_{\phi\phi}(n, r)\\
\Delta_{\phi\pi}(n, r)\\
\Delta_{\pi\pi}(n, r)\\
\end{pmatrix}
=\mathcal{M}(n)\mathcal{M}^{-1}(n_{\rm eq})
\begin{pmatrix}
\Delta_{\phi\phi}(n_{\rm eq}, r)\\
\Delta_{\phi\pi}(n_{\rm eq}, r)\\
\Delta_{\pi\pi}(n_{\rm eq}, r)\\
\end{pmatrix}\,,
\end{equation}
which shows the same time evolution as in~(\ref{eq: 2pt coincident matter+CC}).
Fig.~\ref{fig: 2pt coinc matter+CC} shows the comparison between the full result for the coincident correlators~(\ref{eq: 2pt coincident matter+CC}) computed using the technique explained here against the approximation of neglecting time-ordering in the matrix exponentials of~(\ref{eq: 2pt T-ordered}). The values of the cosmological parameters chosen are given in the caption of the figure. For the most relevant correlator $\Delta_{\phi\phi}(N)$ the difference between the two methods at the current epoch ($N_0\simeq8.1$ e-folds since matter-radiation equality) is less than $2\%$ level. Thus the approximation without time-ordering works quite well, at the few percent level. The deviation grows to $6\%$ in the future at $N=12$.

\section{Effect of non-zero initial conditions in inflation}
\label{app: non-zero ICs}

Here we investigate the consequences of initial conditions for the inflationary era inherited from a pre-inflationary epoch. 
If $P_\phi(t_{\rm in},k)$ is the power spectrum of the field at the beginning of inflation inherited from the pre-inflationary epoch (where $t_{\rm in}$ is cosmological time at the beginning of inflation and $k$ is a comoving wavenumber), then its 2-point function is
\begin{equation}
\label{2-pt correlator pre-inflation}
\Delta_{\phi\phi}(t_{\rm in},r)=\frac{1}{2\pi^2}\int_{k_0}^{\mu a_{\rm in}H_I}\frac{dk}{k}j_0(kr)P_\phi(t,k)\,.
\end{equation}

Let us suppose that the field power spectrum follows a distribution of the form
\begin{equation}
\label{power spectrum pre-inflation}
P_\phi(t_{\rm in},k)=\left(\frac{k}{a_{\rm in}}\right)^2\left[n_0\left(\frac{a_{\rm in}T_{\rm in}}{k}\right)^\sigma+\frac12\right]\,.
\end{equation}
The equation above is obtained naturally by supposing that the average occupation number for a mode with momentum $k$ follows a thermal-like distribution $\bar{n}_k=n_0\left(\frac{a_{\rm in}T_{\rm in}}{k}\right)^\sigma$, where we allow for a generic power $\sigma$ and $T_{\rm in}$ is a generic mass scale parameter that plays the role of a temperature (but it is not necessarily the physical temperature).

The term $1/2$ in eq.~(\ref{power spectrum pre-inflation}) is the zero point energy. Notice that, apart from this $1/2$ addend, the case $\sigma=2$ corresponds to a scale-invariant spectrum.

Using~(\ref{power spectrum pre-inflation}) into~(\ref{2-pt correlator pre-inflation}) and approximating the Bessel function as a Heaviside $\theta$ function, we get for $\sigma<2$ that
\begin{equation}
\label{2-pt correlator pre-inflation:2}
\Delta_{\phi,\phi}(t_{\rm in},r)=\frac{(\mu H_I)^2}{2\pi^2}d^2(r)\left[\frac{n_0}{2-\sigma}\left(\frac{T_{\rm in}}{\mu H_I d(r)}\right)^\sigma+\frac14\right]\,.
\end{equation}
The spatial dependence has been encoded in the function $d(r)$ given by
\begin{equation}
\label{spatial 2-pt}
d(r)\simeq 
\begin{cases}
      1 & \text{if $ r< \frac{1}{\mu a_{\rm in}H_I}$} \\
      \frac{1}{\mu a_{\rm in}H_I r} & \text{if $\frac{1}{\mu a_{\rm in} H_I}<r\lesssim\frac{1}{k_0}$}\,,
\end{cases} 
\end{equation}
where we have truncated distances up to $1/k_0$ since $k_0$ is the IR cutoff.
We want to follow the evolution of 2-pt correlators during inflation.

As for the initial conditions of $\Delta_{\phi,\pi}$ and $\Delta_{\pi,\pi}$ we suppose $\Delta_{\phi,\pi}(t_{\rm in},r)=0$ and $\Delta_{\pi,\pi}(t_{\rm in},r)=R\Delta_{\phi,\phi}(t_{\rm in},r)$, where we introduced a parameter $R$ as the ratio between the amplitude of the initial power spectrum of the canonical momentum $\hat{\pi}$, compared to that of the field $\hat{\phi}$.

Evolving these initial conditions, at the end of a phase of de Sitter inflation, the (coincident for simplicity) 2-point functions $\Delta_{(\rm 2)}=\left(\Delta_{\phi,\phi}, \Delta_{\phi,\pi}, \Delta_{\pi,\pi}\right)$ are in row vector form
\begin{equation}
\label{non-zero ICs: inflation 2pt}
\Delta_{(\rm 2)}^{\rm T}(N_I)=\left(1+\frac{R}{9}\right)\frac{(\mu H_I)^2}{2\pi^2}e^{8|\xi|N_I}\left[n_0\left(\frac{T_{in}}{\mu H_I}\right)^\sigma\left(\frac{1}{2-\sigma}+\frac{1}{\tau+8|\xi|}\right)+\frac14+\frac{1}{16|\xi|}\right]\left(1, 8|\xi|, 16\xi^2\right)\,,
\end{equation}
where the parameter $\tau$ includes both the possibilities of the ``temperature-like" parameter $T$ being constant during inflation (due to some energy refilling akin to~\cite{Berera:1995ie}) in which case $\tau=0$, or a more natural decreasing $T\propto 1/a$ (when no significant particle production occurs) in which case $\tau=\sigma$.

Comparing eq.~(\ref{non-zero ICs: inflation 2pt}) with the result~(\ref{sol_deSitter 2pt 5}) (to be taken at coincidence $r=0$ and with $c_s=1$) obtained in inflation with zero initial conditions, we see that in the new case (with non-zero initial conditions) correlators increase by a factor of
\begin{equation}
\label{ratio new old}
\frac{\Delta_2^{({\rm new})}}{\Delta_2^{({\rm old})}}=\left(1+\frac{R}{9}\right)\mu^2\left[1+4|\xi|+16|\xi|n_0\left(\frac{T_{in}}{\mu H_I}\right)^\sigma\left(\frac{1}{2-\sigma}+\frac{1}{\tau+8|\xi|}\right)\right]\,.
\end{equation}

\newpage

\end{document}